\documentstyle[12pt]{article}
\input epsfig

\begin{document}
\baselineskip=20.5pt
\def\beqra{\begin{eqnarray}} \def\eeqra{\end{eqnarray}}
\def\beqast{\begin{eqnarray*}} \def\eeqast{\end{eqnarray*}}
\def\beq{\begin{equation}}      \def\eeq{\end{equation}}
\def\be{\begin{enumerate}}   \def\ee{\end{enumerate}}

\def\fnote#1#2{\begingroup\def\thefootnote{#1}\footnote{#2}\addtocounter
{footnote}{-1}\endgroup}



\def\gam{\gamma}
\def\Gam{\Gamma}
\def\la{\lambda}
\def\eps{\epsilon}
\def\La{\Lambda}
\def\si{\sigma}
\def\Si{\Sigma}
\def\al{\alpha}
\def\Th{\Theta}
\def\th{\theta}
\def\tnu{\tilde\nu}
\def\vphi{\varphi}
\def\del{\delta}
\def\Del{\Delta}
\def\ab{\alpha\beta}
\def\om{\omega}
\def\Om{\Omega}
\def\mn{\mu\nu}
\def\mun{^{\mu}{}_{\nu}}
\def\kap{\kappa}
\def\rsi{\rho\sigma}
\def\beal{\beta\alpha}
\def\til{\tilde}
\def\rta{\rightarrow}
\def\eqv{\equiv}
\def\nab{\nabla}
\def\pa{\partial}
\def\sit{\tilde\sigma}
\def\ul{\underline}
\def\indt{\parindent2.5em}
\def\nd{\noindent}
\def\rsi{\rho\sigma}
\def\beal{\beta\alpha}
\def\caa{{\cal A}}
\def\cb{{\cal B}}
\def\cac{{\cal C}}
\def\cd{{\cal D}}
\def\ce{{\cal E}}
\def\cf{{\cal F}}
\def\cg{{\cal G}}
\def\cah{{\cal H}}
\def\ci{{\cal I}}
\def\cj{{\cal{J}}}
\def\ck{{\cal K}}
\def\cl{{\cal L}}
\def\cm{{\cal M}}
\def\cn{{\cal N}}
\def\cO{{\cal O}}
\def\cp{{\cal P}}
\def\car{{\cal R}}
\def\cs{{\cal S}}
\def\ct{{\cal{T}}}
\def\cu{{\cal{U}}}
\def\cv{{\cal{V}}}
\def\cw{{\cal{W}}}
\def\cx{{\cal{X}}}
\def\cy{{\cal{Y}}}
\def\cz{{\cal{Z}}}
\def\asymptotic{{_{\stackrel{\displaystyle\longrightarrow}
{x\rightarrow\pm\infty}}\,\, }} 
\def\asymptext{\raisebox{.6ex}{${_{\stackrel{\displaystyle\longrightarrow}
{x\rightarrow\pm\infty}}\,\, }$}} 
\def\epsilim{{_{\textstyle{\rm lim}}\atop
_{~~~\epsilon\rightarrow 0+}\,\, }} 
\def\omegalim{{_{\textstyle{\rm lim}}\atop
_{~~~\om^2\rightarrow 0+}\,\, }} 
\def\xlimp{{_{\textstyle{\rm lim}}\atop
_{~~x\rightarrow \infty}\,\, }} 
\def\xlimm{{_{\textstyle{\rm lim}}\atop
_{~~~x\rightarrow -\infty}\,\, }} 
\def\asymptoticp{{_{\stackrel{\displaystyle\longrightarrow}
{x\rightarrow +\infty}}\,\, }} 
\def\asymptoticm{{_{\stackrel{\displaystyle\longrightarrow}
{x\rightarrow -\infty}}\,\, }} 

\def\raisenot{\raise .5mm\hbox{/}}
\def\nota{\ \hbox{{$a$}\kern-.49em\hbox{/}}}
\def\notA{\hbox{{$A$}\kern-.54em\hbox{\raisenot}}}
\def\notb{\ \hbox{{$b$}\kern-.47em\hbox{/}}}
\def\notB{\ \hbox{{$B$}\kern-.60em\hbox{\raisenot}}}
\def\notc{\ \hbox{{$c$}\kern-.45em\hbox{/}}}
\def\notd{\ \hbox{{$d$}\kern-.53em\hbox{/}}}
\def\notbd{\ \hbox{{$D$}\kern-.61em\hbox{\raisenot}}} 
\def\note{\ \hbox{{$e$}\kern-.47em\hbox{/}}}
\def\notk{\ \hbox{{$k$}\kern-.51em\hbox{/}}}
\def\notp{\ \hbox{{$p$}\kern-.43em\hbox{/}}}
\def\notq{\ \hbox{{$q$}\kern-.47em\hbox{/}}}
\def\notW{\ \hbox{{$W$}\kern-.75em\hbox{\raisenot}}}
\def\notz{\ \hbox{{$Z$}\kern-.61em\hbox{\raisenot}}}
\def\notpa{\hbox{{$\partial$}\kern-.54em\hbox{\raisenot}}}
\def\fo{\hbox{{1}\kern-.25em\hbox{l}}}  
\def\rf#1{$^{#1}$}
\def\bx{\Box}
\def\tr{{\rm Tr}}
\def\rmtr{{\rm tr}}
\def\dgg{\dagger}
\def\lag{\langle}
\def\rag{\rangle}
\def\bmid{\big|}
\def\vlap{\overrightarrow{\La p}} 
\def\lrta{\longrightarrow} \def\lrar{\raisebox{.8ex}{$\longrightarrow$}}
\def\ON{{\cal O}(N)}
\def\UN{{\cal U}(N)}
\def\bdPh{\mbox{\boldmath{$\dot{\!\Phi}$}}}
\def\bPh{\mbox{\boldmath{$\Phi$}}}
\def\bPhs{\bPh^2}
\def\sef{S_{eff}[\sigma,\pi]}
\def\sigx{\sigma(x)}
\def\pix{\pi(x)}
\def\bph{\mbox{\boldmath{$\phi$}}}
\def\bphs{\bph^2}
\def\ex{\BM{x}}
\def\exs{\ex^2}
\def\xdot{\dot{\!\ex}}
\def\y{\BM{y}}
\def\ys{\y^2}
\def\ydot{\dot{\!\y}}
\def\pat{\pa_t}
\def\pax{\pa_x}
\def\hp{{\pi\over 2}}

\renewcommand{\thesection}{\arabic{section}}
\renewcommand{\theequation}{\thesection.\arabic{equation}}

\begin{flushright}
\end{flushright}

\vspace*{.1in}
\begin{center}
  \Large{\sc All about the Static Fermion Bags in the Gross-Neveu Model}
\normalsize

\vspace{36pt}
{\large Joshua Feinberg\fnote{*}{{\it e-mail: joshua@physics.technion.ac.il}}}\\
\vspace{12pt}
 {\small \em Department of Physics,}\\ 
{\small \em Oranim-University of Haifa, Tivon 36006, Israel}\fnote{**}{permanent address}\\
{\small and}\\
{\small \em Department of Physics,}\\
{\small \em Technion - Israel Institute of Technology, Haifa 32000 Israel}

\vspace{.6cm}

\end{center}

\begin{minipage}{5.8in}
{\abstract~~~~~
We review in detail the construction of {\em all} stable static fermion 
bags in the $1+1$ dimensional Gross-Neveu model with 
$N$ flavors of Dirac fermions, in the large $N$ limit. In addition to the 
well known kink and topologically trivial solitons (which correspond, 
respectively, to the spinor and antisymmetric tensor representations of 
$O(2N)$), there are also threshold bound states of a kink and a 
topologically trivial soliton: the heavier topological solitons (HTS). 
The mass of any of these newly discovered HTS's is the sum of masses 
of its solitonic constituents, and it corresponds to the tensor product 
of their $O(2N)$ representations. Thus, it is marginally 
stable (at least in the large $N$ limit). Furthermore, its mass is 
independent of the distance between the centers of its constituents, which 
serves as a flat collective coordinate, or a modulus. There are no 
additional stable static solitons in the Gross-Neveu model. We provide 
detailed derivation of the profiles, masses and fermion number contents of 
these static solitons. For pedagogical clarity, and in order for this paper 
to be self-contained, we also included detailed appendices on 
supersymmetric quantum mechanics and on reflectionless potentials in one 
spatial dimension, which are intimately related with the theory of static 
fermion bags. In particular, we present a novel simple explicit formula for 
the diagonal resolvent of a reflectionless Schr\"odinger operator with an 
arbitrary number of bound states. In additional appendices we summarize the 
relevant group representation theoretic facts, and also provide a simple 
calculation of the mass of the kinks.}

\end{minipage}

\vspace{28pt}

\vfill
\pagebreak

\setcounter{page}{1}

\section{Introduction}

Many years ago, Dashen, Hasslacher and Neveu (DHN) \cite{dhn}, and following 
them Shei \cite{shei}, used inverse scattering analysis \cite{faddeev} to find 
static fermion-bag \cite{sphericalbag, shellbag} soliton solutions to the 
large-$N$ saddle point equations of the Gross-Neveu (GN) \cite{gn} and of the 
$1+1$ dimensional, multi-flavor Nambu-Jona-Lasinio (NJL) \cite{njl} models.
In the GN model, with its discrete chiral symmetry, a topological soliton,
the so called Callan-Coleman-Gross-Zee (CCGZ) kink \cite{ccgz}, was 
discovered prior to the work of DHN. In this paper we will concentrate 
exclusively on the GN model.

One version of writing the action of the $1+1$ dimensional GN model is 
\beq
S=\int d^2x\,\left\{\sum_{a=1}^N\, \bar\psi_a\,\left(i\notpa-\si
\right)\,\psi_a 
-{1\over 2g^2}\,\si^2\right\}\,,
\label{auxiliary}
\eeq
where the $\psi_a\,(a=1,\ldots,N)$ are $N$ flavors of massless Dirac 
fermions, with Yukawa coupling to the scalar\footnote{The fermion bag 
solitons in these models arise, as is well known, at the level of the 
effective action, after integrating the fermions out, and not at the level 
of the action (\ref{auxiliary}).} auxiliary field $\si(t,x)$.

An obvious symmetry of (\ref{auxiliary}) with its $N$ Dirac spinors is $U(N)$.
Actually, (\ref{auxiliary}) is symmetric under the larger group $O (2N)$. 
It is easy to see this\cite{dhn} in a 
concrete representation for $\gamma$ matrices, which we choose as 
the Majorana representation  
\beq\label{majorana}
\gam^0=\si_2\;,\; \gam^1=i\si_3\quad {\rm and} \quad 
\gam^5=-\gam^0\gam^1=\si_1\,.
\eeq
(Henceforth, in this paper we will use this representation for  
$\gam$ matrices in all explicit calculations.)

In order to expose the $O (2N)$ symmetry, we write each Dirac spinor as
\beq\label{majoranaspinor}
\psi_a=\phi_a+i\chi_a
\eeq
where $\phi_a$ and $\chi_a$ are hermitean 
two-component Majorana spinors.
In terms of (\ref{majorana}) and (\ref{majoranaspinor}) we can write
the lagrangian in (\ref{auxiliary}) (up to surface terms) 
as\footnote{Here we use the common notation where an overdot stands for a 
time derivative and a prime stands for a spatial derivative.} 
\beqra\label{explicitsymmetry}
\cl =\sum_{a=1}^N\left[i(\phi_a^T\dot\phi_a + 
\chi_a^T\dot\chi_a) -i(\phi_a^T\si_1\phi_a' + 
\chi_a^T\si_1\chi_a')-\si\,(\phi_a^T\si_2\phi_a + 
\chi_a^T\si_2\chi_a)\right]-{\si^2\over 2g^2}
\eeqra
which is hermitean and non-vanishing, because all spinors are Grassmann 
valued. Evidently, (\ref{explicitsymmetry}) is invariant under orthogonal 
transformations of the $2N$ Majorana spinors $\phi_a$ and $\chi_a$. 

The fact that the symmetry group of (\ref{auxiliary}) is $O (2N)$ rather than 
$ U(N)$, indicates that (\ref{auxiliary}) is invariant against 
charge-conjugation. In the representation (\ref{majorana}), charge-conjugation
is realized simply by complex conjugation of the spinor\footnote{The matrices 
$\gam^0, \gam^1$ are pure imaginary. Thus by taking the complex conjugate 
of the Dirac equation $\left[i\notpa - e\notA-\si (x)\right]\,\psi = 0$
we conclude that $\left[i\notpa + e\notA-\si (x)\right]\,\psi^* = 0$.}
\beq\label{chargeconjugation}
\psi^c(x) = \psi^*(x)\,.
\eeq
Thus, if $\psi = e^{-i\om t} u(x)$ is an eigenstate of the Dirac equation 
\beq\label{diraceq}
\left[i\notpa-\si (x)\right]\,\psi = 0\,,
\eeq
with energy $\om$, then $\psi^*(x) = e^{i\om t} u^*(x)$ is an energy 
eigenstate of (\ref{diraceq}), with 
energy $-\om$. Therefore, the GN model (\ref{auxiliary}) is invariant against 
charge conjugation, and energy eigenstates of (\ref{diraceq}) come in 
$\pm\om$ pairs.

The remarkable discovery DHN made was that all self-consistent static bag 
configurations in the GN model were {\em reflectionless}. That is, the 
static $\sigx$'s that solve the saddle point equations of the GN model are 
such  that the reflection coefficient of the Dirac equation (\ref{diraceq})
vanishes identically for {\bf all} momenta. In other words, a 
fermion wave packet impinging on one side of the potential $\si (x)$ will be 
totally transmitted (up to phase shifts, of course).

We note in passing that besides their role in soliton theory \cite{faddeev}, 
reflectionless potentials appear in other diverse areas of theoretical 
physics \cite{rosner, blackhole, zwiebach, jg}. For a review, which discusses 
reflectionless potentials (among other things) in the context of 
supersymmetric quantum mechanics, see \cite{cooper}.

Since the works of DHN and of Shei, these fermion bags were discussed in the 
literature several other times, using alternative methods \cite{others}. 
For a recent review on these and related matters (with an emphasis on 
the relativistic Hartree-Fock approximation), see \cite{thies}.

In many of these treatments, one solves the variational, saddle point
equations by performing mode summations over energies and phase shifts. 
An alternative to such summations is to solve the saddle point equations 
by manipulating the resolvent of the Dirac operator as a whole, with the 
help of basic tools of Sturm-Liouville operator theory. The resolvent 
of the Dirac operator takes care of mode summation automatically.

Some time ago, such an alternative to mode summation techniques 
was developed \cite{josh1, josh2}, which was based on the Gel'fand-Dikii 
(GD) identity \cite{gd} (an identity obeyed by the diagonal resolvent of 
one-dimensional Schr\"odinger operators, see e.g. (\ref{dikii}))
\footnote{For a simple derivation of the GD identity, see \cite{josh2, fz}.}, 
to study fermion bags in the 
GN model\cite{josh1} as well as other problems \cite{josh2}. In a nut-shell, 
the method of \cite{josh1,josh2} is based on the fact that in certain models, 
the explicit form of the saddle-point equations for one-dimensional (or quasi 
one-dimensional) static, space-dependent field condensates, suggests
certain parameter dependent ansatzes for the diagonal resolvent of the 
Dirac (or Klein-Gordon) operator. This explicit construction of the 
diagonal resolvent is based on the GD identity as well as on simple 
dimensional analysis. All subsequent manipulations with these expressions 
involve the space dependent condensates directly, and given such an 
ansatz for the diagonal resolvent, one can construct a static 
space-dependent solution of the saddle-point equations in a straightforward 
manner, bypassing the need to work with the scattering data and the so 
called trace identities that relate them to the space dependent condensates.

That method was applied in \cite{josh1, fz, reflectionless} to study static 
fermion bags in the GN and NJL models and reproduced the 
static bag results of DHN and of Shei. It was also used in \cite{massivegn} 
to make approximate variational calculations of static fermion bags in the 
massive GN model \cite{massivegn}. Similar ideas were later used in 
\cite{stone} to calculate the free energy of inhomogeneous superconductors.

In this paper we study the entire spectrum of stable static fermion bags in 
the GN model in the large $N$ limit, thus extending the ``static part'' of 
the work\cite{dhn} of DHN. We find that in addition to the well known kink and 
topologically trivial solitons, which correspond, respectively, to the spinor
and antisymmetric tensor representations of $O(2N)$, the GN model bears also 
threshold bound states of a kink and a topologically trivial soliton - the
heavier topological solitons (HTS)\cite{heavykink}. The mass of any of 
these newly discovered HTS's is the sum of masses of its solitonic 
constituents, and it corresponds to the tensor 
product of their $O(2N)$ representations. Thus it is 
marginally stable (at least in the large $N$ limit). Furthermore, the mass 
of an HTS is independent of the distance between the centers of its 
constituents, which thus serves as a flat collective coordinate, or a 
modulus. There are no additional stable {\em static} solitons in the 
Gross-Neveu model.

The rest of the paper is organized as follows: In Section 2 we recall some 
basic facts about the GN model and its dynamics. In particular, we define 
the effective action $S_{eff}[\si]$, obtained from (\ref{auxiliary}) after 
integrating the fermions out, and derive the generic saddle-point equation 
for $\si (x,t)$. 

In Section 3 we focus on the simpler problem 
of finding extremal static $\sigx$ configurations, and as we mentioned 
earlier, DHN have already shown\cite{dhn} that the extremal static 
configurations are necessarily reflectionless. 
Our analysis employs extensively one dimensional supersymmetric quantum 
mechanics and the theory of reflectionless potentials.
We apply these formalisms to construct a generic static 
reflectionless background $\sigx$ (subjected to the appropriate boundary 
conditions at $x=\pm\infty$), which supports a given number of bound 
states of the Dirac equation (\ref{diraceq}) at some given energies, and 
the diagonal resolvent of the Dirac operator in 
(\ref{diraceq}) associated with it. We then use 
this explicit resolvent to calculate the energy functional associated with 
this reflectionless background, as well as its fermion number spectrum. A 
reflectionless background $\sigx$ depends on a 
finite set of real parameters, and the effective action evaluated at such 
a $\sigx$ configuration is an ordinary function of these parameters. By 
solving the ordinary extremum problem for this function we identify all static 
extremal fermion bags in the GN model. 

After listing all extremal static $\sigx$ configurations in Section 3, we 
identify, in Section 4, the {\em stable} configurations, namely, those 
extremal bags which are protected by the conservation of 
topological charge and $O(2N)$ quantum numbers against 
decaying into lighter fermion bags. These are the well 
known CCGZ kink and DHN solitons, and in addition, our newly discovered 
marginally stable HTS.

Some general background material, as well as many technical details are 
relegated to the Appendices. In Appendix A we 
discuss (in some detail) the general properties of the resolvent of the Dirac 
operator in a static $\sigx$ background and its relation with 
supersymmetric quantum mechanics. As is well known, the Dirac equation 
in any such background is equivalent to a pair of two isospectral 
Schr\"odinger equations in one dimension, namely, a realization of 
one-dimensional supersymmetric quantum mechanics. We use this underlying 
supersymmetry to express all four entries of the space-diagonal Dirac 
resolvent (i.e., the resolvent evaluated at coincident spatial coordinates) 
in terms of a single function: the diagonal resolvent of one of the two 
isospectral Schr\"odinger operators. Furthermore, using the underlying 
supersymmetry of the Dirac equation, we prove that the expectation 
value of the spatial component of the fermion current operator in the static 
$\sigx$ background vanishes identically, as should be expected on physical 
grounds. We also prove an identity relating the expectation values of the 
fermion density and the pseudoscalar density operators in the $\sigx$ 
background, and show that it has a simple interpretation in terms of 
bosonization. This appendix follows, in part, our recent discussion in 
\cite{reflectionless}.

In Appendix B we summarize some useful 
properties of reflectionless Schr\"odinger hamiltonians and their resolvents. 
Following that summary of results from the literature, we use them 
to derive an explicit simple representation for the diagonal resolvent of a 
reflectionless Schr\"odinger operator with a prescribed number of bound 
states, which we have not encountered in the literature. We then 
use the formalism reviewed in that appendix to derive an explicit 
formula for a $\sigx$ with a prescribed set of bound states. We also work out
explicitly the cases of backgrounds with a single bound state and with 
two bound states, which are the cases most relevant to this work.

Appendix C contains some technical details concerning the derivation in 
Section 2 of the saddle point equation in the static case.

In Appendix D we discuss the quantization of a multiplet of 
$2N$ Majorana fields in a given static $\sigx$ background and the related 
$O(2N)$ multiplet structure of bound states of fermions trapped in that 
background. We show there that each bound state in the corresponding 
Dirac-Majorana equation, at non-zero frequency, gives rise to an irreducible
factor of an $O(2N)$ antisymmetric tensor, and that the zero mode, which 
exists if and only if $\sigx$ has non-trivial topology, gives rise to a 
single factor of the spinor representation. Also included in that appendix 
are the necessary facts concerning the spinorial and antisymmetric tensor 
representations of $O(2N)$.

In Appendix E, we give a simple derivation of the mass formula
of the CCGZ kink. Finally, in Appendix F, we present an alternative proof
of Eq. (\ref{potentiallystable}), which is the basis for the complete 
classification of stable static solitons.

\pagebreak

\section{Dynamics of the Gross-Neveu Model}
\setcounter{equation}{0}

We now recall some basic facts about the dynamics of the GN model.
The partition function associated with (\ref{auxiliary}) 
is\footnote{From this point to the end of the 
paper flavor indices are usually suppressed.}
\beq
\cz=\int\,\cd\si\,\cd\bar\psi\,\cd\psi \,\exp i\, 
\int\,d^2x\left\{\bar\psi
\left(i\notpa-\si\right)\psi-{1\over 
2g^2}\,\si^2\right\}
\label{partition}
\eeq
Integrating over the grassmannian variables leads to 
$\cz=\int\,\cd\si\,\exp \{iS_{eff}[\si]\}$
where the bare effective action is
\beq
S_{eff}[\si] =-{1\over 2g^2}\int\, d^2x 
\,\si^2-iN\, 
\tr\log\left(i\notpa-\si\right)
\label{effective}
\eeq
and the trace is taken over both functional and Dirac indices.

The theory (\ref{effective}) has been studied in the limit 
$N\rightarrow\infty$ with $Ng^2$ held fixed\cite{gn}. In this limit 
(\ref{partition}) is governed by saddle points of (\ref{effective}) 
and the small fluctuations around them. (In this paper, as in \cite{dhn}, we 
will consider only the leading term in the $1/N$ expansion, and thus 
{\em will not} compute the effect of the fluctuations around the saddle 
points.) The most general saddle point condition reads
\beqra
{\del S_{\em eff}\over \del \si\left(x,t\right)}  &=&
-{\si\left(x,t\right)\over g^2} + iN ~{\rm tr} \left[\langle x,t | 
{1\over i\notpa
-\si} | x,t \rangle \right]= 0\,.
\label{saddle}
\eeqra

In particular, the non-perturbative vacuum of (\ref{auxiliary}) is 
governed by the simplest large $N$ saddle points of the path integral 
associated with it, where the composite scalar operator $\bar\psi\psi$ 
develops a space-time independent expectation value. The relevant object to 
discuss in this context is the effective potential $V_{eff}$ 
associated with (\ref{auxiliary}), namely, the value of  $-S_{eff}$ for 
space-time independent $\si$ configurations per unit time per unit 
length.  $V_{eff} (\si)$ has two degenerate (absolute) minima at  
$\si= \pm m \neq 0$, where $m$ is fixed by the (bare) gap equation\cite{gn}
\beq
-m + iNg^2\,{\rm tr}\int
{d^2k\over\left(2\pi\right)^2}{1\over\notk-m}
= 0
\label{bgap}
\eeq
which yields the fermion dynamical mass
\beq
m = \Lambda\,e^{-{\pi\over Ng^2\left(\Lambda\right)}}\,.
\label{mass}
\eeq
Here $ \Lambda$ is an ultraviolet cutoff. The mass $m$ must be a 
renormalization group invariant. Thus, the model is asymptotically 
free. We can get rid of the cutoff at the price of introducing an 
arbitrary renormalization scale $\mu$. The renormalized
coupling $g_R\left(\mu\right)$ and the cutoff dependent bare
coupling are then related through 
\beq\label{renormalizedg}
\Lambda\,e^{-{\pi\over Ng^2\left(\Lambda\right)}} = 
\mu\,e^{1-{\pi\over Ng_R^2\left(\mu\right)}}\,,
\eeq
in a convention where 
$Ng_R^2\left(m\right) = \pi$. Trading the dimensionless coupling 
$g_R^2$ for the dynamical mass scale $m$ represents the well known 
phenomenon of dimensional transmutation.

In terms of $m$, we can write down the renormalized effective potential 
(in the large $N$ limit) as 
\beq\label{veff}
V_{eff} (\si) = {N\over 4\pi} \si^2\,\log {\si^2\over em^2}\,.
\eeq
It is evidently symmetric under $\si\rightarrow -\si$, which generates 
(together with $\psi\rightarrow\gam_5\psi$ in (\ref{auxiliary})) the discrete 
(or ${\bf Z}\!\!\!{\bf Z}_2$) chiral symmetry of the GN model.

This discrete symmetry is dynamically broken by the non-perturbative vacuum, 
and thus there is a kink solution \cite{dhn,ccgz,josh1}, the CCGZ kink 
mentioned above,
$\sigx = m\,{\rm tanh}(mx)$, interpolating between the two degenerate minima 
$\si =\pm m$ of (\ref{veff}) at  $x= \pm \infty$.
Therefore, topology insures the stability of these kinks.

The CCGZ kink is one example of non-trivial excitations of the 
vacuum, which are described semiclassically by large $N$ saddle points of 
the path integral over (\ref{auxiliary}) at which $\si$ develops space 
dependent, or even space-time dependent expectation values. These expectation 
values are the space-time dependent solution of (\ref{saddle}), and have 
analogs in other field theories \cite{jg,cjt}. 
Saddle points of this type are important also in discussing the large 
order behavior\cite{bh,dev} of the ${1\over N}$ expansion of the path 
integral over (\ref{auxiliary}).

These saddle points describe sectors of (\ref{auxiliary}) that include 
scattering states of the (dynamically massive) fermions in 
(\ref{auxiliary}), as well as a rich collection of bound states thereof. 
These bound states result from the strong infrared interactions, 
which polarize the vacuum inhomogeneously, causing 
the composite scalar $\bar\psi\psi$ field to form finite action 
space-time dependent condensates. We may regard these condensates as one 
dimensional fermion bags \cite{sphericalbag,shellbag} that trap the 
original fermions (``quarks") into stable finite action extended entities 
(``hadrons").

Finding explicit space-time dependent 
solutions of (\ref{saddle}) is a very difficult problem. In \cite{dhn}, DHN 
managed to guess such a set of solutions which oscillate in time (and thus 
are not merely boosts of static bags). Finding a systematic method to 
generate space-time dependent solutions of equations like (\ref{saddle}) 
is, of course, one of the basic goals of quantum field theory.

\subsection{Static Inhomogeneous $\sigx$ Backgrounds}
In this paper we focus on the easier problem of solving (\ref{saddle}) 
for static inhomogeneous condensates. 

Static solutions of (\ref{saddle}) are subjected to certain spatial 
asymptotic boundary conditions. For the usual physical reasons, we set 
boundary conditions on our static background fields such that $\sigx$ starts 
from one of its vacuum expectation values $\si=\pm m$ at $x=-\infty$, wanders 
along the $\si$ axis, and then relaxes back to one of its vacuum expectation 
values at $x=+\infty$: 
\beq
\si\asymptotic \pm m \quad\quad ,\quad\quad \si'\asymptotic 0\,. 
\label{boundaryconditions}
\eeq
These four possible asymptotic behaviors determine the topological 
nature of the condensate. The two cases in which the asymptotic values are 
of opposite signs are topologically non trivial. The CCGZ kink (where, 
by definition, $\si(\infty) =m$ ) and antikink ($\si(\infty)= -m$)
are two such examples. The other two cases, i.e., the cases with asymptotic
values of equal signs, are topologically trivial.

The topological charge associated with these boundary conditions is simply
\beq\label{topological}
q = {1\over 2m}\left(\si(\infty)-\si(-\infty)\right) = 
{1\over 2m}\int\limits_{-\infty}^{\infty}\pa_x\sigx\,,
\eeq
and can take on values $q=\pm 1$ for the topologically non trivial 
configurations, and $q=0$ for the topologically trivial ones.

As typical of solitonic configurations, we expect that $\sigx$ tends to its
asymptotic boundary values (\ref{boundaryconditions}) at an 
exponential rate which is determined, essentially, by the mass gap $m$ of 
the model. It is in such static backgrounds $\sigx$ that we have to 
invert the Dirac operator and calculate its resolvent
in seeking static solutions of (\ref{saddle}).

In addition to the CCGZ kink, there are also topologically trivial 
inhomogeneous condensates. These topologically trivial condensates
are stable because of the binding energy released by the trapped fermions, 
and therefore cannot form  without such binding. Thus, they are stable due 
to dynamics. In contrast, the stability of the CCGZ kink is guaranteed by 
topology already. It can form without binding fermions.

Note in passing, that the $1+1$ dimensional NJL model, with its continuous 
symmetry, does not have a topologically stable soliton solution. Thus, 
unlike the GN model, the solitons arising in the NJL model can be 
stabilized only by binding fermions \cite{shei,fz}. This description agrees 
with the general physical picture drawn in \cite{mackenzie}.

\subsubsection{The Energy of a Static Configuration and Supersymmetry}
In order to study the extremum condition (\ref{saddle}) on $S_{\it eff}$
around a static space dependent background $\sigx$, we need to invert the 
Dirac operator 
\beq
D\equiv \om\gam^0 + i\gam^1\pax -\sigx\,.
\label{dirac}
\eeq
(We have naturally transformed $i\notpa-\sigx$ to the $\om$ plane, since 
$\sigx$ is static.) In particular, we have to find the diagonal resolvent of 
(\ref{dirac}) in that background.

There is an intimate connection between the spectral theory of the Dirac 
operator (\ref{dirac}) and one dimensional supersymmetric 
quantum mechanics \cite{cooper,josh1}. This connection is explained in 
detail Appendix A.
In particular, the topological charge $q$ (\ref{topological})
coincides with the Witten index of the one dimensional supersymmetric 
quantum mechanics (see Eq. (\ref{wittenindex})).

The desired diagonal resolvent $\langle x\,|iD^{-1} | x\,\rangle$ 
is defined in (\ref{diagonal}) in Appendix A. It is shown there, that 
due to the underlying supersymmetric quantum mechanics, 
its four entries 
\beqra\label{diagresolventintext}
\langle x\,|-iD^{-1} | x\,\rangle &\equiv& \left(\begin{array}{cc} A(x) & 
B(x) \\{}&{}\\ C(x) & 
D(x)\end{array}\right)
\eeqra
are completely determined, in a simple way, in terms of any one of the 
off-diagonal entries, $B$ or $C$.

For example, in terms of $B$, one finds (see Eqs. (\ref{abcd}), (\ref{ADeq})
and (\ref{BtoCrel}))
\beqra\label{allfromB}
A(x)~=~D(x) &=& i{\left[\pax+2\sigx\right]B\left(x\right)\over 2\omega}
\nonumber\\{}\nonumber\\
-\om^2 C(x) &=&  {1\over 2}B'' + \si B' + (\si' + \om^2 ) B\,.
\eeqra

A quick way to introduce this underlying supersymmetry is the following: 
From the elementary identity 
\beq\label{gamfive}
\gam_5 (i\notpa-\si)\gam_5 = - (i\notpa+\si)
\eeq
it follows that $\tr\log (i\notpa-\si) = {1\over 2}
\tr\log\left[-(i\notpa-\si)(i\notpa+\si)\right]$. Thus, an alternative 
representation of (\ref{effective}) is \cite{josh1} 
\beq
S_{eff}[\si] =-{1\over 2g^2}\int\, d^2x 
\,\si^2-i{N\over 2}\, \tr\log\left(\pa_\mu^2 + \si^2 -i\gam^\mu\pa_\mu\si
\right)\,.
\label{effective1}
\eeq
If, in addition, $\si$ is time independent, (\ref{effective1}) may be 
further simplified to 
\beq
S_{eff}[\si] =-{T\over 2g^2}\int dx \,\si^2
-i{NT\over 2}\,\int {d\om\over 2\pi}\,
\left[\tr\log (H_b - \om^2) + \tr\log (H_c - \om^2)\right]\,,
\label{effective2}
\eeq
where $T=\int dt$ is a large temporal infrared cutoff, and 
\beq\label{hbhc}
H_b =-\pax^2 + \si^2 -\si' \quad\quad {\rm and} \quad\quad
H_c =-\pax^2 + \si^2 +\si' 
\eeq
are the pair of isospectral Schr\"odinger operators defined in (\ref{bcops}). 
In other words, they have the same spectrum of bound state energies,
save, possibly, the existence of a bound state at $\om^2=0$, which, if it 
exists, appears in the spectrum of only one of the operators. 
As explained in subsection A.1.1 of Appendix A, a bound state at $\om^2=0$ 
exists only in topologically non-trivial $\sigx$ backgrounds. Due to the 
zero-mode mismatch of the spectra of $H_b$ and $H_c$, we will keep 
both $\tr\log (H_b - \om^2)$ and $\tr\log (H_c - \om^2)$ explicitly 
in our formulas for a while. We will take advantage of the isospectrality 
of $H_b$ and $H_c$ in the positive part of the spectrum at the appropriate 
places.

The operators $H_b$ and $H_c$ may be identified as the hamiltonians of the 
bosonic sector and of the fermionic sector of a one dimensional 
supersymmetric quantum mechanical system.

According to (\ref{inversehbhc}), the off-diagonal entries 
$B(x)$ and $C(x)$ in (\ref{diagresolventintext}) are essentially the 
diagonal resolvents of $H_b$ and $H_c$: 
\beqra
{B(x)\over \om} &=& \langle x |\,{1\over H_b - \omega^2 }
\,| x\rangle\nonumber\\{}\nonumber\\
-{C(x)\over \om} &=& \langle x |\,{1\over H_c -\omega^2 }
\,| x\rangle\,.
\label{inversehbhctext}
\eeqra

From (\ref{effective2}) we may express the energy functional $\ce[\sigx]$ of 
a static configuration $\sigx$ as
\beq\label{energyfunctional}
\ce [\sigx]  = -{S_{eff}[\si]\over T} = {1\over 2g^2}\int dx \,\si^2
+i{N\over 2}\,\int{d\om\over 2\pi}\,
\left[\tr\log (H_b - \om^2) + \tr\log (H_c - \om^2)\right]\,.
\eeq
This expression is divergent. We regulate it, as usual, by subtracting from it 
the divergent contribution of the vacuum configuration $\si^2=m^2$ and by 
imposing a UV cutoff $\Lambda$ on $\om$. Thus, the regulated (bare) energy 
functional associated with $\sigx$ is 
\beqra
\ce^{reg} [\sigx]  &=& {1\over 2g^2}\int dx \,(\si^2 -m^2)
+i{N\over 2}\,\int\limits_{|\om|<\Lambda} {d\om\over 2\pi}\,
\left[\tr\log (H_b - \om^2) -\tr\log (H_{_{VAC}} - \om^2)\right]  
\nonumber\\{}\nonumber\\ 
&+& i{N\over 2}\,\int\limits_{|\om|<\Lambda} {d\om\over 2\pi}\,
\left[\tr\log (H_c - \om^2) -\tr\log (H_{_{VAC}} - \om^2)\right]\,,
\label{sigenergy}
\eeqra
where 
\beq\label{vacham}
H_{_{VAC}} = -\pax^2 + m^2
\eeq
is the hamiltonian corresponding to the vacuum configuration. 
We are not done yet, as the integrals over $\om$ in (\ref{sigenergy}) 
still diverge logarithmically with the UV cutoff $\Lambda$. However, as 
will be clear from the explicit calculations in the next section, this 
divergence is canceled by the logarithmic $\Lambda$ dependence of the 
bare coupling $g^2$, as determined by (\ref{mass}) and (\ref{renormalizedg}).

The renormalized quantity $\ce [\sigx]$ thus defined in (\ref{sigenergy}) is
the mass of the static fermion bag.

\pagebreak

\section{Extremal Static Fermion Bags}
\setcounter{equation}{0}

Now that we have written the energy-functional (\ref{sigenergy}) of a
static configuration, or a fermion bag, our next step is to identify those 
fermion bags on which (\ref{sigenergy}) is extremal. 

The energy functional (\ref{sigenergy}) is, in principle, a complicated and 
generally unknown functional of $\sigx$ and of its derivatives. Thus,   
the extremum condition ${\del\ce [\si]\over \del\sigx} =0$, as a functional
equation for $\sigx$, seems untractable. The considerable complexity of the 
functional equations that determine the extremal $\sigx$ configurations is 
the source of all difficulties that arise in any attempt to solve the model 
under consideration.

DHN found a way around this difficulty. They have discovered that the 
extremal static $\sigx$ configurations are necessarily reflectionless 
\cite{dhn}. Let us briefly recall the arguments of \cite{dhn}. DHN used 
inverse scattering techniques \cite{faddeev} to express the energy functional 
$\ce [\si]$  (\ref{sigenergy}) in terms of 
the so-called ``scattering data'' associated with, e.g., the hamiltonian
$H_b$ in (\ref{hbhc}), and thus with $\sigx$. The scattering data associated 
with $H_b$ are \cite{faddeev} the reflection amplitude $r(k)$ of $H_b$ 
(where $k$ is the momentum of the scattering state), the number $K$ of bound 
states in $H_b$ and their corresponding energies $0\leq\om_n^2\leq m^2\,,(
n=1,\cdots K)$, and also additional $K$ parameters $\{c_n\}$,
where $c_n$ has to do with the normalization of the $n$th bound state 
wave function $\psi_n$. 

More precisely, The $n$th bound state wave function, with energy $\om_n^2$,
must decay as $\psi_n(x)\sim {\rm const.}\exp -\kappa_n x$ as 
$x\rightarrow\infty$, where  
\beq\label{kappaomega}
0<\kappa_n = \sqrt{m^2 -\om_n^2}\,.
\eeq
If we impose that $\psi_n (x)$ be normalized, this will determine the 
constant coefficient as $c_n$. (With no loss of generality, we may take 
$c_n>0$.)

Thus, to summarize, in the inverse scattering technique, one trades the 
independent variables $\sigx$ for the scattering data. Then, one looks 
for scattering data that extremize $\ce [\si]$.

The key point is that $\ce [\si]$ depends on $r(k)$ only through certain 
dispersion integrals involving $\log (1-|r(k)|^2)$. Thus, the saddle point 
condition 
\beq\label{reflectionextremum}
{\del \ce[\si]\over \del r^*(k)} = 0\,,\quad\quad k\geq 0
\eeq 
is almost trivially solved by 
\beq\label{reflectionlessolution}
r(k) =0\,.
\eeq
Moreover, there seem to be no other solutions of (\ref{reflectionextremum}). 
In other words, extremal static $\sigx$ configurations are necessarily 
reflectionless.

This restriction on $\sigx$ is very powerful, since once the reflection
amplitude, i.e., the {\em function} $r(k)$, is eliminated out of the 
scattering data, all that remains is a discrete set of $2K$ real 
parameters: 
\beq\label{scatteringdatatext}
0\leq \om_1 < \om_2 < \cdots < \om_K < m\quad\quad ,\quad\quad
 c_1,c_2,\cdots, c_K\,.
\eeq
Thus, the space of all reflectionless backgrounds $\sigx$ is parametrized 
by a discrete set of real parameters. We still have to extremize the energy 
functional $\ce [\si]$ with respect to these parameters. 

Note, that due to the elementary fact, that the spectrum of the one 
dimensional Schr\"odinger operator $H_b$ cannot be degenerate, all the 
$\om_n$'s must be different from each other. Thus, the inequalities 
in (\ref{scatteringdatatext}) are strict. (See also the remark following 
(\ref{energies}).) 

We see that the formidable problem of finding the extremal 
$\sigx$ configurations
of the energy functional $\ce [\si]$ (\ref{sigenergy}), is reduced to the 
simpler problem of extremizing an ordinary function 
$\ce(\om_n, c_n) = \ce\left[\si (x; \om_n, c_n)\right]$ 
with respect to the 2$K$ parameters $\{c_n, \om_n\}$ that determine the 
reflectionless background $\sigx$. If we solve this ordinary 
extremum problem, we will be able to calculate the mass of the fermion bag.

In Appendix B we summarized (in some detail) the necessary facts about 
reflectionless potentials and the diagonal resolvents associated with them.
(We will need these resolvents below.) In particular, 
Eqs. (\ref{explicitsigmaqzero}) and (\ref{sigmaqone}) of Appendix B 
tell us how the 2$K$ parameters determine $\sigx$ explicitly. For the 
particular examples with $K=1$ and $K=2$, see 
(\ref{explicitsigmaqzerosinglebs}) and (\ref{sigmaqone2bs}), respectively.

\subsection{Extremal Static Reflectionless Fermion Bags}

The energy functional $\ce[\si]$, evaluated on a reflectionless 
$\sigx$ configuration, is an ordinary function of the parameters 
(\ref{scatteringdatatext}) which define $\sigx$. This function (when
defined with the renormalized $\ce$ in (\ref{sigenergy}))
\beq\label{massfunction}
M(\om_n, c_n) = \ce[\si(x;\om_n, c_n)]
\eeq
is the mass of the corresponding fermion bag. Thus, the extremal bags are 
determined by solving 
\beq\label{extremalbags}
{\pa M\over \pa \om_n} =0\quad\quad, \quad\quad {\pa M\over \pa c_n} =0\,.
\eeq

Let $\alpha$ be any one of the $2K$ parameters in (\ref{scatteringdatatext}).
Then, 
\beq\label{derivative}
{\pa M\over \pa\alpha} = \int\limits_{-\infty}^\infty dx {\del \ce[\si]
\over \del\sigx}\,{\pa\sigx\over\pa\alpha}\,,
\eeq
where  ${\del \ce[\si]\over \del\sigx}$ is evaluated at the appropriate
reflectionless $\sigx$.

\subsubsection{The Extremum Conditions on the Mass $M(\om_n,c_n)$, Its 
Flat Directions, and Collective Coordinates.}

In order to calculate ${\del \ce[\si]\over \del\sigx}$ around a generic 
$\sigx$ we start with the elementary identity 
\beq\label{variationlogdet}
\del \tr\log\left( -\pax^2 + V-\om^2\right) = \int\limits_{-\infty}^\infty
dx\,\langle x | {1\over -\pax^2 + V-\om^2} | x\rangle\,\del V(x)\,,
\eeq
and apply it to (\ref{sigenergy}), with $H_b$ and $H_c$ defined in 
(\ref{hbhc}). For convenience, let us record here the corresponding 
potentials,
\beq\label{vbvc}
V_b = \si^2 -\si' \quad\quad {\rm and} \quad\quad
V_c =\si^2 +\si'\,. 
\eeq
Thus, using (\ref{inversehbhctext}) (or (\ref{inversehbhc})), which relate
the diagonal resolvents of $H_b$ and $H_c$, respectively, to the entries 
$B(x)$ and $C(x)$ in the diagonal resolvent (\ref{diagresolventintext}),
we arrive at 
\beq\label{varenergy1}
\del \ce[\si] = {1\over g^2}\int dx \,\si\del\si  
+i{N\over 2}\,\int {d\om\over 2\pi}\,\int dx \,
\left[\left({B(x)\over \om}\right) \del V_b(x) - 
\left({C(x)\over \om}\right) \del V_c(x)\right]\,.
\eeq

By substituting (\ref{varenergy1}) into (\ref{derivative}) we obtain 
\beq\label{derivative1}
{\pa M\over \pa\alpha} = \int\limits_{-\infty}^\infty dx 
\,\left\{{1\over 2g^2}\,{\pa\si^2\over\pa\alpha} + 
i{N\over 2}\,\int {d\om\over 2\pi}\,
\left[\left({B(x)\over \om}\right) {\pa V_b(x)\over\pa\alpha} - 
\left({C(x)\over \om}\right) {\pa V_c(x)\over\pa\alpha}\right]\right\}\,.
\eeq

As was mentioned above, the diagonal resolvent of the Dirac operator 
in reflectionless $\sigx$ backgrounds was evaluated in Appendix B 
(see section B.2). Thus, our next step is to substitute $B(x)$ and $-C(x)$ 
from (\ref{Bresolvent}) in (\ref{derivative1}). 
Let us do this explicitly for $B(x)$. According to (\ref{Bresolvent})
\beq\label{Bresolventtext}
B(x) = \langle x |\,{\om\over H_b - \omega^2 }| x\rangle = 
{\om\over 2\sqrt{m^2-\om^2}}\left(1 -2\sum_{n=1}^K 
{\kappa_n\psi_n^2\over \om^2 - 
\om_n^2}\right)\,,
\eeq
where $\psi_n$ are the (normalized) bound state wave 
functions of $H_b$ (and we have also used $k=i\sqrt{m^2-\om^2}$). Thus,
\beqra\label{Bcontribution1}
&&\int\limits_{-\infty}^\infty dx\,
\left({B(x)\over \om}\right) {\pa V_b(x)\over\pa\alpha} = 
{1\over 2\sqrt{m^2-\om^2}}\,\int\limits_{-\infty}^\infty dx\,
\left(1 -2\sum_{n=1}^K {\kappa_n\psi_n^2\over \om^2 - 
\om_n^2}\right){\pa V_b(x)\over\pa\alpha} =\nonumber\\{}\nonumber\\
&&{1\over 2\sqrt{m^2-\om^2}}\,
\left[\int\limits_{-\infty}^\infty dx\,{\pa V_b(x)\over\pa\alpha}
-2\sum_{n=1}^K {\kappa_n\langle \psi_n | {\pa V_b\over\pa\alpha} 
|\psi_n\rangle\over \om^2 - \om_n^2}\right]\,.
\eeqra
From first order perturbation theory we know that 
\beq\label{firstorderpt}
\langle \psi_n | {\pa V_b\over\pa\alpha} |\psi_n\rangle = 
\langle \psi_n | {\pa H_b\over\pa\alpha} |\psi_n\rangle = 
{\pa \om_n^2\over \pa\alpha}\,.
\eeq
Thus, we may simplify (\ref{Bcontribution1}) further and obtain
\beq\label{Bcontribution}
\int\limits_{-\infty}^\infty dx\,
\left({B(x)\over \om}\right) {\pa V_b(x)\over\pa\alpha} = 
{1\over 2\sqrt{m^2-\om^2}}\,
\left[\int\limits_{-\infty}^\infty dx\,{\pa V_b(x)\over\pa\alpha}
-2\sum_{n=1}^K {\kappa_n  \left({\pa \om_n^2\over \pa\alpha}\right)
\over \om^2 - \om_n^2}\right]
\eeq
Recall that $H_b$ and $H_c$ are isospectral, save, possibly, a zero energy 
ground state, which can appear in the spectrum of only one of these 
operators. Since the zero energy ground state (if it exists) is the only 
difference in the energy spectra of $H_b$ and $H_c$, and since, in any case,
it obviously does not contribute to the sum in (\ref{Bcontribution}),
we conclude that the contribution of $-C(x)$ to (\ref{derivative1}) is the 
same as (\ref{Bcontribution}), but with $V_c$ instead of $V_b$ in the first 
term in (\ref{Bcontribution}).

Thus, combining the contributions of of $B(x)$ and $-C(x)$ into 
(\ref{derivative1}) we obtain 
\beq\label{derivative2}
{\pa M\over \pa\alpha} = \left({1\over 2g^2} + 
iN\,\int {d\om\over 2\pi}\,{1\over 2\sqrt{m^2-\om^2}}\right)
\,\int\limits_{-\infty}^\infty dx \,{\pa\si^2\over\pa\alpha} 
-iN\sum_{n=1}^K \,\int {d\om\over 2\pi}\,{1\over \sqrt{m^2-\om^2}}
{\kappa_n  \left({\pa \om_n^2\over \pa\alpha}\right)\over \om^2 - \om_n^2}
\eeq
(where in the first term we have used $V_b+V_c= 2\si^2$).

Consider now the integrals over $\om$ in (\ref{derivative2}). 
We will show in Appendix C (see (\ref{sigap})) that the bare gap equation may 
be written as 
$${m\over g^2} +\,iN~\int{d\om\over 2\pi}\,{m\over \sqrt{m^2-\omega^2}} = 0 
\,.$$ Thus, the first term in (\ref{derivative2}) vanishes, and 
\beq\label{derivative3}
{\pa M\over \pa\alpha} = 
-iN\sum_{n=1}^K \,\int {d\om\over 2\pi}\,{1\over \sqrt{m^2-\om^2}}
{\kappa_n  \left({\pa \om_n^2\over \pa\alpha}\right)\over \om^2 - \om_n^2}\,.
\eeq
The remaining integral over $\om$ is UV-convergent. Thus, 
renormalization of the coupling constant $g^2$, based on the gap 
equation (see (\ref{mass})), also renders (\ref{derivative2}) finite.

Next, observe from (\ref{derivative3}), that if $\alpha$ is one 
of the coefficients $c_k$ in (\ref{scatteringdatatext}), then 
\beq\label{flat}
{\pa M\over \pa c_k} = 0
\eeq
identically! This does not produce any condition on the $c_k$'s. 
The energy functional $\ce[\si]$ (\ref{sigenergy}), evaluated
on a reflectionless $\si (x;\om_n, c_n)$, is independent of the 
coefficients $c_k$ that do affect the shape of $\sigx$. The $c_k$'s are 
thus {\em flat directions} of $\ce[\si]$ in the space
of all reflectionless $\sigx$ configurations. In fact, we show in 
Appendix B, that the $c_k$'s (or more precisely, their logarithms) are 
collective translational coordinates of the fermion bag $\sigx$ (see e.g., 
(\ref{explicitsigmaqzerosinglebs}) and (\ref{sigmaqone2bs})), which is 
a familiar concept in soliton and instanton physics. One of these coordinates,
corresponds, of course, to global translations of the bag as a whole.

The full effective action functional $S_{eff}[\si (x,t)]$ contains, in 
principle, all information on dynamics of space-time dependent $\si (x,t)$ 
bags, or solitons. However, it is a complicated and almost intractable
object. Some progress may be achieved, perhaps, by trying to extend the 
periodic time-dependent solution of (\ref{saddle}) which was {\em guessed} 
by DHN in \cite{dhn}. As another small step towards 
understanding time dependent soliton dynamics from $S_{eff}$, one might 
consider elevating the $c_k$'s, which determine
the shape of static reflectionless bags, into slowly varying
functions of time, and thus study soliton dynamics in the framework of
an adiabatic approximation, in the spirit of 
\cite{collectivecoordinates, tdlee}.

Let us return to (\ref{derivative3}), and consider the remaining possibility
$\alpha=\om_n$. In this case we obtain
\beq\label{derivativefinal}
{\pa M\over \pa\om_n} = 
-iN\,\int {d\om\over 2\pi}\,{1\over \sqrt{m^2-\om^2}}
{2\kappa_n\om_n\over \om^2 - \om_n^2}\,,
\eeq
which has to vanish for extremal bags, and thus produces an equation to
determine $\om_n$. Note that (\ref{derivativefinal}) involves only one 
particular $\om_n$. Thus, the extremal $\om_n$'s are determined independently 
of each other, except for the constraint that they cannot coincide.

We observe from (\ref{derivativefinal}) that $\om_1=0$ is a possible 
solution of \footnote{According to calculations in the next 
subsection, the integral $\int {d\om\over 2\pi\om^2}\,{1\over 
\sqrt{m^2-\om^2}}$ (along the appropriate contour) is finite. Thus, the 
right-hand side of (\ref{derivativefinal}) vanishes at $\om_1 = 0$.} 
${\pa M\over \pa\om_n} =0 $. Thus, there are extremal bags for which $H_b$ 
(or $H_c$) has a zero energy ground state. As explained in Appendix A 
(see subsection A.1.1), a zero mode $\om_1 = 0 $ occurs as a solution of 
(\ref{derivativefinal}) if and only if $\sigx$ has non-trivial topology. 
All other solutions $\om_n^2$ of ${\pa M\over \pa\om_n} =0 $ must be strictly 
positive.

In our calculations so far we used DHN's result that extremal static $\sigx$
configurations are necessarily reflectionless, and looked for the extremal 
reflectionless configurations. As a consistency check of our calculations, 
we verify in Appendix C that vanishing of the right hand side of 
(\ref{derivativefinal}) guarantees that the corresponding 
reflectionless $\sigx$ is indeed extremal among all 
possible static configurations.

\subsubsection{Solution of the Extremum Conditions and Computation of the 
Soliton's Mass.}
In order to determine the extremal $\sigx$ configurations, we have to 
determine the $\om_n$'s from (\ref{derivativefinal}). With no loss of 
generality we will assume henceforth that $\om_n > 0$. 

In order to determine the non-vanishing $\om_n$'s, we have to evaluate 
the (UV finite) integral 
\beq\label{integral}
I(\om_n) = \int_{\cac} {d\om\over 2\pi i} \,{1\over \sqrt{m^2-\om^2}}
{1\over \om^2 - \om_n^2}\,.
\eeq
To this end we have to choose the proper contour ${\cal C}$, and thus we 
have to invoke our understanding of the physics of fermions: The 
Dirac equation $(i\notpa -\sigx)\psi=0$ (Eq. (\ref{diraceq})) in the 
background $\sigx$ under consideration has a pair of charge-conjugate bound 
states at $\pm\om_n$. These are the simple poles in (\ref{integral}). 
The continuum states appear as the two cuts along the real axis with branch 
points at $\pm m$ (see Figure 1).

The bound states at $\pm\om_n$ are to be considered together due to 
the charge conjugation invariance of the GN model, as we discuss in 
Appendix D (see in particular subsection D.2.2). Due to Pauli's principle, 
we can populate each of the bound states $\pm\om_n$ with up to 
$N$ (non-interacting) fermions. In such a typical multiparticle state, the 
negative frequency state is populated by $N-h_n$ fermions and the positive 
frequency state contains $p_n$ fermions. In the parlance of Dirac's hole 
theory, we have thus created a many fermion state, with $p_n$ {\em particles} 
and $h_n$ {\em holes} occupying the pair of charge-conjugate bound states 
at energies $\pm\om_n$. Let us name this many body state a $(p_n,h_n)$ 
configuration.

Since in this paper we are interested only in solitons in their ground state, 
we must take all states in the Dirac sea completely filled, and no positive 
energy scattering states.

Mathematically, we thus have to let ${\cal C}$ enclose the cut on the 
negative $\om$ axis $N$ times, and then go $N-h_n$ times around the pole at 
$-\om_n$, and $p_n$ times around the pole at $\om_n$, as shown in 
Figure 1.
\vspace{24pt}
\par
\hspace{0.5in} \epsfbox{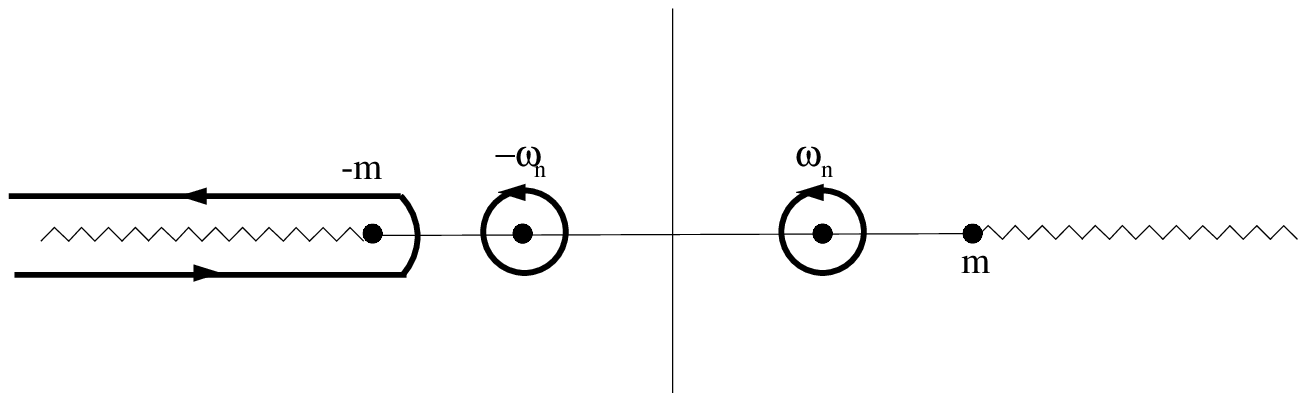}
\par
\baselineskip 12pt
{\footnotesize {\bf Fig.~1:} The contour ${\cal C}$ in the complex $\om$
plane in Eq. (\ref{integral}). The continuum states 
appear as the two cuts along the real axis with branch points at $\pm m$, 
and the bound states are the poles at $\pm\om_n$. The contour wraps $N$ 
times around the cut on the negative $\om$ axis, and then go $N-h_n$ times 
around the pole at $-\om_n$, and $p_n$ times around the pole at $\om_n$.}
\vspace{24pt}
\par
\baselineskip 20pt   
Thus, the integral (\ref{integral}) depends also on $h_n$ and $p_n$. 
In this way, we obtain\footnote{This expression, for $h_n = 0$, is quoted,
e.g., in \cite{dhn, josh1}.}
\beq\label{integral11}
I(\om_n, p_n, h_n ) = {1\over 2\om_n\kappa_n}
\left({2N\over \pi}\,{\rm arctan}{\om_n\over\kappa_n} + p_n - 
( N - h_n)\right)
\eeq
(with $\kappa_n=\sqrt{m^2-\om_n^2}$). The first term in (\ref{integral11})
is the contribution of the part of ${\cal C}$ wrapped around the negative 
cut, and the other two terms come, respectively, from the poles at $\om_n$ 
and $-\om_n$. The function $I(\om_n, p_n, h_n ) $ depends on $h_n$ and 
$p_n$ only through their sum
\beq\label{nu}
\nu_n = p_n + h_n\,,
\eeq
the total number of particles and holes in the pair of charge conjugate 
bound states at $\pm\om_n$. Pauli's Principle allows all
$0\leq \nu_n\leq 2N$. However, 
it turns out that the allowed range for $\nu_n$ in extremal 
configurations is restricted to only half that range: 
Substituting (\ref{integral11}) into (\ref{derivativefinal}) we obtain 
\beq\label{derivativefinalfinal}
{\pa M\over \pa\om_n} = 2\kappa_n\om_n I(\om_n, \nu_n) = 
\left({2N\over \pi}\,{\rm arctan}{\om_n\over\kappa_n} + \nu_n -N\right)\,.
\eeq
Thus, the extremal value $\om_n^*$ satisfies 
\beq\label{extremal1}
{\om_n^*\over \kappa_n^*} = {\om_n^*\over \sqrt{m^2 - \om_n^{*2}} }
= \cot\left({\pi\nu_n\over 2 N}\right)
\geq 0\,,
\eeq
rendering 
\beq\label{extremalomega}
\om_n^* = m\,{\rm cos}\,\left({\pi\nu_n\over 2N}\right)\,.
\eeq
Thus, since $\om_n^*\geq 0$, we must have\footnote{This 
can be already seen from (\ref{derivativefinalfinal}). From 
the actual computation of the contribution of the part of ${\cal C}$ 
wrapped around the negative cut, the first term in  (\ref{integral11}), it is
clear that ${\rm arctan}{\om_n\over\kappa_n}\geq 0$. Thus, 
${\pa M\over \pa\om_n} = 0$ has a solution only in the range (\ref{nurange}).} 
$\nu_n\leq N\,.$

However, the end-points of this range cannot 
be realized physically. Concerning $\nu_n = 0$, note that if the pair of 
energy levels corresponding to (\ref{extremalomega}) are empty, then  
$\om_n^* = m$ plunges into the continuum and ceases to be a bound state. Thus,
the self-consistent, extremal energy levels (\ref{extremalomega}) must trap
fermions in order to exist. This fact will be related to soliton stability 
in Section 4 (see also the discussion around (\ref{bindingenergy}) in this 
section). At the other end-point $\nu_n = N$, we obtain $\om_n^* = 0$. 
As we have already mentioned (recall the discussion in subsection A.1.1 in 
Appendix A, which we have already referred to following (\ref{hbhc}) and 
(\ref{derivativefinal})), $\om_n^*=0$ occurs if and only if $\sigx$ has 
non-trivial topology, and does not depend on dynamical details such as
the value of $\nu_n$. Thus, $\nu_n=N$ cannot occur in a 
topologically trivial soliton. What is perhaps less obvious, is that we cannot
have $\nu_n = N$ in a topologically non-trivial soliton either. To clarify this
assertion, consider an extremal topologically non-trivial soliton with 
$0 = \om_1^* < \om_2^* < \cdots < \om_K^* $. Then, form a sequence of such 
solitons in which $\nu_2$ increases. As $\nu_2$ tends to $N$, $\om_2^*$ tends 
to coalesce with $\om_1^*=0$. This cannot happen, because the spectrum 
$\{\om_n^{*2}\}$ of the one-dimensional Schr\"odinger operator $H_b$ (or 
$H_c$) is not degenerate. 
Putting all these facts together we conclude that $\nu_n$ 
is restricted to the range
\beq\label{nurange}
0 < \nu_n < N\,.
\eeq

We would like at this point to integrate (\ref{derivativefinalfinal})
and find the mass $M(\om_1,\ldots, \om_K)$. To this end 
it is convenient to introduce \cite{dhn} 
the angular parameter $\theta_n$ through the relations 
\beq\label{thetaomkappa}
\om_n = m\,{\rm cos}\,\theta_n\quad\quad {\rm and} \quad\quad 
\kappa_n=m\,{\rm sin}\,\theta_n\,,
\eeq
where positivity of $\om_n$ and $\kappa_n$ implies 
$0\leq\theta_n\leq \frac{\pi}{2}$.
In particular, note that the extremal value $\om_n^*$ in 
(\ref{extremalomega}) corresponds to 
\beq\label{extremaltheta}
\theta_n^* = {\pi\nu_n \over 2N}\,.
\eeq
Recall that all the $\om_n$'s which parametrize a given static soliton, 
must be different from each other (see the paragraph following 
(\ref{scatteringdatatext})). Consequently, all the $\nu_n$ 
must be different from each other, and so must be the $\th_n^*$'s. 

In terms of $\theta_n$, we may rewrite (\ref{derivativefinal}) as 
\beq\label{derivativefinalfinal1}
{\pa M\over \pa\om_n} = - {1\over m\,{\rm sin}\,\theta_n}\,{\pa M\over 
\pa\theta_n} = \nu_n -{2N\theta_n\over \pi}\,. 
\eeq
Finally, we can integrate (\ref{derivativefinalfinal1}) back, to obtain 
$M(\theta_1,\cdots \theta_K)$ explicitly. Thus, 
\beq\label{explicitmass}
M(\theta_1, \cdots, \theta_K) = m\sum_{n=1\atop \om_n > 0}^K\left[\left(
\nu_n -{2N\theta_n\over\pi}\right)\,{\rm cos}\,\theta_n + 
{2N\over\pi}\,{\rm sin}\,\theta_n\right] + 
{\rm  const.}
\eeq
Note that the mass (\ref{explicitmass}) depends only on the $\theta_n$'s 
which correspond to strictly positive $\om_n$'s. The integration constant in 
(\ref{explicitmass}) depends on the topology of $\sigx$, and 
will be determined below according to physical considerations.

Eq. (\ref{explicitmass}) is the mass of a fermion bag with an arbitrary 
reflectionless profile $\sigx$. We are interested only in extremal 
configurations. Substituting (\ref{extremaltheta}) in (\ref{explicitmass}), 
we obtain the mass of the extremal bag as 
\beq\label{explicitextremalmass}
M(\theta_1^*, \cdots, \theta_K^*) = {2Nm\over\pi}\,\sum_{n=1\atop \om_n > 
0}^K \,{\rm sin}\,\left({\pi\nu_n\over 2N} \right) + {\rm  const.}
\eeq

The extremal value (\ref{extremalomega}) of the parameter $\om_n$ is 
determined by the total number $\nu_n$ of particles and holes trapped in 
the bound states of the Dirac equation at $\pm\om_n$, and not by the 
numbers of trapped particles and holes separately. 
As we explain in Appendix D (see subsection D.2.2), this fact is a 
manifestation of the underlying $O(2N)$ symmetry, which treats particles 
and holes symmetrically. It indicates that this pair of bound states 
gives rise to an $O(2N)$ antisymmetric tensor multiplet of rank $\nu_n$ 
of soliton states. The states in this multiplet correspond to all the 
possible $(p_n,h_n)$ configurations, with $\nu_n = p_n+h_n$ fixed, subjected 
to Pauli's principle. There are $\sum_{k=0}^{\nu_n} C_N^k C_N^{\nu_n-k} = 
C^{\nu_n}_{2N} = (2N)!/\nu_n! (2N-\nu_n)!$ such states, precisely the 
dimension of an $O(2N)$ antisymmetric tensor of rank $\nu_n$. Due to 
the restriction (\ref{nurange}) only tensors of ranks $0 < \nu_n < N$ 
are allowed.

The antisymmetric tensors of higher ranks (in the range 
$N <\nu_n \leq 2N$), each of which is dual to an antisymmetric tensor in 
the allowed range (\ref{nurange}), are not realized as {\em extremal} fermion 
bags (their ranks cannot be interpreted as the total number of particles and 
holes trapped by the bag). Note, however, as a mathematical fact, 
that the mass formulas (\ref{explicitmass}) and (\ref{explicitextremalmass}) 
are invariant against the duality transformation of tensorial ranks 
$\nu_n\rightarrow 2N -\nu_n$, performed simultaneously with 
$\theta_n^* \rightarrow \pi- \theta_n^*$. This is equivalent to 
$\om_n^*\rightarrow -\om_n^*$. Thus, a $(p_n, h_n)$ configuration belonging 
to an antisymmetric tensor multiplet of rank $0 < \nu_n < N$ is mapped in this
way onto an $(N - h_n, N - p_n)$ configuration belonging 
to the dual antisymmetric tensor of rank $2N - \nu_n$. 
Evidently, any two dual antisymmetric
tensor multiplets have the same dimension, and make identical contributions 
to the mass formulas. Thus, these duality transformations may be used 
to extend the domain of definition of (\ref{explicitmass}) and 
(\ref{explicitextremalmass}) to all $O(2N)$ antisymmetric tensors $0\leq 
\nu_n \leq 2N$.

The bound state at $\om_1=0$, which arises if and only if $\sigx$ has 
non-trivial topology, behaves 
differently from the previous case. Due to its topological origin, this mode 
is stable and cannot flow away from zero as the other parameters which define
$\sigx$ vary. In particular, it remains unchanged as the number of 
trapped fermions varies, in contrast with (\ref{extremalomega}). Since this 
bound state is unpaired, it can trap up to $N$ fermions. Thus, it gives rise 
to a $2^N$ dimensional multiplet of soliton states, which, as we show 
in subsection D.2.1, is the spinorial representation of $O(2N)$ 
\cite{wittengn}.

Let us determine now the integration constant in (\ref{explicitextremalmass}). 
As was mentioned earlier, this integration constant depends on the topological 
charge carried by the corresponding extremal $\sigx$ configuration. 
We have already mentioned in subsection 2.1, that topologically trivial 
solitons are stable because of the binding energy released by the trapped 
fermions, and cannot form without such binding. Thus, the binding energy 
\beq\label{bindingenergy}
B(\theta_1^*, \cdots, \theta_K^*) = m\sum_{n=1}^K \nu_n  - 
M(\theta_1^*, \cdots, \theta_K^*)
\eeq
of a topologically trivial soliton, which measures the stability of the
soliton, should tend to zero as the number of trapped fermions tends to zero.
In other words, if {\em each} of the pairs of levels $\pm\om_n$ traps a 
small number $\nu_n << N$ of fermions, we expect that  
$M(\theta_1^*, \cdots, \theta_K^*) \simeq m\sum_{n=1}^K \nu_n $. Imposing this 
physical constraint on (\ref{explicitextremalmass}) determines the 
integration constant (in the topologically trivial case) to be zero. 
This cannot be true in the topologically non-trivial sector, since the 
the integration constant in (\ref{explicitextremalmass}) is evidently 
the minimal mass in that sector. The latter is the mass of the topologically 
non-trivial soliton which has a single bound state at $\om_1 =0$, namely the 
CCGZ kink (or antikink). Its mass is 
\beq\label{massccgz}
M_{kink} = {Nm\over\pi}\,,
\eeq
as quoted in \cite{ccgz} (see also Eq. (3.30) in \cite{dhn}), and is 
independent of the number $n_0$ of fermions trapped in the bound state at 
$\om_1=0$, as we mentioned in the previous paragraph. In Appendix E we will 
present our own derivation of (\ref{massccgz}), based on (\ref{sigenergy}) 
and on a simple dimensional argument. In addition, we will provide below 
(see the passage following (\ref{dhnprofile})) a simple physical argument 
for the correctness of (\ref{massccgz}). Thus, gathering all these facts 
together, we conclude that 
\beq\label{solitonmass}
M(\nu_1, \cdots, \nu_K; q) = {2Nm\over\pi}\,\sum_{n=1\atop \om_n > 
0}^K \,{\rm sin}\,\left({\pi\nu_n\over 2N}\right) + 
{Nm\over\pi}\delta_{|q|,1}\,,
\eeq
where $q$ is the topological charge (\ref{topological}) carried by the 
soliton (and all the ranks $\nu_n$ are different from each other). 

From (\ref{massccgz}) and (\ref{solitonmass}) we see that the kink, as well
as other extremal solitons (whose ranks $\nu_n$ are a finite fraction of $N$),
are heavy objects. Their masses are of $\cO (N)$. In fact, this should be 
expected, based on our experience with soliton physics in weakly interacting 
field theories: in this paper we study the GN model (\ref{auxiliary}) in the 
large $N$ limit, in which $N\sim g^{-2}$, and thus all soliton 
masses are of $\cO \left(g^{-2}\right)$.

\subsubsection{Summary: Masses, Profiles and $O(2N)$ Quantum Numbers of 
Extremal Fermion Bags}

We complete the task of determining the extremal static soliton 
configurations by combining the results of the previous discussion and the 
results of the analysis made in sections B.2, B.3 and D.2 in the appendices. 
We summarize our findings as follows:

~~~~~{\em (a) topologically trivial fermion bags, $q=0$}\\
An extremal static, topologically trivial soliton, 
bearing $K$ pairs of bound states, has its $K$ 
$\om_n$'s in (\ref{scatteringdatatext}) determined by 
(\ref{extremalomega}), with the $K$ moduli $c_n$ left as arbitrary 
parameters. Its mass is given by (\ref{solitonmass}) with $q=0$. Its 
profile $\sigx $ is computed in Appendix B (see section B.2) and is given by 
(\ref{explicitsigmaqzero}). According to Appendix D (see subsection D.2.2) we 
should refer to this object more precisely as an $O(2N)$ multiplet of 
solitonic states. This multiplet transforms under $O(2N)$ as the direct 
product of the $K$ individual antisymmetric tensors associated with the 
$K$ $\om_n$'s.

In particular, the case $K=1$ with $\om_1 = m\,{\rm cos}\,
\left({\pi\nu\over 2N}\right)>0$, in which the pair of bound states at 
$\pm\om_1$ trap $\nu_1 = \nu$ fermions, is the so-called
DHN soliton. Its mass is 
\beq\label{dhnmass}
M(\nu) = {2Nm\over\pi}\,{\rm sin}\,\left({\pi\nu\over 2N}\right)
\eeq
and its profile is given by (\ref{explicitsigmaqzerosinglebs}): 
\beq\label{dhnprofile}
\sigx = m + \kappa\, {\rm tanh}(\kappa x) - \kappa\, {\rm tanh}
\left[\kappa x + {1\over 2} \log \left({m+\kappa\over m-\kappa}\right)\right]
\eeq
(where we have set $\si(\infty) = m$ and $x_0 = 0$ in 
(\ref{explicitsigmaqzerosinglebs})), with 
$\kappa = m\,{\rm sin}\,\left({\pi\nu\over 2N}\right)$. These formulas 
agree with the results quoted in Eqs. (3.27) and (3.28) of \cite{dhn}. The 
profile (\ref{dhnprofile}) is that of a bound state of a kink and an 
antikink, with interkink distance 
${1\over 2\kappa} \log \left({m+\kappa\over m-\kappa}\right)$. This interkink 
distance diverges in the limit $\kappa\rightarrow m$, i.e., when 
$\nu \rightarrow N$, the inaccessible end-point of (\ref{nurange}). 
Heuristically, we can understand this divergence as a necessary compromise 
between dynamics and topology: As $\nu$ tends to $N$, 
$\om = m \cos \left({\pi\nu\over 2N}\right)$ tends to zero, which is 
{\em not} in the spectrum of the corresponding Dirac operator 
$i\notpa -\sigx$, since $\sigx$ is topologically trivial. The best
our soliton can do is to split into infinitely separated kink and 
antikink. Probing any finite neighborhood around either of these defects 
may lead us to the illusion that we are in the topologically non-trivial 
sector.

One might perhaps interpret this as a situation in 
which the infinitely separated kink and antikink are bound at threshold. 
(This is subjected to the plausible assumption that there are no long-range 
intersoliton forces in this model, since the mass gap is finite, and there
are no gauge interactions.) Thus, (\ref{dhnmass}) must tend, in the limit 
$\nu \rightarrow N$, to $2M_{kink}$, from which we see that $M_{kink} = 
{Nm\over\pi}$, in accordance with (\ref{massccgz}).

~~~~~{\em (b) topologically nontrivial fermion bags, $q=\pm 1$}\\
An extremal static, topologically non-trivial $\sigx$ 
configuration, with $2K$ parameters (\ref{scatteringdatatext}), has an 
unpaired bound state at $\om_1=0$, and additional $K-1$ pairs of bound 
states at $\pm\om_n$, which are determined by (\ref{extremalomega}). Its 
mass is given by (\ref{solitonmass}) with $|q|=1$, and its profile is 
determined as plus or minus the profile given by 
(\ref{sigmaqone}) (depending on the topological charge $q$ it carries).
As explained in detail in Appendix D (see subsections D.2.1 and D.2.2), it 
gives rise to a multiplet of solitonic states which transform under 
$O(2N)$ as the direct product of one factor of the $2^N$ dimensional 
(reducible) spinor representation of $O(2N)$ (associated with the bound 
state at $\om_1=0$), and the $K-1$ individual antisymmetric tensors 
associated with the remaining non-vanishing $\om_n$'s. In particular, 
the case $K=1$ corresponds to the CCGZ kink (or antikink) with mass 
(\ref{massccgz}) and profile $\sigx = \pm  m {\rm tanh}\,(m x)$ 
(see (\ref{explicitkink})). These kinks are thus pure $O(2N)$ 
isospinors \cite{wittengn}.

\subsection{Fermion Number Content of the Solitonic Multiplets.}

Expressions for the expectation values of fermion bilinear operators in a 
static $\sigx$ background are derived in Appendix A (see Section A.3). 
They are given as dispersion integrals over linear 
combinations of the entries of the diagonal resolvent 
(\ref{diagresolventintext}). In particular, according to 
(\ref{condensate}), (\ref{vac}) and (\ref{fourierdensity}), the regularized
fermion density in a given background $\sigx$ is 
\beq\label{densitytext}
\langle j^0 (x)\rangle_{\rm reg} = 
iN\int {d\om\over 2\pi }\,\left[\left( B(x) - B_{VAC}\right) - 
\left(C(x) - C_{VAC}\right)\right]\,,
\eeq 
where we have subtracted the UV-divergent vacuum contributions.

For evaluating $\langle j^0 (x)\rangle_{\rm reg}$ in reflectionless
$\sigx$ backgrounds, we will substitute, $B(x)$ and $-C(x)$ from 
(\ref{Bresolvent}) (or equivalently, the expression (\ref{Bresolventtext}) 
for $B(x)$, and its analog for $-C(x)$) in (\ref{densitytext}). For 
{\em extremal} reflectionless $\sigx$ backgrounds, the focus of our interest 
in this paper, the non-vanishing $\om_n$'s are determined by 
(\ref{extremalomega}). Thus, for each $\om_n > 0$, we will assume that the 
fermions which occupy the pair of energy levels $\pm\om_n$ form a $(p_n, h_n)$
configuration, with $p_n + h_n = \nu_n$, in accordance with (\ref{nu}). If 
$\om_1 = 0$ (i.e., if $\sigx$ has non-trivial topology), we assume that this 
level is occupied by $0\leq n_0 \leq N$ fermions.

Let us evaluate the contribution of the $B$-term in (\ref{densitytext})
explicitly. From (\ref{Bresolventtext}) and (\ref{vac}) we obtain this 
contribution as 
\beq\label{Bterm}
\sum_{n=1}^K\int_{{\cal C}_n} {d\om\over 2\pi i }\, 
{\om\over \sqrt{m^2-\om^2}} {\kappa_n\psi_{Bn}^2\over \om^2 - \om_n^2}\,, 
\eeq
where $\psi_{Bn}$ is the $n$th bound state wave function of $H_b$, 
and ${\cal C}_n$ is the contour in Fig. 1 (with the obvious exception, that 
if $\om_1=0$, then the contour ${\cal C}_1$ wraps $n_0$ times around the 
pole at $\om=0$). For each $\om_n>0$, the contribution to 
(\ref{Bterm}) coming from the integral around the left cut in Fig. 1 is 
$-\frac{N}{2}\, \psi_{Bn}^2$, the contribution from the pole at $-\om_n$ 
is $\frac{N - h_n}{2}\, \psi_{Bn}^2$, and that from the pole at $\om_n$ is 
$\frac{p_n}{2}\, \psi_{Bn}^2$. Adding these three terms, we obtain
the total contribution of the pair of levels at $\pm\om_n$ to (\ref{Bterm})
simply as $\frac{p_n - h_n}{2}\, \psi_{Bn}^2$. If $\sigx$ has non-trivial 
topology, and carries topological charge $q=1$, then according 
to the discussion following Eqs. (\ref{bzero}), (\ref{czero}),
only $H_b$ has a normalizable zero mode $\psi_{B0}$ , and we must set 
$\om_1=0$ in (\ref{Bterm}). In this case, the contribution to 
(\ref{Bterm}) is $\left(-\frac{N}{2} + n_0\right)\, \psi_{B0}^2$, where the 
first term comes from the integral around the left cut in Fig. 1 (as before), 
and the second term comes from the pole at $\om=0$. Combining all these results
together, we obtain that the $B$-term in (\ref{densitytext}) yields
\beq\label{Bterm1}
\left(-\frac{N}{2} + n_0\right)\, \psi_{B0}^2\,\delta_{q,1} + 
\sum_{n=1\atop \om_n >0}^K \, \frac{p_n - h_n}{2}\, \psi_{Bn}^2\,.
\eeq
Similarly, since $-C(x)$ is given by an expression analogous to 
(\ref{Bresolvent}), the $C$-term in (\ref{densitytext}) yields
\beq\label{Cterm1}
\left(-\frac{N}{2} + n_0\right)\, \psi_{C0}^2\,\delta_{q,-1} + 
\sum_{n=1\atop \om_n >0}^K\, \frac{p_n - h_n}{2}\, \psi_{Cn}^2\,,
\eeq
where $\psi_{Cn}$ is the $n$th bound state wave function of $H_c$. (Recall 
from the discussion following Eqs. (\ref{bzero}), (\ref{czero}) that if the
topological charge $q=-1$, only $H_c$ has a normalizable zero mode $\psi_{C0}
$.) Finally, combining (\ref{Bterm1}) and (\ref{Cterm1}) together, we obtain 
\beq\label{densityfinal}
\langle j^0 (x)\rangle_{\rm reg} = 
\left(-\frac{N}{2} + n_0\right)\, \left(\psi_{B0}^2(x)\,\delta_{q,1} + 
\psi_{C0}^2(x)\,\delta_{q,-1}\right)
+ \sum_{n=1\atop \om_n > 0}^K \, (p_n - h_n)\, {\psi_{Bn}^2(x) + 
\psi_{Cn}^2(x) \over 2}\,.
\eeq 
Integrating over $x$, we obtain the expectation value of the total fermion 
number $N_f$ in the background of the extremal fermion bag, simply as 
\beq\label{Nffinal}
\langle N_f \rangle = \left(n_0 -\frac{N}{2}\right)\, \delta_{|q|,1} 
+ \sum_{n=1\atop \om_n > 0}^K \, (p_n - h_n)\,.
\eeq  
The terms in (\ref{densityfinal}), associated with the positive $\om_n$'s,
have simple physical interpretation: 
\beq\label{valenceNfn}
N_{f,val}^{(n)} = p_n - h_n = \nu_n - 2h_n\,,
\eeq
the number of particles minus the number of holes, is the {\em valence} 
fermion number of the $(p_n,h_n)$ configuration, which occupies the pair of 
bound states at $\pm\om_n$, and 
$\frac{1}{2}\left(\psi_{Bn}^2(x) + \psi_{Cn}^2(x)\right)$
is the probability density to find any of these fermions in a small 
neighborhood of the point $x$. The $(p_n,h_n)$ configuration is a member 
of an $O(2N)$ antisymmetric tensor multiplet of rank $\nu_n < N$ 
(recall (\ref{nurange})). Thus, as we scan through all states in this 
multiplet, we see that $N_{f,val}^{(n)}$ has a symmetric spectrum 
\beq\label{valenceNfnspectrum}
-\nu_n \leq N_{f,val}^{(n)} \leq \nu_n\,,
\eeq
in accordance with charge conjugation invariance. 

Interpretation of the first term in (\ref{densityfinal}), associated with the 
zero mode, is more delicate, and exhibits an interesting physical phenomenon. 
As in the previous case, $\left(\psi_{B0}^2(x)\,\delta_{q,1} + 
\psi_{C0}^2(x)\,\delta_{q,-1}\right)$ is the probability density to find 
a fermion, trapped in the zero mode bound state, in a small 
neighborhood of the point $x$. However, unlike the previous
case, the coefficient of this probability density is the valence number 
$n_0$ of fermions trapped in the zero mode, {\em shifted} by the contribution 
$-\frac{N}{2}$ of the vacuum (the filled Dirac sea). Thus, the fermion 
number associated with the zero mode is 
\beq\label{Nfzeromode}
N_f^{(0)} = n_0 - \frac{N}{2}\,.
\eeq
This quantity coincides with the fermion number operator $N_f^{(spinor)}$
(\ref{Nfoperator}) which we constructed explicitly for the 
$O(2N)$ spinor representation in Appendix D.

Note that for $N$ odd, $N_f^{(0)}$ is not an integer! 
Eq.(\ref{Nfzeromode}) thus exhibits fractional fermion number 
\cite{jackiwrebbi}, a phenomenon which occurs because of the non-trivial 
topology of the background interacting with the fermions, and is independent 
of other details of the background (such as $\sigx$ being reflectionless). 
Furthermore, it is valid for any value of $N$, and has nothing
to do with the large $N$ limit. Charge conjugation invariance of the GN 
model restricts\footnote{The difference of any two eigenvalues of $N_f^{(0)}$ 
must be an integer, of course, and charge conjugation invariance implies 
that if $n$ is an eigenvalue, so is $-n$.} the fractional part of 
$N_f^{(0)}$ to be\footnote{ In other soliton 
bearing quantum field theoretic models, fermion number induced by solitons 
may acquire other rational, or even irrational values 
\cite{goldstonewilczek}.} either 0 or $\frac{1}{2}$. Indeed, it is $N_f^{(0)}$
as defined in (\ref{Nfzeromode}) which acquires a symmetric spectrum 
$-\frac{N}{2}\leq N_f^{(0)}\leq\frac{N}{2}$, as required by charge conjugation 
invariance, and not the valence piece $n_0$ alone.  For more details on 
fractionalization of fermion number in quantum field theory, see \cite{niemi}.

\pagebreak

\section{Investigating Stability of Extremal Static Fermion Bags}
\setcounter{equation}{0}
The extremal static soliton multiplets which we encountered in the previous 
section, correspond, in the limit $N\rightarrow\infty$, to exact eigenstates 
of the hamiltonian of the GN model. However, at large but finite $N$, we 
expect some of these states to become unstable and thus to acquire small 
widths, similarly to the behavior of baryons in QCD with 
a large number of colors \cite{wittenlargeN}. The latter are also solitonic 
objects and are analogous to the ``multi-quark'' bound states of the 
GN model.

Furthermore, we can imagine perturbing the GN action (\ref{auxiliary}) by a 
small perturbation, which is a singlet under all the discrete and 
continuous symmetries of the model (e.g., by adding to (\ref{auxiliary}) a 
term $\epsilon\int d^2 x\,\si^{2n}$), and ask which of the extremal fermion 
bags of the previous section are stable against such perturbations. 

Under these circumstances, all possible decay channels of a given 
soliton multiplet must conserve, in addition to energy and momentum, 
$O(2N)$ quantum numbers and topological charge. 

It turns out that non-trivial results concerning stability 
may be established without getting into all the details of decomposing
$O(2N)$ representations, by imposing a simple necessary 
condition on the spectrum of the fermion number operator $N_f$ in the 
multiplets involved in a given decay channel.
As we have learned so far, a given static soliton multiplet is 
a direct product of $O(2N)$ antisymmetric tensors and, for topologically 
non-trivial solitons, a factor of the spinor representation. 
The decay products of this soliton also correspond to a direct product 
of antisymmetric tensors and spinors. We have also learned that the spectrum 
of $N_f$ in the antisymmetric tensor and spinor
representations is symmetric, namely, $-N_f^{max}\leq N_f\leq N_f^{max}$. When 
we compose two such representations $D_1, D_2$, the spectrum of $N_f$ in the 
composite representation $D_1\otimes D_2$, will obviously have the range 
$-N_f^{max}(D_1) -  N_f^{max}(D_2) \leq N_f (D_1\otimes D_2)\leq 
N_f^{max}(D_1) +  N_f^{max}(D_2)$. In particular, each of the possible 
(i.e., integer or half-integer) eigenvalues in this range, will appear 
in at least one irreducible representation in the decomposition of 
$D_1\otimes D_2$. More generally, the spectrum 
of $ N_f (D_1\otimes D_2\cdots \otimes D_L)$ will have the range $|N_f 
(D_1\otimes\cdots \otimes D_L)|\leq N_f^{max}(D_1) + \cdots  
N_f^{max}(D_L)$.

Consider now a decay process, in which a parent static 
soliton, which belongs to a (possibly reducible) representation 
$D_{\rm parent}$, decays into a bunch of other solitons, such that the 
collection of all irreducible representations associated with the decay 
products is $\{D_1, \ldots, D_L\}$ (in which a given irreducible 
representation may occur more than once). By $O(2N)$ symmetry, the 
representation $D_{\rm parent}$ must occur in the decomposition 
of $D_1\otimes D_2\cdots \otimes D_L$. Thus, according to the discussion 
in the previous paragraph, if this decay process is allowed, we must have 
\beq\label{Nfnecessary}
N_f^{max} (D_{\rm parent}) \leq  N_f^{max}(D_1) + \cdots + N_f^{max}(D_L)\,,
\eeq
which is the necessary condition for $O(2N)$ symmetry we sought for. 
(Obviously, similar necessary conditions arise for the other $N-1$  
components of the highest weight vectors of the representations involved.)
The decay process under consideration must respect energy conservation, i.e.,
it must be exothermic. Thus, we supplement (\ref{Nfnecessary}) by 
the requirement 
\beq\label{energynecessary}
M_{\rm parent} \geq \sum_{\rm products} M_k
\eeq
on the masses  $M_i$ of the particles involved.

For each of the static soliton multiplets discussed in the 
previous section, we will scan through all decay channels 
and check which of these decay channels are {\em necessarily} closed, simply
by requiring that the two conditions (\ref{Nfnecessary}) and 
(\ref{energynecessary}) be mutually contradictory.

The oscillating time dependent DHN solitons \cite{dhn} may appear as decay 
products in some of the channels. However, for a {\em given} assignment of 
$O(2N)$ quantum numbers of the final state, they will occur at a higher mass 
than the final state containing only static solitons\footnote{In \cite{dhn}, 
DHN showed that the oscillating solitons occur in $O(2N)$ hypermultiplets of 
antisymmetric tensors. Such a hypermultiplet is determined by a principal 
quantum number $\nu_p$ ($0<\nu_p\leq N$), which determines the common mass 
${2Nm\over\pi}\sin\,\left( {\pi\nu_p\over 2N}\right)$ of all the states in 
it. For $\nu_p$ even, the hypermultiplet contains antisymmetric tensors of 
ranks $0, 2, \ldots \nu_p$ (each with multiplicity $1$). For $\nu_p$ odd, 
it contains tensors of ranks $1, 3, \ldots, \nu_p$. }.
Thus, if a decay channel into a final state containing only static 
solitons is necessarily closed (in the sense that (\ref{Nfnecessary}) and 
(\ref{energynecessary}) are mutually contradictory), then all decay 
channels into final states with the same $O(2N)$ quantum numbers, which  
contain time dependent solitons, are necessarily closed as well. 
Thus, it is enough to scan only through decay channels into final states
made purly of static solitons\footnote{This statement is surely true if 
DHN's time dependent solitons exhaust all the stable time dependent 
configurations of the model. However, even if there are time 
dependent solitons in the GN model beyond those discovered by DHN, 
it is most likely that they will also be heavier than the corresponding 
static states with the same $O(2N)$ quantum numbers.}.

For all the DHN multiplets ($K=1, q=0$) and for the CCGZ kink multiplet 
($K=1, |q|=1$), we will find in this way that {\em all} decay channels are 
necessarily closed, thus establishing their stability. That these are stable 
multiplets is well known, of course \cite{dhn, ccgz}: they are the lightest 
solitons, given their $O(2N)$ and topological quantum numbers. Our new result 
is the marginal stability of the heavier, topologically non-trivial solitons, 
with a single pair of non-zero bound states ($K=2, |q|=1$) \cite{heavykink}. 
For all other static soliton multiplets, there are always decay 
channels in which (\ref{Nfnecessary}) and (\ref{energynecessary}) are 
mutually consistent. In such cases, more detailed analysis is required
to establish stability or instability of the soliton under consideration. 
Nevertheless, given that the heavier topologically non-trivial soliton 
($K=2, |q|=1$) is already at the threshold of stability, we {\em 
conjecture} that all these other multiplets are unstable.  

\subsection{Investigating Stability of Topologically Trivial Solitons}
Consider the decaying parent soliton to be a topologically trivial static 
soliton with $K$ pairs of bound states at non-zero energies, corresponding 
to the direct product 
of $K$ antisymmetric tensor representations of ranks 
$\tilde\nu_1, \ldots, \tilde\nu_K$. The corresponding angular variables 
$\Theta_1, \ldots, \Theta_K$ are determined by (\ref{extremaltheta}), namely, 
$\Theta_n = {\pi\tilde\nu_n\over 2N}$. Thus, according to (\ref{solitonmass})
and (\ref{valenceNfnspectrum}), the mass of this 
soliton is $M(\tilde\nu_1, \ldots, \tilde\nu_K) = {2Nm\over\pi}
\sum_{n=1}^K\,\sin \Theta_n$, and the maximal fermion number eigenvalue 
occurring in this representation is $N_f^{max}(D_{\rm parent}) = 
\sum_{n=1}^K\,\tilde\nu_n$.

Following the strategy which we laid above, we shall now scan through all 
imaginable decay channels of this parent soliton (into final states of purely 
static solitons), and identify those channels which are necessarily closed.

We first scan through all decay channels into topologically trivial solitons.
Thus, assume that the parent soliton under consideration decays into a 
configuration of lighter topologically trivial solitons, with 
quantum numbers of the direct product of $L$ antisymmetric tensor 
representations $\nu_1,\ldots ,\nu_L$ (and 
corresponding angular parameters $\theta_1,\ldots ,\theta_L$). The way in 
which these $L$ multiplets are arranged into extremal fermion bags is of no 
consequence to our discussion. Thus, we discuss all decay channels consistent 
with these quantum numbers in one sweep.

The necessary conditions (\ref{Nfnecessary}) and (\ref{energynecessary}) imply 
\beq\label{1KintoL}
\sum_{n=1}^K\Th_n \leq \sum_{i=1}^L \th_i \,,\quad\quad \sum_{n=1}^K 
\sin\Th_n \geq \sum_{i=1}^L \sin\th_i\,,
\eeq
where all $0 < \Th_n, \th_i < \hp$. The two pairs of boundary 
hypersurfaces in (\ref{1KintoL}) are 
\beqra\label{1KintoLsurfaces}
&&\Sigma_1, \tilde\Sigma_1: 
\quad\quad\quad\quad \th_1 +\cdots + \th_L = \Th_1+\cdots + 
\Th_K \nonumber\\
&&\Sigma_2, \tilde\Sigma_2: 
\quad \sin\th_1 +\cdots + \sin\th_L = \sin\Th_1 +\cdots + 
\sin\Th_K \,,
\eeqra
where $\Sigma_{1,2}$ are hypersurfaces in $\theta$-space and 
$\tilde\Sigma_{1,2}$ are the corresponding hypersurfaces in $\Theta$-space.

The two necessary conditions in (\ref{1KintoL}) will contradict each other, 
thereby protecting the parent soliton against decaying through the channel 
under consideration, if and only if the hypersurface $\Sigma_2$ lies (in the 
positive orthant) between the origin and the hyperplane $\Sigma_1$. 
When will this happen? 

To answer this question it is useful to split 
\beq\label{sbeta}
\sin\Th_1 +\cdots +  \sin\Th_K = s + \sin\beta\,,\quad {\rm where}
\eeq
\beqra\label{integralfractional}
s &=& {\rm integral~part~of}\,[\sin\Th_1 +\cdots +  \sin\Th_K]\quad 
{\rm and}\nonumber\\
\sin\beta &=& {\rm fractional~part~of}\,[\sin\Th_1 +\cdots +  \sin\Th_K]\,.
\eeqra
Thus, the integer $0\leq s\leq K$, and $0\leq\beta\leq{\pi\over 2}$.

The variables $\theta_i$ are restricted to the $L$ dimensional 
hypercube $[0,{\pi\over 2}]^L$ in the positive orthant. The hypersurface 
$\Sigma_2$ intersects the boundaries of this cube along (curved) subsimplexes
whose vertices are 
\beq\label{intersectionvertices}
\theta_i^{(v)} = {\pi\over 2} \left( \delta_{ii_1} + \delta_{ii_2} + \ldots + 
\delta_{ii_s}\right) + \beta\delta_{ii_{s+1}}\,,\quad (i=1, \ldots, L)
\eeq
with all possible choices of $s+1$ coordinates $i_1, \ldots, i_{s+1}$ out of 
$L$ (as the reader can convince herself or himself by working out explicitly a 
few low dimensional examples). All these vertices lie on the hypersurface 
\beq\label{Lintersectionhypersurface}
\Sigma_{s\beta}:\quad\quad\quad\quad \th_1 +\cdots + \th_L = \beta + 
{\pi\over 2}s\,.
\eeq
The curved hypersurface $\Sigma_2$ is dented towards the origin. In the 
positive orthant it lies between the hypersurface $\Sigma_{s\beta}$ and the 
origin. Thus, the two necessary conditions in (\ref{1KintoL}) will contradict 
each other as required, if and only if $\Sigma_{s\beta}$ lies 
between $\Sigma_1$ and the origin, namely, when the parameters of the parent 
soliton are restricted according to 
\beq\label{contradiction}
\Th_1+\cdots + \Th_K  \geq \beta + {\pi\over 2}s\,.
\eeq
At the same time, these parameters are subjected to (\ref{sbeta}), 
which picks up a particular hypersurface $\tilde\Sigma_2$ in the
space of the $\Theta$'s. Thus, the desired
contradiction of the two necessary conditions in (\ref{1KintoL}) occurs for 
parent solitons which correspond to points on the intersection of 
$\tilde\Sigma_2$ and (\ref{contradiction}). Let us determine this intersection:

The hypersurface $\tilde\Sigma_2$ intersects the boundaries of the 
$K$ dimensional hypercube $[0,{\pi\over 2}]^K$ (the range of the $\Theta_n$'s) 
along (curved) subsimplexes whose vertices are 
\beq\label{Kintersectionvertices}
\Theta_n^{(v)} = {\pi\over 2} \left( \delta_{nn_1} + \delta_{nn_2} + \ldots + 
\delta_{nn_s}\right) + \beta\delta_{nn_{s+1}}\,,\quad (n=1, \ldots, K)
\eeq
with all possible choices of $s+1$ coordinates $i_1, \ldots, i_{s+1}$ out of 
$K$ (similarly to the intersection of $\Sigma_2$ and the boundary of the 
$L$ dimensional cube $[0,{\pi\over 2}]^L$). All these vertices 
lie on the hypersurface 
\beq\label{Kintersectionhypersurface}
\tilde\Sigma_{s\beta}:\quad\quad\quad\quad \Th_1 +\cdots + \Th_K = \beta + 
{\pi\over 2}s\,,
\eeq
which is, of course, the boundary of (\ref{contradiction}). 
Furthermore, $\tilde\Sigma_2$ is dented towards the origin and lies (in the
positive orthant) between the origin and $\tilde\Sigma_{s\beta}$. Thus, we 
conclude that the desired intersection is precisely the set of all 
vertices defined by (\ref{Kintersectionvertices}).

Each such vertex represents a soliton which cannot decay through the channel 
under consideration, and is thus potentially stable. More precisely, all 
these vertices correspond to the same soliton, since the coordinates of 
these vertices are just permutations of each other, and thus all of them 
correspond to the same 
set of parameters, in which 
\beqra\label{potentiallystable}
&&s~ {\rm of~the~} \Theta{\rm 's~ are~ degenerate~ and~ equal~ to~}\hp
\nonumber\\
&&{\rm one~ of~ the~} \Theta{\rm 's~ is~equal~ to~} \beta\,,~ {\rm and}
\nonumber\\ 
&&{\rm the~remaining~} K-(s+1)~ \Theta{\rm 's~ are~ null.}
\eeqra 
In Appendix F we provide an alternative proof of (\ref{potentiallystable}).

Does such a soliton exist? To answer this question let us recall a few 
basic facts: The parent soliton under discussion is topologically trivial. As
such, it must bind fermions to be stabilized, and none of its bound state
energies may vanish. Thus, all the ranks occurring in it must satisfy 
$0 < \tnu_n < N$, in accordance with (\ref{nurange}). Finally, recall that 
all the $\om_n$'s which parametrize a given static 
soliton, must be different from each other (see the paragraph following 
(\ref{extremaltheta})). Gathering all these facts we deduce from 
(\ref{thetaomkappa}) and (\ref{extremaltheta}) that all the $\Theta_n$'s must 
be different from each other, and furthermore, that each 
$\Theta_n \neq 0,{\pi\over 2}$. 

Thus, the only physically realizable parent solitons, which are {\em 
necessarily} stable against the decay channel in question, correspond to 
$s=0$ and $K=s+1 = 1$. These are, of course, the static DHN solitons with 
profile (\ref{dhnprofile}) and mass (\ref{dhnmass}).

A corollary of our analysis so far is that DHN solitons are  stable against 
decaying into $L_f$ free fermions or antifermions\footnote{Strictly speaking, 
this argument is valid only for values of $L_f$ which are a finite fraction 
of $N$, since our mass formula (\ref{solitonmass}) is the leading order in 
the $1/N$ expansion, while removing a finite number of particles from the 
parent soliton is a perturbation of the order $1/N$ relative to 
(\ref{solitonmass}).} (i.e., $L_f$ fundamental $O(2N)$ representations) 
plus $L-L_f$ solitons corresponding to higher antisymmetric tensor 
representations. In particular, it is stable against complete evaporation 
into free fermions. This fact manifests itself in the binding energy 
function (\ref{bindingenergy}),
which for a DHN soliton of rank $\nu = {2N\Theta\over\pi}$ is 
\beq\label{dhnbindingenergy}
B(\Theta) = {2Nm\over\pi}(\Theta - \sin\,\Theta)\,.
\eeq
$B(\Theta)$ is positive and increases monotonically in the physical range 
$0 < \Theta < {\pi\over 2}$, and so does the binding energy per
fermion, ${B(\Theta)\over \nu} = m\left (1 - {\sin\,\Theta\over\Theta} 
\right)$. This is in contrast with binding in nuclei, where the binding 
energy per nucleon saturates. The DHN soliton becomes ever more stable as 
it traps more fermions (up to the maximal number $\nu=N$). This is the 
so-called ``mattress effect'', familiar from the physics of fermion bags.

All the DHN solitons ($K=1, q=0$) have masses below the kink-antikink 
threshold ${2Nm\over\pi}$. Thus they cannot decay into pairs of 
topologically non-trivial solitons. The topologically trivial solitons 
with $K\geq 2$, on the other hand, may have masses above the kink-antikink 
threshold, but there is not much interest in looking for such heavier 
solitons which are stable against decaying into pairs of topological defects, 
since we have already established that they are not necessarily
protected against decaying into topologically trivial solitons\footnote{For 
what it worths, it can be shown, by going through the same analysis as in the 
previous case, that there are no decay channels into pairs of topological 
solitons, against which any of the $K\geq 2, q=0$ solitons 
(with mass above the kink-antikink threshold) is necessarily protected.}.

To summarize, we have thus covered all possible decay channels of a given 
static topologically trivial soliton into a configuration of other 
solitons (static or time dependent).  We have found that the only static 
solitons, which are {\em necessarily} stable against decaying into any of 
these channels (in the sense that (\ref{Nfnecessary}) and 
(\ref{energynecessary}) will always contradict each other), are just the 
DHN solitons. This is of course, well known. It follows almost trivially 
from the fact that if $0\leq\theta_n\leq {\pi\over 2}, \forall n $ and  
$0\leq \sum_n \theta_n \leq {\pi\over 2}$, then  
$\sin \left(\sum_n \theta_n\right) \leq \sum_n \sin\theta_n$.

\subsection{Investigating Stability of Topologically Non-Trivial Solitons}

The CCGZ kink and antikink are the lightest states with topological 
charge $q=\pm 1$. Thus they are stable. Are there any other stable 
topologically non-trivial solitons? 

To answer this question we will apply our strategy to specify all 
topologically non-trivial solitons which are necessarily stable.  
Thus, consider such a generic parent soliton. 
It has a zero energy bound state $\om_1=0$ with its corresponding 
spinorial representation factor, and additional $K-1$ pairs of bound states 
at non-zero $\om_n$'s, with the direct product of their $K-1$ antisymmetric 
tensor representations of ranks $\tilde\nu_1, \ldots, \tilde\nu_{K-1}$. 
The corresponding angular variables $\Theta_1, 
\ldots, \Theta_{K-1}$ are determined by (\ref{extremaltheta}), namely, 
$\Theta_n = {\pi\tilde\nu_n\over 2N}$. Thus, according to 
(\ref{solitonmass}), (\ref{valenceNfnspectrum}) and (\ref{Nfzeromode}), 
the mass of this soliton is $M(\tilde\nu_1, \ldots, \tilde\nu_{K-1}) = 
{Nm\over\pi} + {2Nm\over\pi}\sum_{n=1}^{K-1}\,\sin \Theta_n$, and the 
maximal fermion number eigenvalue occurring in this representation is 
$N_f^{max}(D_{\rm parent}) = {N\over 2} + \sum_{n=1}^{K-1}\,\tilde\nu_n$.

The parent soliton may decay into a final state, which will contain, in 
addition to a lighter soliton with the same topological charge,  
$P$ pairs of topologically non-trivial solitons (with opposite topological 
charges), and a bunch of topologically trivial solitons. The final state of 
the decay is thus most generally a direct product of some $L$ antisymmetric 
tensor representations $\nu_1,\ldots ,\nu_L$ (and corresponding angular 
parameters $\theta_1,\ldots ,\theta_L$), and $2P+1$ spinorial representations.
As in the previous case, the way in which the $L$ antisymmetric tensors are 
arranged into extremal fermion bags is of no consequence to our discussion.

The necessary conditions (\ref{Nfnecessary}) and (\ref{energynecessary}) now 
imply 
\beq\label{spinorKintoL2P}
\sum_{n=1}^{K-1}\Th_n \leq  {\pi\over 2} P + \sum_{i=1}^L \th_i \,,\quad\quad 
\sum_{n=1}^{K-1} \sin\Th_n \geq  P + \sum_{i=1}^L \sin\th_i\,,
\eeq
(where all $0 < \Th_n, \th_i <\hp$). These conditions are identical 
to (\ref{1KintoL}), with $K$ there replaced here by $K-1$ and with $s\geq P$. 
Going through the same analysis as we did for the topologically trivial 
soliton, we conclude that the two necessary conditions (\ref{spinorKintoL2P}) 
contradict each other only when $s=0$ (which means $P=0$), and $K-1 \leq s+1 
=1$.

Thus, in addition to the CCGZ kink ($K=1$), we have discovered a heavier 
stable, topologically non-trivial soliton ($K=2$). As in \cite{heavykink}, we 
shall refer to it as the ``Heavier Topological Soliton'' (HTS), which is 
a bound state of a DHN soliton and a kink or an antikink.
Its mass is 
\beq\label{HTS}
M_{HTS,\nu} = {Nm\over\pi} + {2Nm\over\pi}\sin\left({\pi\nu\over 2N}\right)\,,
\eeq
where $\nu$ is the rank of its antisymmetric tensor factor, which is tensored 
with the $2^N$ dimensional spinor. Thus, $M_{HTS,\nu}$ coincides with 
the sum of masses of a CCGZ kink and a DHN soliton of rank $\nu$. 
Drawing further the analogy between the solitons discussed in this Review 
and baryons in QCD with large $N_{color}$, the HTS would correspond to a
dibaryon.

The profile of this HTS (for boundary conditions corresponding to $q=1$) is 
given by (\ref{sigmaqone2bsalt}) and (\ref{sigmaqone2bsaltfinal}), which we 
repeat here for convenience:
\beqra\label{HTSprofile}
\sigx &=&  m + {2\kappa\over 1 + {m+\kappa\over m-\kappa}
e^{2\kappa (x-y_0)}}
\nonumber\\{}\nonumber\\
&-& 2(m+\kappa) {1+ {m+\kappa\over 
(m-\kappa)^2}\,\kappa\, e^{2m(x-x_0)} + 
{m+\kappa\over (m-\kappa)^2}\,m\,
e^{2\kappa (x-y_0)}
\over 
1+ \left({m+\kappa\over 
m-\kappa}\right)^2 e^{2m(x-x_0)} + 
\left({m+\kappa\over m-\kappa}\right)^2 e^{2\kappa (x-y_0)} + 
\left({m+\kappa\over m-\kappa}\right)^2 
e^{2m (x-x_0)+2\kappa(x-y_0)}}\nonumber\\{}\nonumber\\
&=& -\kappa\tanh\left[\kappa(x-y_0 + R)
\right] \nonumber\\{}\nonumber\\
&+& \om_b{\sinh\left[m(x-x_0) +\kappa(x-y_0)+ 2\kappa R\right] + 
\sinh\left[m(x-x_0) -\kappa(x-y_0)\right]\over
e^{-\kappa R}\cosh\left[m(x-x_0) +\kappa(x-y_0)+ 2\kappa R\right] +
e^{\kappa R}\cosh\left[m(x-x_0) -\kappa(x-y_0)\right]}\,,\nonumber\\{}
\eeqra
with $R = {1\over 2\kappa}\log\left({m+\kappa\over m-\kappa}\right)$
(Eq.(\ref{Rdistance})) and 
$\kappa = \sqrt{m^2-\om_b^2} = m\sin\left({\pi\nu\over 2N}\right)$, as usual. 
The arbitrary real parameters $x_0$ and $y_0$ are related to the 
coefficients $c_1, c_2$ in (\ref{scatteringdatatext}) via 
(\ref{collectivecoordinates}). These are the flat directions, or moduli, 
which we discussed in subsection 3.1.1. We can invoke global translational 
invariance to set one of these collective coordinates, say $x_0$, to zero. 
The remaining parameter $y_0$ is left arbitrary.

In \cite{heavykink} we have studied the profile of 
(\ref{HTSprofile}) in some detail. (See Figures 1-3 in \cite{heavykink}.)
In general, it has the shape of the profiles of a CCGZ kink and a 
DHN soliton superimposed on each other, which move relative to each other 
as we vary the remaining free modulus $y_0$. The latter is essentially the
separation of the kink and DHN soliton. The mass (\ref{HTS}) of the HTS remains
fixed throughout this variation, as we discussed in subsection 3.1.1.

The fact that $\partial M_{HTS,\nu}/\partial y_0 =0$, just means that 
the kink and DHN soliton exert no force on each other, whatever their 
separation is. In other words - they are very loosely bound.
Indeed, it is evident from (\ref{HTS}) that the HTS is 
degenerate in mass with a pair of non-interacting CCGZ kink and a DHN 
soliton of rank $\nu$. If we interpret the latter two as the constituents 
of the HTS, they are bound at threshold. Thus, the HTS is marginally stable 
\cite{heavykink}.

This is a somewhat surprising result, since one would normally expect
soliton-soliton interactions to be of the order $\frac{1}{g^2}\sim N$
in a weakly interacting field theory. This is consistent, of course, with
what one should expect from general $\frac{1}{N}$ counting 
rules \cite{wittenlargeN}, according to which the baryon-baryon interaction 
is of order $N$. Yet, due to dynamical reasons which ellude us at this point, 
the solitonic constituents of the HTS avoid these general considerations and 
do not exert force on each other.

The limit $\nu\rightarrow N$ is of some interest. Strictly speaking, there is
no HTS with $\nu=N$, as was discussed above.
However, it is possible to study HTS's with $\nu$ arbitrarily close to $N$. 
In this case, $\kappa\rightarrow m$, with $R\rightarrow \infty$ 
in (\ref{HTSprofile}). Thus, for $|x|, |x_0|, |y_0| << R$, 
(\ref{HTSprofile}) tends in this limit to 
\beq\label{sigmaqone2bsntoN}
\sigx = m{1-e^{-2m(x-y_0)}-e^{-2m(x-x_0)}
\over 1+e^{-2m(x-y_0)}+e^{-2m(x-x_0)}}\,.
\eeq
In the asymptotic region $1<< m|x_0-y_0|~
(<< mR)$ (\ref{sigmaqone2bsntoN}) simplifies further, and appears as a 
kink  $m\tanh[m(x-x_{\rm max})]$, located at $x_{\rm max} = 
{\rm max}\{x_0, y_0\}$. This clearly has mass $M_{kink} = {Nm\over\pi}$, but 
according to (\ref{HTS}), $M_{HTS,\nu\simeq N}$ should tend to 
${3Nm\over\pi} = 3M_{kink}$. The extra mass $2M_{kink}$ corresponds, of 
course, to the kink-anti-kink pair which receded to spatial infinity.

\pagebreak

%


\newpage
\setcounter{equation}{0}
\setcounter{section}{0}
\renewcommand{\theequation}{A.\arabic{equation}}
\renewcommand{\thesection}{Appendix A:}
\section{Resolvent of the Dirac Operator With Static Background Fields
and Supersymmetry}
\vskip 5mm
\setcounter{section}{0}
\renewcommand{\thesection}{A}

In this Appendix we recall some useful properties of the resolvent of
the Dirac operator $D = \om\gam^0 + i\gam^1\pax -\sigx $
in a static background. In particular, we make connections with 
supersymmetric quantum mechanics\footnote{For a thorough review of 
supersymmetric quantum mechanics see \cite{cooper}.} and use it to express 
all four entries of the diagonal resolvent in terms of a single function. 
This will be useful in evaluating masses of fermion bags.

The presentation in this Appendix is a special case of the
recent discussion in \cite{reflectionless}, which analyzes the more generic
Dirac operator $D\equiv i\notpa-(\sigx+i\pix\gam_5)$. The spectral theory of 
the latter operator gives rise to an interesting generalization of 
supersymmetric quantum mechanic which was discussed in \cite{reflectionless}.
We also mention that supersymmetric quantum mechanics was instrumental in the 
analysis of fermion bags in \cite{josh1}. 

As was discussed in Section 2, our task is to invert the Dirac operator 
(\ref{dirac}), $D = \om\gam^0 + i\gam^1\pax -\sigx $, in a static 
background $\sigx$ subjected to the boundary conditions 
(\ref{boundaryconditions}). We emphasize that inverting (\ref{dirac}) 
has nothing to do with the large $N$ approximation, and consequently our 
results in this section are valid for any value of $N$.

In the Majorana representation (\ref{majorana}) we are using in 
this paper, $\gam^0=\si_2\;,\gam^1=i\si_3$ and $\gam^5=-\gam^0\gam^1=\si_1$,
the Dirac operator (\ref{dirac}) is 
\beq
D =\left(\begin{array}{cc} -\pa_x - \si & -i\omega \\{}&{}\\ 
i\omega & \pa_x - \si\end{array}\right)
=\left(\begin{array}{cc} -Q & -i\omega  \\{}&{}\\ 
i\omega & -Q^\dgg\end{array}\right)\,,
\label{dirac1}
\eeq
where we introduced the pair of adjoint operators 
\beq\label{qqdagger}
Q = \sigx + \pa_x\,,\quad\quad Q^\dgg = \sigx - \pa_x\,.
\eeq

Inverting (\ref{dirac1}) is achieved by solving  
\beq
\left(\begin{array}{cc} -Q & -i\omega  \\{}&{}\\ 
i\omega & -Q^\dgg\end{array}\right)\cdot 
\left(\begin{array}{cc} a(x,y) &  b(x,y) \\{}&{}\\ c(x,y) & 
d(x,y)\end{array}\right)\,=\,-i{\bf 1}\del(x-y)
\label{greens}
\eeq
for the Green's function of (\ref{dirac1}) in a given background 
$\sigx$. By dimensional analysis, we see that the quantities $a,b,c$ and 
$d$ are dimensionless.

\subsection{``Supersymmetry'' in a Bag Background}
The diagonal elements $a(x,y), ~d(x,y)$ in (\ref{greens}) may be 
expressed
in term of the off-diagonal elements as
\beq
a(x,y)={-i\over \omega}Q^\dgg c(x,y)\,,\quad\quad
d(x,y)={i\over \omega} Q b(x,y)
\label{ad}
\eeq
which in turn satisfy the second order partial differential equations
\beqra
&&\left(Q^\dgg Q -\om^2\right){b(x,y)\over\omega}\,=
\left[-\pa_x^2 + \sigx^2 - \si'(x)-\omega^2 \right]{b(x,y)\over \omega}\,
=\,~\del(x-y)\nonumber\\&&{}\nonumber\\
&&\left(Q Q^\dgg -\om^2 \right){c(x,y)\over\omega}\,=
\left[-\pa_x^2 + \sigx^2 + \si'(x)-\omega^2 \right]{c(x,y)\over 
\omega}\,=\,-\del(x-y)\,.\nonumber\\&&{}
\label{bc}
\eeqra

Thus, $b(x,y)/\om$ and $-c(x,y)/\om$ are simply the Green's functions of the 
corresponding Schr\"odinger operators
\beq\label{bcops}
H_b = Q^\dgg Q = \left[-\pa_x^2 + \sigx^2 - \si'(x)\right]\,,\quad\quad
{\rm and}\quad\quad 
H_c = Q Q^\dgg = \left[-\pa_x^2 + \sigx^2 + \si'(x)\right] 
\eeq
in (\ref{bc}), namely, 
\beqra
{b(x,y)\over\om}&=&{\theta\left(x-y\right)b_2(x)b_1(y)
+\theta\left(y-x\right)b_2(y)
b_1(x) \over W_b}
\nonumber\\{}\nonumber\\
{c(x,y)\over\om}&=&-{\theta\left(x-y\right)c_2(x)c_1(y)
+\theta\left(y-x\right)c_2(y
)c_1(x)\over W_c}\,.
\label{bcexpression}
\eeqra
Here $\{b_1(x), b_2(x)\}$ and $\{c_1(x), c_2(x)\}$ are pairs of independent 
fundamental solutions of the two equations 
\beq\label{eigenvalueequations}
H_b b(x)=\om^2 b(x)\quad\quad {\rm and}\quad\quad H_c c(x)=\om^2 c(x)\,, 
\eeq
subjected to the 
boundary conditions 
\beq\label{planewaves}
b_1(x)\,,c_1(x) \asymptoticm A_{b,c}^{(1)}e^{-ikx} \quad\quad ,\quad\quad 
b_2(x)\,,c_2(x)\asymptoticp A_{b,c}^{(2)}e^{ikx}
\eeq
with some possibly $k$-dependent coefficients $A_{b,c}^{(1)}(k), 
A_{b,c}^{(2)}(k)$ and 
with\footnote{We see that if ${\rm Im}k>0$, $b_1$ and $c_1$ decay 
exponentially to the left, and $b_2$ and $c_2$ decay to the right. 
Thus, if ${\rm Im}k>0$, both  $b(x,y)$ and $c(x,y)$ decay as 
$|x-y|$ tends to infinity.} 
\beq\label{kmom}
k=\sqrt{\om^2 -m^2}\,,\quad {\rm Im}k\geq 0\,.
\eeq
The purpose of introducing the (yet unspecified) coefficients 
$A_{b,c}^{(1)}(k),  A_{b,c}^{(2)}(k)$ will become clear later, 
following Eqs. (\ref{btoc}) and (\ref{ctob}) below. 
The boundary conditions (\ref{planewaves}) are consistent, of course, with 
the asymptotic behavior (\ref{boundaryconditions}) on $\si$ due to 
which both $H_b$ and $H_c$ tend to a free particle hamiltonian 
$-\pa_x^2 + m^2$ as $x\rightarrow\pm\infty$.

The wronskians of these pairs of solutions are 
\beqra\label{wronskian}
&& W_b(k) =b_2(x)b_1^{'}(x)-b_1(x)b_2^{'}(x)
\nonumber\\&&{}\nonumber\\
&& W_c(k) = c_2(x)c_1^{'}(x)-c_1(x)c_2^{'}(x)\,.
\nonumber\\&&{}
\eeqra
As is well known, $W_b(k)$ and $W_c (k)$ are independent of $x$.

We comment at this point (for later reference) that in the scattering 
theory of the operators 
$H_b$ and $H_c$ it is more customary to consider pairs of independent 
fundamental solutions $\{b_1(x), b_2(x)\}$ and $\{c_1(x), c_2(x)\}$ for 
which the coefficients $A_{b,c}^{(1)}, A_{b,c}^{(2)}$ in (\ref{planewaves})
are $k$ independent, because, as is well known in the 
literature \cite{faddeev}, with such boundary conditions, the corresponding 
wronskians are proportional (up to a $k$-independent coefficient) 
to $k/t(k)$, where $t(k)$ is the transmission amplitude of the 
corresponding operator $H_b$ or $H_c$. 
Thus, we will refer to the wronskians of pairs of independent fundamental 
solutions with asymptotic behavior (\ref{planewaves}) with 
$A_{b,c}^{(1)}= A_{b,c}^{(2)}=1$ as the {\em standard} wronskians 
$W_b^{st}(k)$ and $W_c^{st}(k)$. Therefore, we can express the wronskians 
(\ref{wronskian}) of pairs of fundamental solutions with asymptotic 
behavior (\ref{planewaves}) as
\beq\label{standard}
W_b(k)= A_{b}^{(1)}(k) A_{b}^{(2)}(k)W_b^{st}(k)
\quad\quad {\rm and} \quad\quad 
W_c(k) = A_{c}^{(1)}(k) A_{c}^{(2)}(k) W_c^{st}(k)\,.
\eeq

We remind the reader that at a bound state energy $\om=\om_b$ (or, 
equivalently, at the corresponding $k_b=i\sqrt{m^2-\om_b^2}$), say, 
of the operator $H_b$, the transmission amplitude $t_b(k)$ of $H_b$ has a 
simple pole, $t_b(k)\sim 1/(k-k_b)$. Thus, $W_b^{st}(k)\sim k/t_b(k)$ 
vanishes as 
\beq\label{vanishwronskian}
W_b^{st} (k) \sim k_b (k- k_b)\quad\quad ,\quad\quad k\sim k_b
\eeq 
near that bound state. This behavior occurs simply 
because at the bound state $k=k_b$, $b_1(x)$ and $b_2 (x)$ are both 
proportional to the bound state wave function (which decays asymptotically as 
${\rm exp} (-\sqrt{m^2 - \om_b^2}|x|)$ ), and are thus linearly dependent. 
Thus, the zeros of $W_b^{st}(k)$ determine the spectrum of bound states of 
$H_b$. The non-standard wronskian $W_b(k)$ will have 
some extra $k$ dependence resulting from the factors 
$ A^{(1)}(k) A^{(2)}(k)$ in (\ref{standard}). 
Similar assertions hold, of course for $H_c$, $W_c^{st}$ and $W_c$.

Substituting the expressions (\ref{bcexpression}) for the off-diagonal 
entries $b(x,y)$ and $c(x,y)$ into (\ref{ad}), we obtain the appropriate 
expressions for the diagonal entries $a(x,y)$ and $d(x,y)$. We do not bother
to write these expressions here. It is useful however to note, that despite 
the $\pax$'s in the $Q$ operators in (\ref{ad}), that act on the 
step functions in (\ref{bcexpression}), neither $a(x,y)$ nor $d(x,y)$ contain 
pieces proportional to $\del(x-y)$\,. Such pieces cancel one another due 
to the symmetry of (\ref{bcexpression}) under $x\leftrightarrow y$.

The factorization (\ref{bcops}) of $H_b = Q^\dgg Q$ and $H_c = Q Q^\dgg$ 
into a product of the two first order differential operators 
$Q$ and $Q^\dgg$ is the hallmark of supersymmetry. 
As is well known, this factorization means that $H_b$ and $H_c$ are 
{\em isospectral}, i.e., have the same eigenvalues, save possibly an 
unmatched zero mode which belongs to the spectrum of only one of the 
operators. 

Let us now recall how this property arises: 
The factorized equations 
\beq\label{bcfactor}
Q^\dgg Q\, b = \om^2 b\,\quad\quad {\rm and}\quad\quad Q Q^\dgg\, c = \om^2 c
\eeq
suggest a map between
their solutions. Indeed, given that $H_b b = \om^2 b $, then clearly
\beq\label{btoc}
c(x) = {1\over \om} Q\, b(x)
\eeq
is a solution of $H_c c = \om^2 c$. The factor ${1\over \om}$ in 
(\ref{btoc}) is relevant for the case in which $b(x)$ is an 
eigenstate of $H_b$ (with eigenvalue $\om^2$). In such a case, $c(x)$ on the 
left hand side of (\ref{btoc}) is an eigenstate of $H_c$ with the same
eigenvalue $\om^2$, and furthermore, due to the factor ${1\over \om}$, it also
has same norm as $b(x)$. Similarly, if $H_c c =\om^2 c$, then 
\beq\label{ctob}
b(x) = {1\over \om } Q^\dgg \,c(x)
\eeq
solves $H_b b =\om^2 b$, and has the same norm as $c(x)$, in case $c(x)$ 
is an eigenstate of $H_c$.

We remark, that in the more generic case of the Dirac operator 
$i\notpa -\sigx-\gam_5\pix$ discussed in \cite{reflectionless}, the factors 
analogous to the factors ${1\over\om}$ in (\ref{btoc}) and (\ref{ctob}) 
(factors of ${1\over \om\pm\pix}$, see \cite{reflectionless}, 
Eqs. (2.12)-(2.14)), arise already at the level of the 
differential equations analogous to (\ref{bcfactor}), and not from 
considerations of the norm of $b(x)$ and $c(x)$.

One particular useful consequence of the mappings (\ref{btoc}) and 
(\ref{ctob}) is that given a pair $\{b_1(x), b_2(x)\}$ of independent 
fundamental solutions of $(H_b-\om^2) b(x) =0$, we can obtain from it 
a pair $\{c_1(x), c_2(x)\}$ of independent 
fundamental solutions of $(H_c-\om^2) c(x) =0$ by using (\ref{btoc}), and 
vice versa. Therefore, with no loss of generality, we henceforth assume,
that the two pairs of independent fundamental solutions $\{b_1(x), b_2(x)\}$
and $\{c_1(x), c_2(x)\}$, are related by (\ref{btoc}) and (\ref{ctob}).

The coefficients 
$A_{b,c}^{(1)}(k), A_{b,c}^{(2)}(k)$ in (\ref{planewaves}) are to be adjusted
according to (\ref{btoc}) and (\ref{ctob}), and this was the purpose of
introducing them in the first place.

It is clear (and of course well known from the literature on 
supersymmetric quantum mechanics), that the mappings 
$b(x)\leftrightarrow c(x)$ break down if either $H_b$ or $H_c$ 
has an eigenstate at zero energy. In case that zero mode exists, it must 
be the ground state of the corresponding operator $H_b$ or $H_c$, since these
operators obviously cannot have negative eigenvalues. 

\subsubsection{Zero Modes and Topology}
Let us determine under which conditions such zero-energy eigenstates exist
and what they look like. Assume, for example, that $H_b$ has 
its ground state $|b_0\rangle$ at zero energy. Our system is defined 
over the whole real axis, thus bound states are strictly square-integrable.  
By assumption $\langle b_0 |H_b| b_0\rangle = \langle b_0 | Q^\dgg Q 
| b_0\rangle = \Big| |Q | b_0\rangle \Big|^2 =0$. Thus, $H_b | b_0\rangle =0$ 
implies the first-order equation 
\beq\label{qbzero}
Q\,b_0(x) = b_0'(x) + \sigx b_0 (x) = 0\,.
\eeq
Thus, 
\beq\label{bzero}
b_0(x) = \exp -\int^x \si(y) dy\,.
\eeq
Similarly, if $H_c$ has its ground state $|c_0\rangle$ at zero energy, 
we have 
\beq\label{czero}
Q^\dgg\,c_0(x) = -c_0'(x) + \sigx c_0 (x) = 0\quad\quad {\rm namely}\quad\quad
c_0(x) = \exp +\int^x \si(y) dy\,.
\eeq
For $\sigx$ subjected to the boundary conditions (\ref{boundaryconditions}),
the wave function (\ref{bzero}) is square-integrable if and only if 
$\si(\infty) = -\si(\-\infty) = m$, i.e., only when 
$\sigx$ carries topological charge $q=1$. Similarly, the wave function 
(\ref{czero}) is square-integrable if and only if 
$\si(\infty) = -\si(\-\infty) = -m$, i.e., only when 
$\sigx$ carries topological charge $q=-1$. Thus, $b_0$ and $c_0$ cannot
be simultaneously square-integrable, and correspondingly, only one of the 
operators $H_b$ and $H_c$ can have a zero energy ground state in a given
topologically non-trivial background $\sigx$. Of course, none of these 
operators has a zero energy state if $\sigx$ is topologically trivial, i.e., 
when $q=0$

To summarize, if $\sigx$ carries topological charge $q=1$, $H_c$ is 
strictly positive and isospectral to $H_b$, except for the normalizable 
zero-mode (\ref{bzero}) of $H_b$. In the opposite case, where $\sigx$ carries 
topological charge $q=-1$, $H_b$ is strictly positive and isospectral to 
$H_c$, except for the normalizable 
zero-mode (\ref{czero}) of $H_c$. In the third case, when $\sigx$ is 
topologically trivial, $q=0$, neither $H_b$ nor $H_c$ 
has a zero energy state, and they are strictly isospectral.

Let us reformulate all this in terms of the Dirac equation (\ref{diraceq}). 
Assuming the existence of a bound state at $\om=0$, $\psi(x,t) = u(x)$, we 
write 
\beq\label{zerodiraceq}
\left[i\gam^1\pax-\sigx \right]\,u(x) = 0\,.
\eeq
This equation may be further reduced as follows: The matrix $i\gam^1$ is 
hermitean with eigenvalues $\pm1 $. (This basis independent property is 
manifest in the particular representation (\ref{majorana}).) Thus, consider 
(\ref{zerodiraceq}) with $i\gam^1 u_{\pm}(x) 
= \pm u_{\pm}(x)$. The corresponding solutions are $u_{\pm}(x) = 
u_{\pm} (0)\,\exp \pm\int\limits_0^x \si (y)\,dy$. If $q=1$, then only 
$u_-(x) = u_-(0) b_0(x)$ is normalizable. Similarly, if $q=-1$, only 
$u_+(x) = u_+(0) c_0(x)$ is normalizable. And if $q=0$, none of these solutions
is normalizable. This is a general feature of the Dirac equation 
(\ref{diraceq}). Thus, as was first noticed in \cite{jackiwrebbi}, 
a fermion which is Yukawa-coupled to a topologically non-trivial scalar 
field (as in (\ref{auxiliary})), will have a {\em single}, unpaired, 
normalizable zero energy mode.

The ``Witten index''\cite{wittenindex} associated with the pair of 
isospectral operators $H_b$ and $H_c$, which in this context may be 
defined\footnote{With this definition, we are identifying $H_b$ and $H_c$, 
respectively, as the hamiltonians of the bosonic and fermionic sectors 
of the supersymmetric system.} 
as 
\beq\label{prewittenindex}
{\rm Witten~index} = {\cal N}_0 (H_b) - {\cal N}_0 (H_c)\,,
\eeq
where ${\cal N}_0 (H_{b,c})$ is the number of normalizable zero modes of 
$H_{b,c}$. This index is of course a topological invariant 
of the space of hamiltonians $H_b$ and $H_c$ defined with configurations 
$\sigx$ that satisfy the boundary conditions (\ref{boundaryconditions}). 
Indeed, it is clear from the discussion above that in our context, 
it coincides with the topological charge (\ref{topological}) of the 
background $\sigx$:
\beq\label{wittenindex}
{\rm Witten~index} = q\,. 
\eeq
Thus, the Witten index  in our system is either 
$\pm 1$ or zero.

If one of the operators $H_b$ or $H_c$ supports a normalizable zero mode,
we say that supersymmetry is unbroken. When none of the 
operators has a normalizable zero mode, we say that supersymmetry is   
broken. Since for our system only one operator can support a normalizable 
zero mode, we may summarize these definitions by saying that for our system
a null Witten index means broken supersymmetry. A non vanishing Witten index,
which in our case can take only the values $\pm 1$, always means unbroken 
supersymmetry. 

Since the Witten index in our model coincides with the topological charge 
(\ref{topological}) $q$, we may rephrase the last paragraph by saying that 
only topologically non-trivial $\sigx$ backgrounds lead to a Dirac operator 
with a normalizable bound state at $\om=0$.

\subsubsection{Wronskians and Isospectrality}
An interesting outcome of the isospectrality of $H_b$ and $H_c$ 
concerns their wronskians. Indeed, from the definition 
(\ref{wronskian}), it follows for pairs of independent fundamental 
solutions $\{b_1(x), b_2(x)\}$ and $\{c_1(x), c_2(x)\}$ which are related 
through (\ref{btoc}) and (\ref{ctob}), that 
\beq\label{wbwc}
{W_c\over\om} = {c_2\pax c_1 - c_1\pax c_2\over \om} = c_1b_2 -c_2b_1 = 
{b_2\pax b_1 - b_1\pax b_2\over \om} = {W_b\over\om}\,,\quad \om\neq 0\,.
\eeq
The wronskians of pairs of independent fundamental solutions of $H_b$ and 
$H_c$ are equal for all $\om\neq 0~$! Eq. (\ref{wbwc}) will play 
an important role in establishing useful properties of the diagonal 
resolvent of the Dirac operator in the next subsection.

With no loss of generality, we may choose 
\beq\label{standardchoice}
A_{b}^{(1)} = A_{b}^{(2)} =1
\eeq 
in (\ref{planewaves}). The coefficients $A_{c}^{(1)}, A_{c}^{(2)}$ are 
then determined by (\ref{btoc}):
\beqra\label{Accoeffs}
A_c^{(1)} &=& {\si(-\infty)-ik\over \om}\nonumber\\{}\nonumber\\
A_c^{(2)} &=& {\si(\infty)+ik\over \om}\,. 
\eeqra
Thus, from (\ref{standard}), (\ref{standardchoice}), (\ref{wbwc}) and 
(\ref{Accoeffs}) we have 
\beq\label{standardb}
W_b (k) = W_b^{st} (k) = 
{(\si(-\infty)-ik)(\si(\infty)+ik)\over \om^2}~W_c^{st}(k) = W_c(k)\,.
\eeq
Using $\om^2=m^2+k^2$, the definition of topological charge $q=(\si(\infty)-
\si(-\infty))/2m$ (\ref{topological}), and the boundary condition $\si(
\pm\infty)^2=m^2$ (\ref{boundaryconditions}), 
we may write (\ref{standardb}) more compactly as 
\beq\label{standardbc}
W_b^{st} (k) = \left(1 + {2mq \over \si(-\infty) + ik}\right)~W_c^{st}(k)\,.
\eeq
Recall from (\ref{vanishwronskian}) that the standard wronskians 
$W_b^{st} (k), W_c^{st} (k)$ vanish, respectively, at the bound states of 
$H_b$ and $H_c$. Thus, {\em contrary} to (\ref{wbwc}), which holds for all
backgrounds $\sigx$, we cannot have $W_b^{st} (k) = W_c^{st} (k)$ when 
$q = \pm 1$, i.e., when only one of the hamiltonians has an unpaired 
normalizable zero energy ground state.  It is 
rewarding to verify that (\ref{standardbc}) is consistent with these facts:

For $q=0$, $H_b$ and $H_c$ are strictly positive and thus must have the 
same bound states. Thus, we expect to find $W_b^{st} (k) = W_c^{st} (k)$, 
which 
is what (\ref{standardbc}) tells us at $q=0$. For $q=1$, $H_b$ and $H_c$ are 
isospectral, save for the unpaired $\om_b^2=0$ ground state of $H_b$. 
$\om_b^2=0$ means $k_b=im$. Thus, we expect to find that $W_b^{st} (k)$ has 
an extra zero at $k=im$, relative to $W_c^{st} (k)$. Indeed, from 
(\ref{standardbc}) with $q=1$ and $\si(-\infty) = -m$,
we find that $W_b^{st} (k) = {k-im\over k+im}W_c^{st}(k)$. 
Finally, for $q=-1$, the roles of $H_b$ and $H_c$ are 
reversed relative to the $q=1$ case, and upon substituting $q=-1$ and 
$\si(-\infty) = m$ in (\ref{standardbc}) we find accordingly
that $W_b^{st} (k) = {k+im\over k-im}W_c^{st}(k)$.

We note in passing that isospectrality of $H_b$ and $H_c$ is consistent 
with the $\gam_5$ symmetry of the system 
of equations in (\ref{greens}), which relates the resolvent of $D$ with that  
of $\tilde D = -\gam_5 D \gam_5 $ (recall (\ref{gamfive})). Due to this 
symmetry, we can map 
the pair of equations $(H_b-\om^2)b(x,y)/\om = \delta (x-y)$ and 
$(H_c-\om^2) c(x,y)/\om = -\delta (x-y)$ (Eqs. (\ref{bc})) on each other by 
\beq\label{bcflip}
b(x,y)\leftrightarrow -c(x,y)\quad {\rm together~ with}\quad  
\si \rightarrow -\si\,.
\eeq 
(Note that under these reflections we also have 
$a(x,y)\leftrightarrow -d(x,y)$, as we can see 
from (\ref{ad}).) The reflection $\si\rightarrow -\si$ flips the signs of 
both asymptotic values $\si(\pm\infty)$ and thus flips the sign of the 
topological charge $q$. It changes the roles of $H_b$ and $H_c$, but 
obviously it cannot change physics. Since this reflection 
interchanges $b(x,y)$ and $c(x,y)$ without affecting the physics, these two 
objects must have the same singularities as functions of $\om$, consistent 
with isospectrality of $H_b$ and $H_c$ (save possibly an unpaired zero mode).

\subsection{The Diagonal Resolvent}

Following \cite{josh2,fz, reflectionless} we define the diagonal resolvent 
$\langle x\,|iD^{-1} | x\,\rangle$ symmetrically as
\beqra
\langle x\,|-iD^{-1} | x\,\rangle &\equiv& \left(\begin{array}{cc} A(x) & 
B(x) \\{}&{}\\ C(x) & 
D(x)\end{array}\right)\nonumber\\{}\nonumber\\{}\nonumber\\
 &=& {1\over 2} \lim_{\epsilon\rightarrow 0+}\left(\begin{array}{cc} 
a(x,y) + a(y,x) &  b(x,y) + b(y,x)\\{}&{}\\ c(x,y) +c(y,x) & d(x,y) + 
d(y,x)\end{array}\right)_{y=x+\epsilon}
\label{diagonal}
\eeqra
Here $A(x)$ through $D(x)$ stand for the entries of the diagonal 
resolvent, which following (\ref{ad}) and (\ref{bcexpression}) have the 
compact representation\footnote{$A, B, C$ and $D$ are obviously functions 
of $\omega$ as well. For notational simplicity we suppress their explicit 
$\omega$ dependence.}
\beqra
B(x)&=&~~{\om b_1(x)b_2(x)\over W_b}\quad\quad , \quad\quad D(x)=i
{\left[\pax+2\sigx\right]B\left(x\right)\over 
2\omega}\,,\nonumber\\
C(x)&=&-{\om c_1(x)c_2(x)\over W_c}\quad\quad , \quad\quad A(x)=i
{\left[\pax-2\sigx\right]C\left(x\right)\over 2\omega}\,.
\label{abcd}
\eeqra
Clearly, from (\ref{bc}) and (\ref{bcops}) we have 
\beqra
{B(x)\over \om} &=& \langle x |\,{1\over H_b - \omega^2 }
\,| x\rangle\nonumber\\{}\nonumber\\
-{C(x)\over \om} &=& \langle x |\,{1\over H_c -\omega^2 }
\,| x\rangle\,.
\label{inversehbhc}
\eeqra
Thus, comparing with \cite{josh1}, we see that $B(x)/\om$ and $-C(x,\om)/\om$
are, respectively, the resolvents 
$R_-(x,\om^2)$ and $R_+(x,\om^2)$ defined in Eqs.(9) and (10) of \cite{josh1}.

The expressions for $A(x)$ and $D(x)$ in terms of $B(x)$ and $C(x)$ in 
(\ref{abcd}) have an interesting property concerning zero modes. Consider, 
for example the case where $H_b$ has a normalizable zero mode. From 
(\ref{inversehbhc}) we expect that as $\om^2\rightarrow 0$, $B(x)/\om$
will be dominated by the pole at $\om_b^2 =0$ with a residue proportional 
to the wave function of the zero energy state (\ref{bzero}) squared,
\beq\label{Batzero}
\omegalim {B(x)\over \om} =  -{b_0^2(x)\over\om^2}\,.
\eeq
How does $D(x)$ behave near that pole? Note from (\ref{bzero}) that 
$(\pax + 2\sigx)b_0^2(x)=0$. Thus, the operator $(\pax + 2\sigx)$, which 
appears in the expression of $D(x)$ in terms of $B(x)$ in (\ref{abcd}),
projects the zero mode $b_0(x)$ out of $D(x)$! Similarly, 
the operator $(\pax - 2\sigx)$, which appears in the expression of $A(x)$ 
in terms of $C(x)$ in (\ref{abcd}), projects the zero mode $c_0(x)$ 
out of $A(x)$. $A(x)$ and $D(x)$ do not have a pole at $\om^2=0$, no matter 
what value the topological charge $q$ takes on.

Since $B(x)/\om$ and $-C(x,\om)/\om$ are, respectively, the diagonal 
resolvents of the Schr\"odinger operators $H_b$ and $H_c$, each one of
them satisfies its appropriate Gel'fand-Dikii \cite{gd} identity:
\beqra\label{dikii}
-2B\pax^2 B + (\pax B)^2  + 4B^2 (\si^2 - \si'-\omega^2) &=& \om^2\nonumber\\
{}\nonumber\\
-2C\pax^2 C + (\pax C)^2  + 4C^2 (\si^2 + \si'-\omega^2) &=& \om^2\,.
\eeqra
A linearized form of these identities may be obtained by deriving once
and dividing through by $2B$ and $2C$. One obtains
\beqra\label{lindikii}
\pax^3 B - 4\pax B (\si^2 - \si'-\omega^2) -2B (\si^2 - \si')'&=& 0\nonumber\\
{}\nonumber\\
\pax^3 C - 4\pax C (\si^2 + \si'-\omega^2) -2C (\si^2 + \si')'&=& 0\,.
\eeqra
(For a simple derivation of the GD identity, see \cite{josh2, fz}.)

\subsubsection{Relations Among A, B, C, and D}

We now use the supersymmetry of the Dirac operator,
which we discussed in the previous subsection, to deduce some important 
properties of the functions $A(x)$ through $D(x)$.

From (\ref{abcd}) and from (\ref{qqdagger}) we have 
$$ A(x) = i{\pax-2\sigx\over 2\omega}\left(-{\om c_1c_2\over W_c}
\right)={i\over 2W_c}\left(c_2 Q^\dgg c_1 + c_1 Q^\dgg c_2 \right)\,.$$
Using (\ref{ctob}) first, and then (\ref{btoc}), we rewrite this expression as 
$$A(x) = {i\om\over 2W_c}(c_2b_1 + c_1b_2) = {i\over 2W_c}  
\left(b_1 Q b_2  + b_2 Q b_1 \right)\,.$$
Then, using the fact that $W_c=W_b$ (Eq. (\ref{wbwc})) and (\ref{abcd}),
we rewrite the last expression as 
$$A(x) = i{\pax + 2\si \over 2\omega}\left({\om b_1b_2\over W_b}
\right) = i{(\pax + 2\si)B(x) \over 2\omega}\,.$$
Thus, finally, 
\beq\label{ADeq}
A(x)=D(x)\,.
\eeq
Supersymmetry renders the diagonal elements $A$ and $D$ equal. Clearly, 
the projection of zero modes (when they exist) out of $A(x)$ and $D(x)$,
which we discussed following Eq. (\ref{Batzero}), makes $A=D$ possible. 
We certainly cannot have $B$ and $C$ equal, since when one of them has 
a pole at $\om^2=0$, the other does not. (See (\ref{BtoCrel}) and 
(\ref{CtoBrel}) below.)

Due to (\ref{abcd}), $A=D$ is also a first order differential equation
relating $B$ and $C$:
\beq\label{firstorderBC}
\left(\pax+2\sigx\right){B\left(x\right)\over \omega} = 
\left(\pax-2\sigx\right){C\left(x\right)\over \omega}\,.
\eeq
With the identification of the resolvents $B/\om$ and $-C/\om$, 
respectively with $R_-(x,\om^2)$ and $R_+(x,\om^2)$ alluded to above, 
we can write (\ref{firstorderBC}) as 
$$(2\si-\pax)R_+ = (2\si +\pax)R_-\,,$$
which is essentially Eq.(18) of \cite{josh1}. 

We can also relate the off diagonal elements 
$B$ and $C$ to each other more directly.
From (\ref{abcd}) and from (\ref{btoc}) we find
\beqra\label{cfromb}
C(x) = -{\om c_1c_2\over W_c} = -{(Qb_1)(Qb_2)\over \om W_c}\,.
\eeqra
After some algebra, and using (\ref{wbwc}), we can rewrite this as 
$$ -\om^2 C = \si^2 B + \si B' + {\om b_1' b_2' \over W_b}$$
The combination $\om b_1'b_2'/W_b$ appears in $B''=(\om b_1b_2/W_b)''$. 
After using $(H_b-\om^2) b_{1,2}=0$ to eliminate $b_1''$ and $b_2''$ from 
$B''$, we find
$${\om b_1'b_2'\over W_b} = {1\over 2}B'' - \left(\si^2 - \si' -
\om^2 \right)B\,.$$
Thus, finally, we have 
\beq\label{BtoCrel}
-\om^2 C = {1\over 2}B'' + \si B' + (\si' + \om^2 ) B\,.
\eeq

In a similar manner we can prove that 
\beq\label{CtoBrel}
\om^2 B = -{1\over 2}C'' + \si C' + (\si' -\om^2 ) C\,.
\eeq
We can simplify (\ref{BtoCrel}) and (\ref{CtoBrel}) further. 
After some algebra, and using (\ref{abcd}) we arrive at 
\beqra\label{BCfinal}
C(x) &=& {i\over \om}\,\pax D(x) - B(x) = -{1\over 2\om^2}\pax\left[(\pax 
+ 2\sigx)B(x)\right] -B(x)\nonumber\\{}\nonumber\\
B(x) &=& {i\over \om}\,\pax A(x) - C(x) = -{1\over 2\om^2}\pax\left[(\pax 
- 2\sigx)C(x)\right] -C(x)
\eeqra
(which are one and the same equation, since $A(x)=D(x)$). 
Supersymmetry, namely, isospectrality of $H_b$ and $H_c$, enables us 
to relate the diagonal resolvents of these operators, $B$ and $C$, to each 
other. 

Thus, we can use (\ref{abcd}), (\ref{ADeq}) and (\ref{BCfinal}) to 
eliminate three of the entries of the diagonal resolvent in (\ref{abcd}),
in terms of the fourth. 

The relations (\ref{BCfinal}) (or (\ref{BtoCrel}) and (\ref{CtoBrel})) 
were not discussed in \cite{josh1}, but one can verify them, for 
example, for the resolvents corresponding to the kink case 
$\sigx = m\,{\rm tanh}\, mx$ (Eq. (29) in \cite{josh1}), for which 
\beq\label{kinkresolvents}
C=-{\om\over 2\sqrt{m^2-\om^2}}\,,\quad
B=\left[\left({m\, {\rm sech}\, mx\over \om}\right)^2 - 1\right]C\,.
\eeq
Note that the two relations (\ref{BtoCrel}) and (\ref{CtoBrel}) transform 
into each other under 
\beq\label{diagBCflip}
B\leftrightarrow -C\quad {\rm simultaneously~ with}\quad 
\si\rightarrow -\si\,,
\eeq
in consistency with (\ref{bcflip}).

From the relations in (\ref{BCfinal}), it is clear that away from $\om=0$, 
$B$ and $C$ must have the same singularities as functions of $\om^2$.
However, we see, for example from (\ref{BtoCrel}), 
that if $C$ has a pole $1/\om^2$ (which corresponds to a normalizable zero 
mode of $H_c$), then $B$ will not have such a pole, and vice versa, as it 
should be.

\newpage
\subsection{Bilinear Fermion Condensates and Vanishing of the Spatial 
Fermion Current}

Following basic principles of quantum field theory, we may write the most 
generic flavor-singlet bilinear fermion condensate in our static background
as
\beqra\label{condensate}
&&\langle \bar\psi_{a\alpha}(t,x)\,\Gamma_{\alpha\beta}
\,\psi_{a\beta}(t,x)\rangle_{{\rm reg}}
=N\int_{\cac} {d\om\over 2\pi }\,\rmtr\left[\Gamma 
\langle x| {-i\over \om\gam^0 + i\gam^1\pax 
-\si} | x \rangle_{{\rm reg}}\right]
\nonumber\\{}\nonumber\\ 
&&= N\int_{\cac} {d\om\over 2\pi }\,
\rmtr\left\{\Gamma\left[ 
\left(\begin{array}{cc} A(x) & 
B(x) \\{}&{}\\ C(x) & 
D(x)\end{array}\right)\,-\, \left(\begin{array}{cc} A & 
B \\{}&{}\\ C & 
D\end{array}\right)_{_{VAC}}\right]\right\}\,,
\eeqra
where we have used (\ref{diagonal}). The integration contour is specified
in Section 3 (see Figure 1). 
Here $a=1,\cdots ,N$ is a flavor index, and the trace is taken over 
Dirac indices $\alpha, \beta$. As usual, we regularized this condensate by 
subtracting from it a short distance divergent piece embodied here by the
diagonal resolvent
\beq\label{vac}
\langle x\,|-iD^{-1} | x\,\rangle_{_{VAC}} = 
\left(\begin{array}{cc} A & 
B \\{}&{}\\ C & 
D\end{array}\right)_{_{VAC}} = 
{1\over 2 \sqrt{m^2-\omega^2}}
\left(\begin{array}{cc} i \si_{_{VAC}} & \omega 
\\{}&{}\\ -\omega  &  i \si_{_{VAC}}\end{array}\right)
\eeq
of the Dirac operator in a vacuum configuration $\si_{_{VAC}} = 
\pm m$.

In our convention for $\gam$ matrices (\ref{majorana}) we have 
\beq\label{diagresolvent}
\left(\begin{array}{cc} A(x) & 
B(x) \\{}&{}\\ C(x) & 
D(x)\end{array}\right)\, = {A(x)+D(x)\over 2}{\bf 1} + 
{A(x)-D(x)\over 2i}\gam^1 + i{B(x)-C(x)\over 2}\gam^0 + 
{B(x)+C(x)\over 2}\gam_5\,.
\eeq

An important condensate is the expectation value of the fermion current 
$\langle j^\mu (x)\rangle $. In particular, consider its spatial component.
In our static background $\sigx $, it 
must, of course, vanish identically
\beq\label{j1vanish}
\langle j^1 (x)\rangle =0\,.
\eeq
Thus, substituting $\Gamma = \gam^1$ in (\ref{condensate}) and using 
(\ref{diagresolvent}) we find 
\beq\label{fourierspat}\langle j^1 (x)\rangle = 
iN\int_{\cac} {d\om\over 2\pi }\,\left[A(x)-D(x)\right]\,.
\eeq
But we have already proved that $A(x)=D(x)$ in {\em any} 
static background $\sigx$ (Eq.(\ref{ADeq})). Thus, each
frequency component of $\langle j^1 \rangle$ vanishes separately, and 
(\ref{j1vanish}) holds identically. 
It is remarkable that the supersymmetry of the Dirac operator
guarantees the consistency of any static $\sigx$ 
background.

Expressions for other bilinear condensates may be derived in a similar 
manner (here we write the unsubtracted quantities). 
Thus, substituting $\Gamma = \gam^0$ in (\ref{condensate}) and using 
(\ref{diagresolvent}), (\ref{ADeq}) and (\ref{BCfinal}), 
we find that the fermion density is 
\beq\label{fourierdensity}
\langle j^0 (x)\rangle = 
iN\int_{\cac} {d\om\over 2\pi }\,\left[B(x)-C(x)\right] = 
iN\int_{\cac} {d\om\over 2\pi }\,{2\om B(x) -i\pax D(x)\over  \om }\,.
\eeq
Similarly, the scalar and pseudoscalar condensates are 
\beq\label{fourierscalar}
\langle \bar\psi (x)\psi (x)\rangle = 
N\int_{\cac} {d\om\over 2\pi }\,\left[A(x)+D(x)\right] = 
2N\int_{\cac} {d\om\over 2\pi }\, D(x)\quad\quad \,,
\eeq
and 
\beq\label{fourierpseudoscalar}
\langle \bar\psi (x)\gam^5\psi (x)\rangle = 
N\int_{\cac} {d\om\over 2\pi }\,\left[B(x)+C(x)\right] =
iN\pax\,\int_{\cac} {d\om\over 2\pi }\,{D(x)\over  \om }\,.
\eeq

We can also derive a simple relation between the fermion density 
(\ref{fourierdensity}) and the pseudoscalar condensate 
(\ref{fourierpseudoscalar}). From (\ref{firstorderBC}) we have 
$\pax (B-C) = -2\si (B+C)$. 
Substituting this result into (\ref{fourierpseudoscalar}) and comparing with 
(\ref{fourierdensity}) we deduce that 
\beq\label{densityandpseudoscalar}
\pax\, \langle j^0 (x)\rangle = -2i\sigx\,\langle \bar\psi (x)
\gam^5\psi (x)\rangle\,.
\eeq
We may verify (\ref{densityandpseudoscalar}) easily for the kink 
$\sigx = m\,\tanh\,mx$ by substituting (\ref{kinkresolvents}) in 
(\ref{fourierdensity}) and (\ref{fourierpseudoscalar}). 

The result (\ref{densityandpseudoscalar}) has a simple interpretation in 
terms of bosonization. Since (\ref{densityandpseudoscalar}) stems essentially 
from the properties of the diagonal resolvent (\ref{diagonal}) of the Dirac 
operator $i\notpa -\sigx $, it is independent of the $O(2N)$ symmetry of the 
GN model, and also independent of the dynamics of the auxiliary field 
$\sigx $. Thus, we need to consider only bosonization of the action 
\beq\label{psiaction}
S_\psi = \int d^2x\,\bar\psi\left(i\notpa -\sigx\right)\psi
\eeq
of a {\em single} Dirac fermion $\psi$, corresponding to the Abelian case  
\cite{mandelstam}. The relevant bosonization operator identities are 
\beqra\label{boso}
i\bar\psi\notpa\psi &\rightarrow &\frac{1}{2}\,\left(\pa_\mu\phi\right)^2
\nonumber\\{}\nonumber\\
\bar\psi\gam^\mu\psi &\rightarrow & {\epsilon^{\mu\nu}\pa_\nu\phi\over 
\sqrt{\pi}}\nonumber\\{}\nonumber\\
\bar\psi\psi &\rightarrow & \mu\,{\rm cos} (2\sqrt{\pi}\phi)\quad\quad 
{\rm and}\nonumber\\{}\nonumber\\
i\bar\psi\gam^5\psi &\rightarrow & \mu\,{\rm sin} (2\sqrt{\pi} \phi)\,,
\eeqra
where $\phi$ is the corresponding boson,
and $\mu$ is an arbitrary scale parameter.

According to these formulas, the bosonized form of $S_\psi$ is 
\beq\label{phiaction}
S_{\phi} = \int d^2x\,\left[ \frac{1}{2} (\pa_\mu\phi)^2 - 
\mu\si\cos\,(2\sqrt{\pi}\phi)\right]\,,
\eeq
which leads to the equation of motion 
\beq\label{phieom}
\bx\phi - 2\mu\sqrt{\pi}\si\sin\,(2\sqrt{\pi}\phi) = 0\,.
\eeq
In the static limit, the latter equation becomes 
\beq\label{boso1}
\pax^2 \phi (x) + 2\mu\sqrt{\pi}\sigx \,{\rm sin}\, (2\sqrt{\pi}\phi (x)) = 0
\,,
\eeq
which is easily recognized as the bosonized form of 
(\ref{densityandpseudoscalar}) (up to a trivial multiplication of 
(\ref{boso1}) by a factor $N$). Thus, the {\em identity}
(\ref{densityandpseudoscalar}) is equivalent to the equation of motion 
of the bosonic field $\phi$ in the static limit. This of course, should be 
expected on physical grounds. After all, Eq. (\ref{densityandpseudoscalar}) 
is a consequence of (\ref{ADeq}) (or (\ref{firstorderBC})), which implies 
$\langle j^1(x)\rangle = 0$ in our static background $\sigx$, 
which leads to $\pa_0\phi =0$, according to the second equation in 
(\ref{boso}).

\pagebreak


\newpage
\setcounter{equation}{0}
\setcounter{section}{0}
\renewcommand{\theequation}{B.\arabic{equation}}
\renewcommand{\thesection}{Appendix B:}
\section{Reflectionless Schr\"odinger Operators and their Resolvents}
\vskip 5mm
\setcounter{section}{0}
\renewcommand{\thesection}{B}

In this Appendix we gather (without derivation) some useful 
properties of reflectionless Schr\"odinger hamiltonians\cite{faddeev} 
and their resolvents.  A particularly useful summary of the theory of 
inverse scattering in one dimension and of reflectionless Schr\"odinger 
operators is given in the first reference in
\cite{rosner} (which we partly follow here).

Following that summary of results from the literature, we use one particular 
result, namely, the explicit formulas for a pair of fundamental independent 
solutions of the Schr\"odinger equation, to derive an 
explicit simple representation for the diagonal resolvent of a 
reflectionless Schr\"odinger operator with a prescribed set of bound 
states, which we have not encountered in the literature. 

\subsection{The Potential and Wave Functions}

Consider the Schr\"odinger equation 
\beq\label{schrodeq}
\left(-\pax^2 + V(x)\right) b(x) = k^2 b(x)
\eeq
on the whole real axis, where $V(x)$ is a {\em reflectionless} potential, 
namely, the reflection amplitude $r(k)$ of (\ref{schrodeq}) vanishes 
identically. Reflectionless potentials are bounded, and tend to an 
asymptotic constant value $V(\pm\infty) = V_0$ at an exponential rate. 
With no loss of generality  we will set $V_0=0$. Thus, $V(x)$ satisfies the 
boundary conditions 
\beq\label{potential}
V(x)\asymptotic 0\,.
\eeq

Since $V(x)$ is bounded and tends to its asymptotic value (\ref{potential}) at 
an exponential rate, it can support only a finite number of bound states.
Let us assume that the reflectionless $V(x)$ has $K$ bound states at energies 
\beq\label{energies}
-\kappa_1^2 < -\kappa_2^2 \cdots < -\kappa_K^2 < 0\,.
\eeq
Eq.(\ref{energies}) is the spectrum of bound states of a one-dimensional 
Schr\"odinger operator, which cannot be degenerate. Thus all inequalities 
in (\ref{energies}) are strict.

The reflectionless potential $V(x)$ is uniquely 
determined \cite{faddeev} by the asymptotic behavior of its $K$ bound 
state wave functions only. (In order to determine a generic reflectionful 
potential, one also requires, of course, the reflection amplitude $r(k)$.)

A bound state wave function in a potential which tends to its asymptotic 
behavior (\ref{potential}) at an exponential rate, must also decay 
exponentially. Thus, from (\ref{schrodeq}), the $n$-th bound 
state wave function $\psi_n(x)$, which corresponds to bound state energy 
$-\kappa_n^2$, decays asymptotically as 
\beq\label{asymptoticdecay}
\psi_n(x)\asymptoticp = c_n \exp -\kappa_n x\,,
\eeq
with some real coefficient $c_n$. The parameter $c_n$ is determined from 
the requirement that $\psi_n$ be normalized to unity. With no loss of 
generality we take $c_n$ to be positive. For convenience, let us name the 
right hand side of 
(\ref{asymptoticdecay}) as 
\beq\label{lambdadef}
\lambda_n(x)  = c_n \exp -\kappa_n x\,.
\eeq

The asymptotic behavior of $\psi_n(x)$ is  
determined by two positive parameters $\kappa_n $ and $c_n$, and thus,
there are $2K$ parameters at our disposal: 
\beq\label{parameters}
\kappa_1 > \kappa_2 > \cdots \kappa_K > 0\quad\quad  {\rm and}\quad\quad
c_1, c_2, \cdots c_K\,.
\eeq 
These are the ``scattering data'' alluded to in Section 3 
(see (\ref{scatteringdatatext})). These parameters determine 
the reflectionless potential $V(x)$ with $K$ bound states (\ref{energies})
neatly as 
\beq\label{hirota}
V(x) = -2{\pa^2\over \pa x^2}\log \det A\,,
\eeq
where $A (x)$ is the symmetric $K\times K$ matrix 
\beq\label{amatrix}
A_{mn} = \del_{mn} + {\lambda_m  \lambda_n \over \kappa_m + \kappa_n}\,.
\eeq

It can be shown that the $K$ dimensional vectors $\psi_n$ and $\lambda_n$
are related by 
\beq\label{lambdapsi}
\sum_{n=1}^K A_{mn}\,\psi_n = \lambda_m\,.
\eeq
We can solve this linear equation and thus obtain an explicit formula for 
the $\psi_n(x)$. The desired solution is
\beq\label{psin}
\psi_n (x) =  -{1\over \lambda_n}\,{\det A^{(n)}\over \det A}\,,
\eeq
where $A^{(n)} (x)$ is the $K\times K$ matrix obtained from (\ref{amatrix}), 
by replacing the $n$-th column of $A$ by its derivative.

Inverse scattering theory also provides us with explicit expressions 
for the pair of fundamental independent solutions of (\ref{schrodeq}).
Let $\{b_1(x), b_2(x)\}$ be a pair of independent fundamental solutions
of (\ref{schrodeq}) subjected to the (standard, or canonical) 
asymptotic boundary conditions 
\beq\label{standardplanewaves}
b_1(x) \asymptoticm e^{-ikx} \quad\quad ,\quad\quad 
b_2(x) \asymptoticp e^{ikx}
\eeq
(i.e., Eq. (\ref{planewaves}) with $A_{b}^{(1)}=A_{b}^{(2)}=1$).
Then, inverse scattering theory tells us that for ${\rm Im}k\geq 0$, 
$b_1$ and $b_2$ are given 
by\footnote{The functions
$b_1$ and $b_2$ are, respectively, the functions $\phi_2$ and 
$\phi_1$ in the first reference cited in \cite{rosner}.} 
\beqra\label{fundamental}
b_1(x)&=&{1\over t(k)}e^{-ikx}\left[1+i\sum_{n=1}^K {\lambda_n(x)\psi_n(x)
\over k-i\kappa_n}
\right]
\nonumber\\{}\nonumber\\
b_2(x)&=& e^{ikx}\left[1-i\sum_{n=1}^K {\lambda_n(x)\psi_n(x)\over k+i\kappa_n}
\right]\,,
\eeqra
where 
\beq\label{ak}
{1\over t(k)} = {i\over 2k} \left[b_2(x)b_1^{'}(x)-b_1(x)b_2^{'}(x)\right]
\eeq
is the inverse of the transmission amplitude $t(k)$ of (\ref{schrodeq}). 
It is proportional to the wronskian 
\beq\label{W}
W=b_2(x)b_1^{'}(x)-b_1(x)b_2^{'}(x)
\eeq
of $b_1$ and $b_2$, as was mentioned in Appendix A,
\beq\label{at}
{1\over t(k)} = {iW\over 2k}\,.
\eeq

According to inverse scattering theory, the transmission amplitude $t(k)$ 
of the reflectionless potential with $K$ bound states which correspond to 
the data (\ref{parameters}) is simply 
\beq\label{transmissionamplitude}
t(k) = \prod_{n=1}^K {k+i\kappa_n\over k-i\kappa_n}\,.
\eeq
This transmission amplitude is a pure phase, as it should be, since the 
potential $V(x)$ is reflectionless. Note the simple poles of $t(k)$ 
at the bound states, or equivalently, the simple zeros of $1/t(k)$, which 
arise, as we have already discussed (in connection with 
(\ref{vanishwronskian})), since $1/t(k)$ is proportional to the 
wronskian of the two fundamental solutions, and both these solutions become 
proportional to the bound state wave function, and thus linear dependent.

Thus, as $k$ tends to one of its bound state values, say at $k=i\kappa_m$,
the sum in the square brackets in the expression for $b_1(x)$ in 
(\ref{fundamental}) has a pole, which dominates the sum . But at the 
same time, $1/t(k)$ vanishes linearly as $k-i\kappa_m$ (up to some constant
one can calculate from (\ref{transmissionamplitude})).
This is, of course, a general feature of scattering theory, and is not 
particular to reflectionless potentials. Thus, at the limit 
$k\rightarrow i\kappa_m$ the expression for $b_1(x)$ in (\ref{fundamental}) 
collapses simply to 
\beq\label{boneatpole}
b_1(x) = {\rm const.}\,\psi_m(x)
\eeq
(where we used $ e^{\kappa_m x}\lambda_m(x)=c_m$). To see that at 
$k=i\kappa_m$, the other fundamental solution $b_2(x)$ in 
(\ref{fundamental}) is also proportional to $\psi_m (x)$ already requires 
the special properties of 
reflectionless potentials. Indeed, at $k=i\kappa_m$ , the expression
for $b_2(x)$ may be written 
\beq\label{btwoatpole}
b_2(x) = e^{-\kappa_m x}\left[1-{1\over \lambda_m}
\sum_{n=1}^K \left( A_{mn}-\delta_{mn}\right)\psi_n(x)\right] = {1\over c_m}
\psi_m(x)\,,
\eeq
where we have used (\ref{lambdadef}) and (\ref{lambdapsi}). $b_1(x)$ and 
$b_2(x)$ are indeed both proportional to the bound state wave function at 
$k=i\kappa_m$.

We can also use the explicit forms of the fundamental solutions 
(\ref{fundamental}) and of the transmission amplitude 
(\ref{transmissionamplitude}) to derive an explicit formula
for the integral over the potential $V(x)$. According to (\ref{hirota}),
the definition of the matrix $A^{(n)}$, and (\ref{psin}), we have 
\beq\label{integral1}
\int\limits_{-\infty}^\infty V(x)\, dx = 
2\left[{\pax\det A(x)\over \det A(x)}\right]_{\infty}^{-\infty} = 
-2\!\!\!\!\xlimm\left(\sum_{n=1}^K \lambda_n (x) \psi_n (x)\right)\,.
\eeq
Here we used the fact that $\xlimp \lambda_n (x) \psi_n (x) =0$,
obviously. However, $\xlimm\lambda_n (x) \psi_n (x)$ is a constant,
which has to be determined. In fact, in (\ref{integral1}) we need only the 
sum of these constants. 
This sum can be extracted from (\ref{fundamental}). Indeed, from 
(\ref{standardplanewaves}) and from (\ref{fundamental}), we observe that 
\beq\label{asymptoticsum}
t(k) =  1+i\!\!\!\!\xlimm \left(\sum_{n=1}^K {\lambda_n(x)\psi_n(x)\over k-i
\kappa_n}\right)\,.
\eeq
Thus, combining (\ref{asymptoticsum}) and (\ref{integral1}), 
we finally arrive at the sum-rule 
\beq\label{integralfinal}
\int\limits_{-\infty}^\infty V(x)\, dx = 2i \int_{{\cal C}} 
{dk\over 2\pi i} t(k)\,,
\eeq
where the contour ${\cal C}$ encircles all the poles of $t(k)$ in 
(\ref{transmissionamplitude}) in the upper half $k$-plane.\\

~~~~~~{\em The Diagonal Resolvent}\\
We now use (\ref{fundamental}) to derive a simple formula 
(Eq. (\ref{finalresolvent}) below) for the Green's 
function of the reflectionless hamiltonian 
\beq\label{hamilton}
H = -\pax^2 + V(x)
\eeq
in (\ref{schrodeq}). We have not encountered the formula we are about to 
derive in the literature.

In a similar manner used to derive the Green's functions in 
(\ref{bcexpression}), we may express the Green's function of $H$ in terms of 
the fundamental solutions (\ref{fundamental}) as
\beq\label{greensofh}
G(x,y;k) = \langle x |{1\over H-k^2 } | y \rangle = 
{\theta\left(x-y\right)b_2(x)b_1(y)
+\theta\left(y-x\right)b_2(y)
b_1(x) \over W}\,.
\eeq
Then the diagonal resolvent of $H$,  
$$R(x; k^2) = G(x,x;k) = \epsilim {G(x,x+\epsilon;k)
+G(x+\epsilon,x;k)\over 2}$$ 
(compare with Eq. (\ref{diagonal})) is 
\beq\label{diagonalresolvent}
R(x; k) = {b_1(x)b_2(x)\over W} = {i\over 2k}
\left[1+i\sum_{n=1}^K {\lambda_n\psi_n\over 
k-i\kappa_n}\right]\left[1-i\sum_{m=1}^K {\lambda_m\psi_m\over k+i\kappa_m}
\right]\,.
\eeq

The last expression may be simplified considerably 
by simple algebraic manipulations. First, let us multiply 
the factors in (\ref{diagonalresolvent}). Thus, 
\beq\label{algebra1}
R(x; k) = {i\over 2k}
\left[1+i\sum_{n=1}^K \left( {1\over k-i\kappa_n} - {1\over k+i\kappa_n} 
\right)\lambda_n\psi_n + \sum_{m,n=1}^K {\lambda_m\lambda_n\psi_m\psi_n\over
(k-i\kappa_n)(k+i\kappa_m)}\right]\,.
\eeq
Then, concentrate on the last term in (\ref{algebra1}). Using (\ref{amatrix}),
we can write it as
\beq\label{algebra2}
{\lambda_m\lambda_n\psi_m\psi_n\over (k-i\kappa_n)(k+i\kappa_m)} = 
-i(A_{mn}-\del_{mn})
\left( {1\over k-i\kappa_n} - {1\over k+i\kappa_m}\right)\psi_m\psi_n\,.
\eeq
Then, summing the last expression over $m,n$ and using (\ref{lambdapsi})
we obtain
\beq\label{algebra3}
\sum_{m,n=1}^K {\lambda_m\lambda_n\psi_m\psi_n\over (k-i\kappa_n)(k+i\kappa_m)}
= -i\sum_{n=1}^K \left( {1\over k-i\kappa_n} - {1\over k+i\kappa_n} 
\right)\lambda_n\psi_n -2\sum_{n=1}^K {\kappa_n\psi_n^2\over k^2 + \kappa_n^2}
\,.
\eeq
Finally, substituting (\ref{algebra3}) in (\ref{algebra1})
we arrive at the desired simpler formula for the diagonal resolvent:
\beq\label{finalresolvent}
R(x; k) = {i\over 2k}\left(1 -2\sum_{n=1}^K {\kappa_n\psi_n^2\over k^2 + 
\kappa_n^2}\right)\,.
\eeq
As one trivial consistency check on (\ref{finalresolvent}), note from 
(\ref{finalresolvent}) that $R(x;k)$, as a function of energy 
$E=k^2$, has a simple pole at each of the bound state energies 
$E_n=-\kappa_n^2$, with residue $-\psi_n^2$, as it should 
be.\footnote{Note that (\ref{finalresolvent}) is not the most obvious way 
to write a function of $x$ and $k^2$ with these properties (e.g., the 
function similar to (\ref{finalresolvent}), but
with all prefactors $i\kappa_n/\sqrt{k^2}$ removed, has these properties, 
and also the correct large $k$ behavior $i/2k$, and is much simpler). 
Thus, we really had to go through all the steps 
of the derivation to obtain (\ref{finalresolvent}).}

Also note that (\ref{finalresolvent}), being the diagonal resolvent of 
(\ref{hamilton}) with potential (\ref{hirota}), must satisfy the GD identity 
\beq\label{gdid}
-2RR'' + R'^2 + 4R^2 (V(x)- k^2) =1\,. 
\eeq
In fact, in \cite{josh1}, the analysis went the other way around, and 
used (\ref{gdid}) to derive (\ref{finalresolvent}) in the 
case of a single bound state, $K=1$. 

As yet another consistency check on (\ref{finalresolvent}), note that as 
$|x|\rightarrow\infty$, it tends to $R=i/2k$, which is indeed the 
asymptotic solution of (\ref{gdid}) in view of (\ref{potential}).

\subsection{Application to the Dirac Operator}

In order to make contact with the discussion in the text, let us identify 
the hamiltonian (\ref{hamilton}) with one of the operators $H_b$ or $H_c$ 
in (\ref{bcops}). With no loss of generality we chose to identify it
with $H_b = -\pa_x^2 + \sigx^2 - \si'(x)$. Actually, due to the boundary 
conditions (\ref{boundaryconditions}), the potential $\sigx^2 - \si'(x)$
tends asymptotically to $m^2$ and not to zero. Thus, in order to comply with 
(\ref{potential}), we consider $H=H_b-m^2$, so that the eigenvalue equation
(\ref{eigenvalueequations}) $H_b b=\om^2 b$ turns into $(H_b-m^2)b =
(\om^2 -m^2) b$, which coincides with (\ref{schrodeq}), due to $\om^2=
k^2+m^2$ (\ref{kmom}). At the bound state with $k^2=-\kappa_n^2$ we thus 
have $\om_n^2 = m^2-\kappa_n^2$. Thus, from (\ref{energies}) the 
bound state energies of $H_b$ are
\beq\label{energieshb}
0\leq \om_1^2 < \om_2^2 \cdots < \om_K^2 < m^2\,.
\eeq
Since (\ref{energies}) is non-degenerate, all the $\om_n^2$ must be 
different from each other.

In the topologically trivial sector of the model, $q=0$,
$H_b= -\pa_x^2 + \sigx^2 - \si'(x)$ and $H_c= -\pa_x^2 + \sigx^2 + \si'(x)$ 
are strictly isospectral, and also reflectionless. 
Thus, we have two different reflectionless potentials, with the same set of 
bound state energies (\ref{energieshb}). Clearly, it is the additional $K$ 
parameters $c_1, \cdots c_K$ that remove the ambiguity between these
two isospectral potentials. The parameters $c_1, \cdots, c_K$ 
parametrize a $K$ dimensional family of reflectionless potentials with 
given $K$ bound state energies (\ref{energieshb}). 
As the $c_n$ vary, the system wanders around in that family space without 
changing its bound state energies. In particular, $H_b$ and $H_c$ should 
correspond to two points in the $K$ dimensional space of the $c_n$'s. 
Let us denote these points, respectively, as $\{c_n (H_b)\}$ and  
$\{c_n (H_c)\}$. It will be interesting to find a representation of
the supersymmetric transformation $H_b\leftrightarrow H_c$ as 
an invertible map on the $K$ dimensional vectors 
\beq\label{ctransformation}
\{c_n (H_b)\}\leftrightarrow \{c_n (H_c)\}\,.
\eeq 
Computation of this transformation in the simplest case, i.e., when 
there is a single bound state, $K=1$, is given in section B.3 (see
(\ref{cbcctrans})).

Finally, from (\ref{inversehbhc}) we have 
\beq\label{Bresolvent}
B(x) = \langle x |\,{\om\over H_b - \omega^2 }| x\rangle = 
{i\om\over 2k}\left(1 -2\sum_{n=1}^K {\kappa_n\psi_n^2\over \om^2 - 
\om_n^2}\right)\,.
\eeq
We can then use (\ref{BtoCrel}) to determine $C(x)$. (Alternatively, due to 
(\ref{inversehbhc}) and isospectrality of $H_b$ and $H_c$, we
can just write an expression for $-C(x)$ similar to (\ref{Bresolvent})
with $\psi_n$ being the bound state wave functions of $H_c$.)

\subsubsection{The Condensate $\sigx$}

Let us now determine the condensate $\sigx$. 
We will consider the cases of zero topological charge and non-zero 
topological charge separately. 

~~~~~{\em (a) topologically trivial condensates, $q=0$}\\
In this case, none of the operators $H_b$ and $H_c$ has a normalizable
zero mode. Thus, none of the $\om_n$'s in (\ref{energieshb}) vanishes.
(Equivalently, none of the $\kappa_n$'s in (\ref{energies}) can be equal to 
$m$.) Since $H_b$ does not have a normalizable zero mode, the fundamental 
solutions (\ref{fundamental})
at $\om^2=0$, i.e., at $k=im$, 
\beqra\label{fundamentalatim}
b_1(x)_{|_{k=im}} &=&{1\over t(im)}e^{mx}\left[1+\sum_{n=1}^K 
{\lambda_n(x)\psi_n(x)\over m-\kappa_n}\right]
\nonumber\\{}\nonumber\\
b_2(x)_{|_{k=im}} &=& e^{-mx}\left[1-\sum_{n=1}^K 
{\lambda_n(x)\psi_n(x)\over m+\kappa_n}
\right]\,,
\eeqra
are not square-integrable functions. 
They are the two independent solutions of 
$Q^\dgg Q b_{1,2} =0$. According to (\ref{bzero}), one of these solutions 
must be $b_0(x) = \exp -\int^x \si(y) dy$. Thus, if we knew 
$b_0$, we could determine 
\beq\label{sigmafrombzero}
\sigx = -{d\over dx} \log b_0(x)\,.
\eeq
Clearly, 
\beq\label{bzeroasympt}
b_0(x)\asymptoticp e^{-\si (\infty) x}\,.
\eeq
On the other hand, 
\beq\label{fundamentalasymptotic}
b_1(x)_{|_{k=im}} \asymptoticp {e^{mx}\over t(im)}\quad\quad {\rm and}
\quad\quad  b_2(x)_{|_{k=im}} \asymptoticp e^{-mx}\,,
\eeq
since in that limit, the terms proportional to $\lambda_n(x)\psi_n(x)$ in 
(\ref{fundamentalatim}) are negligible compared to $1$. 
Now, identification of $b_0(x)$ with one of the functions in 
(\ref{fundamentalatim}) is just a matter of comparing (\ref{bzeroasympt})
and (\ref{fundamentalasymptotic}). Thus, if $\si (\infty) = -m$, then 
$b_0(x)=b_1(x)_{|_{k=im}}$, and if $\si (\infty) = m$, then 
$b_0(x)=b_2(x)_{|_{k=im}}$. Therefore, we conclude that 
\beqra\label{explicitsigma1}
\sigx &=& -m -{d\over dx}\log\left[1+\sum_{n=1}^K {\lambda_n(x)
\psi_n(x)\over m-\kappa_n}\right]\,,\quad\quad \si(\infty) = -m\nonumber\\
{}\nonumber\\
\sigx &=& ~~~m -{d\over dx}\log\left[1-\sum_{n=1}^K {\lambda_n(x)
\psi_n(x)\over m+\kappa_n}\right]\,,\quad\quad \si(\infty) = m\,,
\eeqra
or more compactly, 
\beq\label{explicitsigmaqzero}
\sigx = \si(\infty) -{d\over dx}\log\left[1-\sum_{n=1}^K {\lambda_n(x)
\psi_n(x)\over \kappa_n +\si(\infty)}\right]\,.
\eeq
This is manifestly a configuration belonging to the $q=0$ sector, 
since the expression on the right hand side of (\ref{explicitsigmaqzero})
tends to $\si(\infty)$ as $x\rightarrow\pm\infty$, i.e., on both sides of 
the one dimensional world.\\

~~~~~{\em (b) the $q=1$ sector}\\
According to the discussion following Eqs. (\ref{bzero}), (\ref{czero}),
in the $q=1$ sector, only $H_b$ has a normalizable zero mode. 
According to (\ref{bzero}) this normalizable 
ground state is $b_0(x) = \exp -\int^x \si(y) dy$, and thus must be 
just the lowest energy state $\psi_1$ in (\ref{psin}), corresponding to 
$\om_1^2=0$, i.e., $\kappa_1=m$. Thus, from (\ref{psin}) and
(\ref{sigmafrombzero}) we find 
\beq\label{sigmaqone}
\sigx = -{d\over dx} \log \psi_1(x) =  -{d\over dx} \log \left(
 -{1\over \lambda_1}\,{\det A^{(1)}\over \det A}\right)\,.
\eeq

Let us make the side remark, that due to the discussion which led to
(\ref{boneatpole}) and (\ref{btwoatpole}), the present case can be 
considered as a singular limit of the previous case (a) (with 
$\si (\infty) = m $, as appropriate to the sector $q=1$), simply by 
taking the second equation in (\ref{explicitsigma1}) at $\kappa_1
\rightarrow m$. 
Following the same steps as in the derivation of (\ref{btwoatpole}) with 
$\kappa_m=\kappa_1 = m $ we end up with 
\beq\label{sigmaqonelternative}
\sigx = m -{d\over dx}\log 
{\psi_1\over \lambda_1} =  -{d\over dx} \log \psi_1(x)\,
\eeq
i.e., back to (\ref{sigmaqone}).\\

~~~~~{\em (c) the $q=-1$ sector}\\
According to the discussion following 
Eqs. (\ref{bzero}), (\ref{czero}), in the $q=-1$ sector, only $H_c$ has a 
normalizable zero mode. If it has $K-1$ additional bound states at positive 
energies, they will be isospectral with those of $H_b$. 
Thus, in that case $H_c$ has $K$ bound states at energies 
\beq\label{energieshczero}
0=\om_1^2 < \om_2^2 \cdots < \om_K^2 < m^2\,.
\eeq
Determination of a $\sigx$ configuration that carries topological charge 
$q=-1$ can be reduced to the previous case, of determination of the
configuration $-\sigx$ that
carries $q=1$. 
A sign flip $\sigx\rightarrow -\sigx$ obviously interchanges $H_b$ and $H_c$. 
Thus, $-\sigx$ will give rise to an $H_b$ with bound state spectrum 
(\ref{energieshczero}), and we will be able to determine this $-\sigx$ by 
the method of case (b), from (\ref{sigmaqone}).

So far, we have assumed $K\geq 2$ in (\ref{energieshczero}). If $K=1$, 
the zero mode is the only 
bound state of $H_c$, and $H_b$ cannot have bound states at all. In that case, 
it is therefore the free particle hamiltonian
$H_b = -\pax^2 +m^2$. Thus, $\si^2-\si' =m^2$ and $\sigx$ is the CCGZ antikink
$\sigx = -m {\rm tanh}\, mx$.

\subsection{Reflectionless Potentials with One and Two Bound States}

We end this appendix by working out the cases of reflectionless
potentials with one and two bound states, which are the two 
cases relevant to our discussion of stable static bags in the GN model.

\subsubsection{A Single Bound State, $K=1$}

Let us set the bound state energy at $\om_b^2<m^2$ 
(with the corresponding $\kappa^2=m^2-\om_b^2$). We also have a single 
parameter $c>0$. 
Then, from (\ref{lambdadef}), (\ref{amatrix}), (\ref{psin})
and (\ref{hirota}) we find 
\beq\label{kone}
\lambda (x) = c\, e^{-\kappa x}\quad\quad ,\quad\quad A(x) = 1 + 
{c^2 e^{-2\kappa x}\over 2\kappa}
\eeq
and 
\beq\label{singlebs}
\psi_b(x) =  -{1\over \lambda}\,{A'(x)\over A(x)}= 
\sqrt{{\kappa\over 2}}\,{\rm sech}\,\left(\kappa (x-x_0)\right)\,
\eeq
where $x_0$ is determined from 
\beq\label{xzeromodulus}
e^{-\kappa x_0} = \sqrt{2\kappa}/c\,.
\eeq 
$\psi_b$ is of course normalized to 1. 
From (\ref{hirota}) we find the potential 
$\si^2-\si' = m^2 + V(x) = m^2 -2\pax^2 
\log A(x) = m^2 -2\kappa^2\,{\rm sech}^2\left(\kappa (x-x_0)\right)$, 
which is a potential of the P\"oschl-Teller type.

The role of the parameter $c$ is explicit in these
formulas: it just shifts the center of the potential, and thus cannot affect 
the energy of the bound state. Thus, $\log c$ is essentially the 
translational collective mode of the static soliton $\sigx$, which is obvious 
from the explicit expression (\ref{explicitsigmaqzerosinglebs}) below.

The diagonal resolvent is found from (\ref{Bresolvent}) and is given by
\beq\label{Bresolventsinglebs}
B(x) = {i\om\over 2k}\left(1 -2{\kappa\psi_b^2\over \om^2 - 
\om_b^2}\right) = {i\om\over 2k}\left[1 -\kappa^2 {{\rm sech}^2 
\left(\kappa (x-x_0)\right)\over \om^2 - \om_b^2}\right]\,,
\eeq
in agreement with the expression found in \cite{josh1}.

Let us determine the corresponding $\sigx$. Since $\om_b>0$, we have to use 
(\ref{explicitsigmaqzero}). We find
\beqra\label{explicitsigmaqzerosinglebs}
&&\sigx = \si(\infty) -{d\over dx}\log\left[1-{\lambda(x)\psi_b(x)\over 
\kappa +\si(\infty)}\right]\nonumber\\{}\nonumber\\
&& = \si(\infty) + \kappa\, {\rm tanh}\left[\kappa 
(x-x_0)\right] - \kappa\, {\rm tanh}\left[\kappa \left(x-x_0\right) + 
{1\over 2} \log \left({m+\kappa\over m-\kappa}\right)\right]\,,\nonumber\\
\eeqra
in agreement with the result quoted in Eq. (3.28) of \cite{dhn}.
It has the profile of a bound state of a kink and an antikink, with 
interkink distance ${1\over 2\kappa} 
\log \left({m+\kappa\over m-\kappa}\right)$. It thus carries topological
charge $q=0$.

Let us now find the representation of the supersymmetric transformation 
$H_b\leftrightarrow H_c$, alluded to in the previous subsection, as a map 
(\ref{ctransformation}) between the two normalization constants 
$c (H_b)$ and  $c (H_c)$ in (\ref{ctransformation}). Since the transformation
$H_b\leftrightarrow H_c$ is achieved simply by $\sigx\leftrightarrow -\sigx$,
we have to find a pair of normalization constants $c (H_b)$ and  $c (H_c)$,
which will yield, upon substitution into (\ref{explicitsigmaqzero})
and (\ref{kone}) (with boundary conditions $\si(\infty;b) = 
m = - \si(\infty;c)$ in the latter equation), two $\sigx$ configurations
of opposite signs. The calculation is straightforward, and we find 
\beq\label{cbcctrans}
c^2 (H_c) = c^2 (H_b) {m-\kappa\over m+\kappa}\,,
\eeq
or equivalently, from (\ref{xzeromodulus}),
\beq\label{x0bx0c}
\kappa x_0(H_c) = \kappa x_0 (H_b) + \frac{1}{2}\log\, 
{m-\kappa\over m+\kappa}\,.
\eeq
It is easy to verify, by substituting (\ref{x0bx0c}) in 
(\ref{explicitsigmaqzerosinglebs}), followed by $ \si(\infty) = 
m \rightarrow -m$, that $\si(x - x_0(H_b)) = - \si(x - x_0(H_c))$, 
as required. Furthermore, note that if $x_0 = x_0 (H_b)$ in the first 
hyperbolic tangent in (\ref{explicitsigmaqzerosinglebs}), the second 
hyperbolic tangent there is shifted by $x_0(H_c)$.

As $\kappa$ tends to $m$, $\om_b^2$ tends to zero. An eigenvalue $\om_b^2 = 
0$ cannot occur in the topologically trivial sector. Thus, the limit 
$\kappa\rightarrow m$ in (\ref{explicitsigmaqzerosinglebs}) is singular, and 
the kink and antikink become infinitely separated, as we discussed following 
(\ref{dhnprofile}).

In the sector with topological charge $q=1$, we have $\om_b^2 = 0$ in the 
spectrum of $H_b$, and we should use (\ref{sigmaqone}) for $\sigx$. Thus, from 
(\ref{sigmaqone}) and (\ref{singlebs}) with $\kappa = m$ we find 
\beq\label{explicitkink}
\sigx = -{d\over dx} \log \psi_1(x) =  m {\rm tanh}\,\left(m (x-x_0)\right)\,,
\eeq
which is of course the CCGZ kink. In the CCGZ kink background, $H_b$ has a 
single bound state at $\om_b=0$ and no other bound states. Thus, $H_c$ has no
bound states at all, and is thus the free particle hamiltonian 
$H_c = -\pax^2 +m^2$. Indeed, with (\ref{explicitkink}), one has $\si^2 + \si'
=m^2$ for the potential of $H_c$. The CCGZ antikink was already mentioned at
the end of case (c) in the previous subsection.

\subsubsection{Two Bound States, $K=2$}

We now determine the reflectionless potentials with two bound states.
Let us set the bound state energies (\ref{energieshb}) at 
$0\leq \om_1^2 < \om_2^2 <m^2$, with the corresponding 
$0 < \kappa_2 < \kappa_1 \leq m$. 
We also need $c_1, c_2$, the two positive parameters in 
(\ref{asymptoticdecay}). The case relevant for our discussion of stable
static fermion bags with two bound states is, of course, the case  
$0 = \om_1 < \om_2$. We will discuss it at the end of this subsection.

Then, from (\ref{lambdadef}) and (\ref{amatrix}) we construct 
\beqra\label{atwo}
A= \left(\begin{array}{cc} 
1 + {c_1^2\over 2\kappa_1}e^{-2\kappa_1 x} &  
{c_1c_2\over \kappa_1+\kappa_2} e^{-(\kappa_1 + \kappa_2) x} 
\\{}&{}\\ 
{c_1c_2\over \kappa_1+\kappa_2} e^{-(\kappa_1 + \kappa_2) x} 
 & 1 + {c_2^2\over 2\kappa_2}e^{-2\kappa_2 x}\end{array}\right)\,.
\eeqra
We will also need 
\beqra\label{aonetwo}
\!\!\! A^{(1)} = \left(\begin{array}{cc} 
-c_1^2\,e^{-2\kappa_1 x} &  
{c_1c_2\over \kappa_1+\kappa_2} e^{-(\kappa_1 + \kappa_2) x} 
\\{}&{}\\ 
-c_1c_2\,e^{-(\kappa_1 + \kappa_2) x} 
 & 1 + {c_2^2\over 2\kappa_2}e^{-2\kappa_2 x}\end{array}\right)\,,~
A^{(2)}= \left(\begin{array}{cc} 
1 + {c_1^2\over 2\kappa_1}e^{-2\kappa_1 x} &  
-c_1c_2\,e^{-(\kappa_1 + \kappa_2) x} 
\\{}&{}\\ 
{c_1c_2\over \kappa_1+\kappa_2} e^{-(\kappa_1 + \kappa_2) x} 
 & -c_2^2\,e^{-2\kappa_2 x}\end{array}\right)\,.\nonumber\\{}
\eeqra
The determinants of these matrices are 
\beqra\label{determinants}
F(x) &=& \det\,A~ = 1 + {c_1^2\over 2\kappa_1}e^{-2\kappa_1 x}+ {c_2^2\over 
2\kappa_2}e^{-2\kappa_2 x} + {c_1^2 c_2^2\over 4\kappa_1\kappa_2}
\left({\kappa_1-\kappa_2\over \kappa_1+\kappa_2}\right)^2 e^{-2
(\kappa_1 + \kappa_2) x}\nonumber\\{}\nonumber\\
F_1(x)&=& \det\,A^{(1)} = -c_1^2 e^{-2\kappa_1 x}\left( 1 + 
{c_2^2\over 2\kappa_2} {\kappa_1-\kappa_2\over \kappa_1+\kappa_2} 
e^{-2\kappa_2 x}\right)\nonumber\\{}\nonumber\\
F_2(x)&=& \det\,A^{(2)} = -c_2^2 e^{-2\kappa_2 x}\left( 1 + 
{c_1^2\over 2\kappa_1} {\kappa_2-\kappa_1\over \kappa_2+\kappa_1} 
e^{-2\kappa_1 x}\right)\,.   
\eeqra

The potential $\si^2-\si'$ can be found from (\ref{hirota}) and 
(\ref{determinants}) as 
\beq\label{twostatespotential}
\si^2 -\si' = m^2 + V(x) = m^2 -2{d^2\over dx^2}\,\log F(x)\,.
\eeq
We will not need an explicit expression for $V(x)$.

The bound state wave functions may be found from (\ref{psin})
\beqra\label{wavefunctions}
\psi_1(x) &=& -{1\over \lambda_1(x)}{F_1(x)\over F(x)}\nonumber\\{}\nonumber\\
&=&{c_1 e^{-\kappa_1 x}\,\left( 1 + 
{c_2^2\over 2\kappa_2} {\kappa_1-\kappa_2\over \kappa_1+\kappa_2} 
e^{-2\kappa_2 x}\right)\over 
1 + {c_1^2\over 2\kappa_1}e^{-2\kappa_1 x}+ {c_2^2\over 
2\kappa_2}e^{-2\kappa_2 x} + {c_1^2 c_2^2\over 4\kappa_1\kappa_2}
\left({\kappa_1-\kappa_2\over \kappa_1+\kappa_2}\right)^2 e^{-2
(\kappa_1 + \kappa_2) x}}\nonumber\\{}\nonumber\\
\psi_2(x) &=& -{1\over \lambda_2(x)}{F_2(x)\over F(x)}\nonumber\\{}\nonumber\\
&=&{c_2 e^{-\kappa_2 x}\,\left( 1 + 
{c_1^2\over 2\kappa_1} {\kappa_2-\kappa_1\over \kappa_2+\kappa_1} 
e^{-2\kappa_1 x}\right)\over 
1 + {c_1^2\over 2\kappa_1}e^{-2\kappa_1 x}+ {c_2^2\over 
2\kappa_2}e^{-2\kappa_2 x} + {c_1^2 c_2^2\over 4\kappa_1\kappa_2}
\left({\kappa_1-\kappa_2\over \kappa_1+\kappa_2}\right)^2 e^{-2
(\kappa_1 + \kappa_2) x}}\,.
\eeqra

The parameters $c_1, c_2$ appear in the potential and wave functions 
(\ref{twostatespotential}) and (\ref{wavefunctions}) explicitly. They 
parametrize a 2 dimensional family of reflectionless potentials with 
given two bound state energies $\om_1^2<\om_2^2$. In other words, as $c_1$ 
and $c_2$ vary, the profiles of $V(x)$, $\psi_1(x)$ and $\psi_2(x)$ (and in 
particular, their ``center of gravity'') change
accordingly, with the energies fixed. Thus, as in the single bound state 
case, $c_1$ and $c_2$ represent collective coordinates of the 
soliton $\sigx$. This is manifest, for example, in the explicit expression 
(\ref{sigmaqone2bs}) below, for the $\sigx$ corresponding to 
$\kappa_1=m>\kappa_2=\kappa$. 

One particular choice of $c_1$ and $c_2$ leads to a symmetric 
potential $V(-x)=V(x)$. From (\ref{hirota}) and (\ref{determinants})
we see that this will occur when ${d^2\over dx^2} \log {F(-x)\over F(x)} =0$,
i.e., at $F(-x) = e^{ax + b} F(x)$ for some constants $a,b$. 
Setting $x=0$ in the last equation tells us that $b=0$. Thus, an even 
potential occurs when 
\beq\label{Fevenpotential} 
F(-x) = e^{ax} F(x)\,.
\eeq
Making this demand on $F(x)$ in (\ref{determinants}), leads, after some 
algebra, to the condition 
\beq\label{csforevenpotential}
{c_1^2\over 2\kappa_1} = {c_2^2\over 2\kappa_2} = {\kappa_1+\kappa_2
\over \kappa_1-\kappa_2}\,.
\eeq
For these values of $c_1$ and $c_2$ we obtain from (\ref{determinants}) 
\beq\label{Fsymmetric}
F(x) = 1 + {\kappa_1+\kappa_2\over \kappa_1-\kappa_2}
\left( e^{-2\kappa_1 x} + e^{-2\kappa_2 x}\right) + 
e^{-2(\kappa_1 + \kappa_2) x}\,.
\eeq
It is straightforward to check that (\ref{Fsymmetric}) satisfies 
(\ref{Fevenpotential}) with $a=2(\kappa_1 + \kappa_2)$, 
\beq\label{Fevenpotential1} 
F(-x) = e^{2(\kappa_1 + \kappa_2) x} F(x)\,.
\eeq
In this case, we can read the ground state wave function off 
(\ref{wavefunctions}). It is
\beq\label{groundstateeven}
\psi_1(x) = 
{c_1 e^{-\kappa_1 x}\,\left( 1 + e^{-2\kappa_2 x}\right)\over F(x)}\,.
\eeq
Similarly, the excited state is 
\beq\label{excitedeven}
\psi_2(x) = 
{c_2 e^{-\kappa_2 x}\,\left( 1 - e^{-2\kappa_1 x}\right)\over F(x)}\,.
\eeq
Thanks to (\ref{Fevenpotential1}), $\psi_1(-x)=\psi_1(x)$ is symmetric, and 
$\psi_2(-x) = -\psi_2(x)$ is antisymmetric, as it should be. \\

{\em Avoiding Degeneracy}~~~~It is amusing to observe how this 
one-dimensional quantum system avoids degeneracy in its bound state 
spectrum, as it must. (From now on, we consider generic 
backgrounds and do not assume a symmetric $V(x)$.) 
If we attempt to construct a
reflectionless potential with two degenerate bound states, i.e., with 
$\kappa_1=\kappa_2=\kappa$, we observe from 
(\ref{wavefunctions}) that 
\beq\label{degeneratewavefunctions}
{\psi_1(x)\over c_1} = {\psi_2(x)\over c_2} = 
{e^{-\kappa x}\over 1 + 
{c_1^2 + c_2^2\over 2\kappa}e^{-2\kappa x}}\,.
\eeq
The two wave functions become linearly dependent, and thus, there is no 
degeneracy. Rather, the system loses one bound state, since now 
\beq\label{degenerateF}
F(x) = 1 + 
{c_1^2 + c_2^2\over 2\kappa}e^{-2\kappa x}
\eeq
coincides with $A(x) = 1 + {c^2 e^{-2\kappa x}\over 2\kappa} $ in 
(\ref{kone}) of the single bound state case, with $c^2 = c_1^2 + c_2^2$.\\ 

{\em The Case $0=\om_1^2 < \om_2^2$}~~~~Rather than continuing the 
general discussion, let us specialize at this 
point in the reflectionless potential which is relevant
for our discussion of stable fermion bags with two bound states, namely, 
the case with $0=\om_1^2 < \om_2^2$. This corresponds to setting
$\kappa_1=m > \kappa_2$. Let us also denote $\kappa_2=\kappa$
in (\ref{atwo}) through (\ref{wavefunctions}) above. In particular, the 
ground state wave function is 
\beq\label{groundstate2bs}
\psi_1(x) =  {c_1 e^{-m x}\,\left( 1 + 
{c_2^2\over 2\kappa} {m-\kappa\over m+\kappa} 
e^{-2\kappa x}\right)\over 
1 + {c_1^2\over 2m}e^{-2m x}+ {c_2^2\over 
2\kappa}e^{-2\kappa x} + {c_1^2 c_2^2\over 4m\kappa}
\left({m-\kappa\over m +\kappa }\right)^2 e^{-2
(m + \kappa) x}}\,.
\eeq
The wave function $\psi_1(x)$ is the normalizable zero mode of $H_b$. 
Thus, we are in the $q=1$ topological sector. In this sector
we can determine $\sigx$ from (\ref{sigmaqone}). Thus, substituting 
(\ref{groundstate2bs}) in (\ref{sigmaqone}),
we obtain 
\beqra\label{sigmaqone2bs}
\sigx &=& -{d\over dx} \log \psi_1(x) =  
m + {2\kappa\over 1 + {2\kappa\over c_2^2}{m+\kappa\over m-\kappa}
e^{2\kappa x}}\nonumber\\{}\nonumber\\
&-& 2(m+\kappa) {1+ {2m\kappa\over c_1^2}\,{m+\kappa\over 
(m-\kappa)^2} e^{2mx} + {2m\kappa\over c_2^2}\,{m+\kappa\over (m-\kappa)^2}
e^{2\kappa x}
\over 
1+ {2m\over c_1^2}\,\left({m+\kappa\over 
m-\kappa}\right)^2 e^{2mx} + {2\kappa\over c_2^2}\,
\left({m+\kappa\over m-\kappa}\right)^2 e^{2\kappa x} + 
{4m\kappa\over c_1^2c_2^2}\left({m+\kappa\over m-\kappa}\right)^2 
e^{2(m+\kappa) x}}\,.\nonumber\\{}
\eeqra
One can readily check that the asymptotic values of $\sigx$ are
$\si (\infty) = m$ and $\si(-\infty) = m + 2\kappa -2(\kappa+m) = -m$. 
Thus, $\sigx$ indeed carries topological charge $q=1$.

By varying the parameters $c_1, c_2$ (and keeping $\kappa$ fixed), we can 
modify its shape (e.g., translate its center of gravity), without affecting 
the bound state energies. Eq. (\ref{sigmaqone2bs}) represents a two parameter
family of isospectral kink backgrounds. Thus, as we have asserted earlier, 
$c_1$ and $c_2$ are related to the collective coordinates of the soliton 
$\sigx$. In this case, we have two such translational collective coordinates, 
$x_0$ and $y_0$, given by 
\beq\label{collectivecoordinates}
e^{2mx_0} = {c_1^2\over 2m}\quad\quad {\rm and}\quad\quad e^{2\kappa y_0} 
= {c_2^2\over 2\kappa}\,,
\eeq
in complete analogy with (\ref{xzeromodulus}). In terms of $x_0$ and $y_0$, 
we can write (\ref{sigmaqone2bs}) in a slightly less cluttered form as
\beqra\label{sigmaqone2bsalt}
\sigx &=&  m + {2\kappa\over 1 + {m+\kappa\over m-\kappa}
e^{2\kappa (x-y_0)}}
\nonumber\\{}\nonumber\\
&-& 2(m+\kappa) {1+ {m+\kappa\over 
(m-\kappa)^2}\,\kappa\, e^{2m(x-x_0)} + 
{m+\kappa\over (m-\kappa)^2}\,m\,
e^{2\kappa (x-y_0)}
\over 
1+ \left({m+\kappa\over 
m-\kappa}\right)^2 e^{2m(x-x_0)} + 
\left({m+\kappa\over m-\kappa}\right)^2 e^{2\kappa (x-y_0)} + 
\left({m+\kappa\over m-\kappa}\right)^2 
e^{2m (x-x_0)+2\kappa(x-y_0)}}\,.\nonumber\\{}
\eeqra

We may simplify (\ref{groundstate2bs}) and (\ref{sigmaqone2bsalt}) further as 
follows. In terms of the distance scale\footnote{Note that (\ref{Rdistance}) 
coincides with the interkink distance in (\ref{explicitsigmaqzerosinglebs}).}
\beq\label{Rdistance}
R = {1\over 2\kappa}\log\left({m+\kappa\over m-\kappa}\right)
\eeq
we rewrite (\ref{groundstate2bs}) as 
\beq\label{groundstate2bssimplified}
\psi_1(x) = \sqrt{2m} {\cosh\left[\kappa(x-y_0 + R)\right]\over 
e^{-\kappa R}\cosh\left[m(x-x_0) + \kappa(x-y_0) + 2\kappa R\right]
+e^{\kappa R}\cosh\left[m(x-x_0) - \kappa(x-y_0)\right]}\,.
\eeq
Thus, we obtain the somewhat more transparent expression 
\beqra\label{sigmaqone2bsaltfinal}
\sigx &=&  -{d\over dx} \log \psi_1(x) = -\kappa\tanh\left[\kappa(x-y_0 + R)
\right] \nonumber\\{}\nonumber\\
&+& \om_b{\sinh\left[m(x-x_0) +\kappa(x-y_0)+ 2\kappa R\right] + 
\sinh\left[m(x-x_0) -\kappa(x-y_0)\right]\over
e^{-\kappa R}\cosh\left[m(x-x_0) +\kappa(x-y_0)+ 2\kappa R\right] +
e^{\kappa R}\cosh\left[m(x-x_0) -\kappa(x-y_0)\right]}\nonumber\\{}
\eeqra
for $\sigx$, where we used $(m+\kappa)e^{-\kappa R} = (m-\kappa)e^{\kappa R} 
= \sqrt{m^2-\kappa^2} = \om_b$.

With no loss of generality, we can set, of course, one of the collective
coordinates to zero. In \cite{heavykink} we have studied the profile of 
(\ref{sigmaqone2bsalt}) in some detail. (See Figures 1-3 in \cite{heavykink}.)

We can also tune the parameters $c_1$ and $c_2$ in (\ref{sigmaqone2bs}) (for
purely mathematical reasons and without apparent physical motivation), such 
that $V_b(x) = \si^2(x) -\si'(x)$ be an even function (and $\sigx$ be 
an odd function). Thus, using 
(\ref{csforevenpotential}) and (\ref{collectivecoordinates}),  
we set $2mx_0 = 2\kappa y_0 = 2\kappa R $ 
in (\ref{sigmaqone2bsaltfinal}), and obtain
\beq\label{oddsigmaqone2bsaltfinal}
\sigx = -\kappa\tanh (\kappa x) 
+ \om_b{\sinh\left[(m+\kappa)x\right] + 
\sinh\left[(m-\kappa)x\right]\over
e^{-\kappa R}\cosh\left[(m+\kappa)x\right] +
e^{\kappa R}\cosh\left[(m-\kappa)x\right]}\,.
\eeq
Then, if for example, we set $\kappa = m/2$ in (\ref{oddsigmaqone2bsaltfinal})
we see that it simplifies considerably, and we obtain
\beq\label{deformedkink}
\sigx = m\, {\rm tanh}\, (mx/2)\,,
\eeq
i.e., a deformation of the CCGZ kink $m\,{\rm tanh}\, mx$. It is clearly 
spread out in space as twice as much as the CCGZ kink.

\pagebreak


\newpage
\setcounter{equation}{0}
\setcounter{section}{0}
\renewcommand{\theequation}{C.\arabic{equation}}
\renewcommand{\thesection}{Appendix C:}
\section{Reflectionless $\sigx$'s Are Extremal}
\vskip 5mm
\setcounter{section}{0}
\renewcommand{\thesection}{C}

As a consistency check of our calculations in Section 3, we will verify 
now that vanishing of the right hand side of (\ref{derivativefinal}) guarantees
that the corresponding $\sigx$ is indeed extremal among all possible 
admissible static configurations (i.e., a solution of 
${\del \ce[\si]\over \del\sigx} =0$), and not just an extremum over the 
space of reflectionless configurations:
Substituting $V_b$ and $V_c$ from (\ref{vbvc}) into (\ref{varenergy1}), 
and integrating by parts over $\del\si'(x)$, we obtain  
\beq\label{varenergy2}
\del\ce [\si] = \int dx \,\del\sigx \left\{
{\sigx\over g^2}
+N\,\int{d\om\over 2\pi}\left[{i(2\sigx + \pax)B(x)\over 2\om}
+{i(-2\sigx +\pax)C(x)\over 2\om}\right]\right\}
\,.
\eeq
With the help of (\ref{abcd}) we recognize the two expressions
in the square brackets in (\ref{varenergy2}), respectively, as 
$D(x)$ and $A(x)$, the two diagonal entries of the diagonal resolvent 
(\ref{diagresolventintext}). Thus, from (\ref{varenergy2}) we conclude that 
\beq\label{saddleenergy1}
{\del\ce [\si]\over\del\sigx} = {\sigx\over g^2}
+N\,\int{d\om\over 2\pi}\left(A(x) + D(x)\right)\,.
\eeq
Finally, we may simplify (\ref{saddleenergy1}) further, by invoking 
(\ref{ADeq}), which tells us that $A(x)=D(x)$. Thus, we may write the 
variational derivative as 
\beq\label{saddleenergy}
{\del\ce [\si]\over\del\sigx} = {\sigx\over g^2}
+2N\,\int{d\om\over 2\pi} D(x)\,.
\eeq
As a consistency check, we obtain (\ref{saddleenergy}) in the next subsection 
as the static limit of the general extremum condition (\ref{saddle}).

For later use, let us record here (\ref{saddleenergy}), evaluated 
at the vacuum condensate $\si=\si_{_{VAC}} = \pm m$. Thus, 
\beq
{\si_{_{VAC}}\over g^2} +\,2N~\int{d\om\over 2\pi}\,D_{_{VAC}}= 0\,,
\label{sigap}
\eeq
where $D_{_{VAC}} = {i \si_{_{VAC}}\over 2 \sqrt{m^2-\omega^2}}$ is 
given in (\ref{vac}) in Appendix A. It is a 
trivial matter to verify from (\ref{vac}) and (\ref{saddle}) 
that (\ref{sigap}) is equivalent to the bare gap equation (\ref{bgap}).

Equations (\ref{varenergy2}) - (\ref{saddleenergy}) are general identities,
valid for {\em any} static $\sigx$. We will now evaluate (\ref{saddleenergy})
at a {\em reflectionless} $\sigx$ and show that its right hand side vanishes. 
We start by substituting the relation
$D(x)=i{\left[\pax+2\sigx\right]B\left(x\right)\over 
2\omega}$ (from (\ref{abcd})), and the expression (\ref{Bresolvent}) for 
$B(x)$ into (\ref{saddleenergy}), and obtain 
\beqra\label{checksaddleenergy}
&&{\del\ce [\si]\over\del\sigx} = 
\left({1\over g^2} + 
iN\,\int {d\om\over 2\pi}\,{1\over \sqrt{m^2-\om^2}}\right)\sigx\nonumber\\
{}\nonumber\\
&&-iN\sum_{n=1}^K \kappa_n \left((\pax +2\si)\psi_n^2\right)
\,\int {d\om\over 2\pi}\,{1\over \sqrt{m^2-\om^2}}\,.
{1 \over \om^2 - \om_n^2}
\eeqra
The first term on the right hand side of this equation vanishes due to 
the gap equation (\ref{sigap}) (compare with (\ref{derivative2})). 
As for the sum in (\ref{checksaddleenergy}), we now show that it vanishes 
term by term. One has to be a little bit more careful in case $H_b$ has its 
ground state at zero energy. Consider first all terms in the sum which 
correspond to positive bound state energies $\om_n^2>0$. Due to 
(\ref{derivativefinal}), each one of these terms is proportional to the 
corresponding ${\pa M\over \pa\om_n}$, which vanishes at an extremal $\om_n$.
If $H_b$ has a zero energy ground state, i.e., if $\om_1^2 =0$, its 
contribution to the sum in (\ref{checksaddleenergy}) vanishes also,
but due to the fact that it is proportional to 
$(\pax + 2\sigx)\psi_1^2$. The latter expression vanishes, since in this 
case $\psi_1(x)$ is proportional to $b_0(x) = \exp -\int^x \si(y)dy$ 
in (\ref{bzero}). In any case, the right hand side of 
(\ref{checksaddleenergy}) vanishes. Thus, we have verified
that a reflectionless $\si (x;\om_n,c_n)$, with parameters $\om_n$ that 
satisfy (\ref{derivativefinal}) (and any set of $c_n$'s), is an extremal
configuration of $\ce[\si]$.

\subsection{A Consistency Check of the Static Saddle Point Equation 
(\ref{saddleenergy})}
As a consistency check on (\ref{saddleenergy}), let us verify 
that the saddle point equation ${\del\ce [\si]\over\sigx}=0$, with the 
variational derivative given by 
(\ref{saddleenergy}), is equivalent to the extremum condition (\ref{saddle}),
evaluated at a static configuration\footnote{Note that (\ref{saddle}) is 
obtained by taking a generic space-time dependent variation 
$\delta \si (x,t)$ of the effective action $S_{eff}$ around a 
static $\sigx$, whereas (\ref{saddleenergy}) is obtained from 
varying the energy functional, which is defined, of course, only over 
the space of static $\sigx$ configurations, and thus allows only  
static variations $\del\sigx$.}.

To this end, observe that for static $\sigx$ backgrounds, we have the 
(divergent) formal relation
\beqra\label{divergentrelation}
\langle x,t |{1\over i\notpa -\si } | x,t \rangle 
&=& \int {d\om\over 2\pi} \langle x| {1\over \om\gam^0 + i\gam^1\pax 
-\si } | x \rangle \nonumber\\{}\nonumber\\
&=&i\int{d\om\over 2\pi}\,
\left(\begin{array}{cc} A(x) & B(x) \\{}&{}\\ C(x) & 
D(x)\end{array}\right)
\eeqra
(see Eq. (\ref{condensate})). 
Thus, we may write the (bare) saddle point equation (\ref{saddle}) for 
static bags as 
\beq
{\del S_{\em eff}\over \del \si\left(x,t\right)}_{|_{\rm static}}  =
-{\si\left(x\right)\over g^2} -\,N~\int{d\om\over 2\pi}\,\left(A(x) + D(x)
\right)= 
0~~~
\label{staticsaddle}
\eeq
which is manifestly equivalent to the saddle point condition obtained by 
equating the right hand side of (\ref{saddleenergy1}) to zero.

From (\ref{condensate}) (with $\Gamma={\bf 1}$), we observe that 
\beqra\label{scalarcondensate}
N~\int{d\om\over 2\pi}\,\left(A(x) + D(x)\right) = 
N\int {d\om\over 2\pi }\,
\rmtr\left(\begin{array}{cc} A(x) & 
B(x) \\{}&{}\\ C(x) & 
D(x)\end{array}\right) = 
\langle \bar\psi_a(t,x)\,\psi_a(t,x)\rangle_{|_{\sigx}}\,.
\eeqra
Thus, (\ref{staticsaddle}) is just the statement that 
\beq\label{psibarpsicondensate}
\langle \bar\psi_a(t,x)\,\psi_a(t,x)\rangle_{|_{\sigx}} = 
-{\si\left(x\right)\over g^2}
\eeq
i.e., that the auxiliary field $\si$ in (\ref{auxiliary}) is proportional 
to the condensate $\langle\bar\psi\psi\rangle$, which is nothing but the 
equation of motion of $\si$ with respect to the action 
$S$ in (\ref{auxiliary}).

\pagebreak


\newpage
\setcounter{equation}{0}
\setcounter{section}{0}
\renewcommand{\theequation}{D.\arabic{equation}}
\renewcommand{\thesection}{Appendix D:}
\section{The $O(2N)$ Energy Multiplets}
\vskip 5mm
\setcounter{section}{0}
\renewcommand{\thesection}{D}

\subsection{Quantization of Majorana Fields in a Background $\sigx$} 
The $O(2N)$ quantum numbers of the bound state multiplets associated with 
the static solitons discussed in Sections 3 and 4 can be determined by 
considering the action\footnote{The presentation in  
subsections D.1 and D.2.1 follows, in part, portions of section 3 of 
\cite{wittengn}. For more details on the representation theory of orthogonal
groups see, e.g., \cite{georgi}.}
\beq
S = \int d^2x\,\left\{\sum_{a=1}^N\, \bar\psi_a\,\left(i\notpa-\si
\right)\,\psi_a \right\} 
\label{auxiliaryD}
\eeq
for $N$ non-interacting Dirac fermions in a static background $\sigx$. 
Following (\ref{explicitsymmetry}), we write (\ref{auxiliaryD}) in the 
explicit $O(2N)$ invariant form 
\beq
S = \int d^2x\,
\sum_{i=1}^{2N}\left[i\xi_i^T\dot\xi_i
- i\xi_i^T\si_1\xi_i'
-(\xi_i^T\si_2\xi_i)\sigx\right]\,,
\label{auxiliaryD2N}
\eeq
in terms of the $2N$ Majorana fermions $\xi_i$, where we used the obvious 
notations 
\beq\label{obviousnotations}
\phi_a = \xi_{2a-1},\quad\quad \chi_a = \xi_{2a}
\eeq
in (\ref{majoranaspinor}).

Quanta created by the Dirac-Majorana fields $\xi_i$ may form non-interacting, 
many-fermion bound states in the external background $\sigx$,   
which fall into multiplets of the $O(2N)$ symmetry group. In this appendix 
we study these multiplets. 

In the context of the large-$N$ limit of the GN model, $\sigx$ is 
determined self-consistently from the static saddle point equation 
(\ref{staticsaddle}). The $O(2N)$ multiplet content associated with that 
soliton then follows from the simple group theoretic considerations presented 
here. 

The Dirac-Majorana equation associated with (\ref{auxiliaryD2N}) is 
\beq\label{diracmajoranaeq}
i\dot\xi - i\si_1\xi'-(\si_2\xi)\sigx = 0\,. 
\eeq
Stationary solutions $f_n(x,t) = e^{-i\om_n t} f_n(x)$ of 
(\ref{diracmajoranaeq}) with $\om_n\neq 0$, are complex. However, since 
(\ref{diracmajoranaeq}) has purely imaginary coefficients, the spinors 
${\rm Re} f_n(x,t)$ and ${\rm Im} f_n(x,t)$ comprise a pair of independent 
solutions
of (\ref{diracmajoranaeq}). This is just the statement, in terms of Majorana 
spinors, that the stationary solutions of the Dirac equation (\ref{diraceq}), 
$[i\notpa-\si (x)]\,\psi = 0$, come in charge-conjugate pairs 
$e^{-i\om_n t}f_n(x)$ and $e^{i\om_n t}f_n^*(x)$, as was discussed in the 
Introduction. In addition, if $\sigx$ is topologically non-trivial, as we 
discussed in subsection A.1.1, (\ref{diracmajoranaeq}) has a non-degenerate 
real zero energy eigenstate $f_0(x)$ \cite{jackiwrebbi}. Thus, we need only 
consider stationary solutions of (\ref{diracmajoranaeq}) with $\om\geq 0$. 

From (\ref{auxiliaryD2N}) we see that the quantum Dirac-Majorana field 
operators $\xi_i(x,t), i=1, \ldots
2N$ satisfy the equal time canonical anticommutation relations 
\beq\label{etcar}
\left\{\xi_{i\alpha}(x,t), \xi_{j\beta}(x',t)\right\} = \frac{1}{2}\delta_{ij}
\delta_{\alpha\beta}\delta (x-x')\,,
\eeq
where $\alpha, \beta$ are spinor indices. As usual, these (hermitean) 
operators are to be expanded \cite{wittengn} in normal modes of 
(\ref{diracmajoranaeq}). This expansion, which is exact for the simple field
theory (\ref{auxiliaryD}), is the semiclassical expansion of the 
fermion field in a self-consistent static background in the GN model. For 
example, in the presence of a topologically non-trivial $\sigx$, with its 
unpaired real zero energy bound state spinor $f_0(x)$, we have the expansion
\footnote{For simplicity, we assume
here that the system lives in a big spatial box, rendering its spectrum 
discrete.} 
\beq\label{expansion}
\xi_{i\alpha}(x,t) = f_{0\alpha} (x) b_i + 
\sum_{\om_n>0}\left\{f_{n\alpha}(x,t) a_{n,i} + 
f_{n\alpha}^*(x,t) a_{n,i}^\dgg\right\}\,.
\eeq
(For topologically trivial backgrounds, 
we will obtain an expansion similar to (\ref{expansion}), with the first 
term $f_{0\alpha} (x) b_i$ excluded.)

In (\ref{expansion}) $a_{n,i}$ and $a_{n,i}^\dgg$ are,
respectively, annihilation and creation operators for a particle of type $i$ 
in the state corresponding to $f_n(x,t)$. Indeed, from (\ref{etcar}) (with
the appropriate normalization of the complete set of stationary states 
$f_n$) one obtains the anticommutation relations 
\beq\label{aacr}
\{a_{n,i}, a_{m,j}^\dgg\} = \delta_{nm}\delta_{ij}\,,
\eeq
and all other anticommutators involving $a$'s or $a^\dgg$'s vanish.

The operators $b_i$ are hermitean and transform according to the vector 
representation of $O(2N)$, as they inherit these properties from the 
$\xi_i$. Moreover, the canonical anticommutation relations (\ref{etcar}) 
(with the appropriate normalization of $f_0$) imply that the $b_i$ satisfy 
the Clifford algebra 
\beq\label{aacr}
\{b_i, b_j\} = 2\delta_{i,j}\,.
\eeq
In other words, the $b_i$ have the transformation properties and the 
anticommutation relations of the gamma matrices of $O(2N)$. 

In terms of the creation and annihilation operators discussed above, the 
(normal ordered) Hamiltonian corresponding to (\ref{auxiliaryD2N}) is 
\beq\label{majoranahamiltonian}
H = \sum_{\om_n>0}\,\om_n \,a_{n,i}^\dgg a_{n,i}\,.
\eeq

\subsection{The $O(2N)$ Multiplets}
\subsubsection{The Spinor Representation}
The multiplet of states associated with topologically non-trivial solitons, 
on which the Clifford generators $b_i$ act, must contain a factor 
transforming according to the spinor representation of $O(2N)$. Since 
according to our discussion in subsection A.1.1 
(and according to \cite{jackiwrebbi}), the Dirac-Majorana equation 
(\ref{diracmajoranaeq}) with a topologically non-trivial $\sigx$ has only 
a single zero energy bound state $f_0(x)$, there is only one such factor of
the spinor representation associated with a given topologically non-trivial 
$\sigx$. For example, CCGZ kinks, for which (\ref{diracmajoranaeq}) has 
only a zero energy bound state, are pure isospinors\cite{wittengn}. 

It will be useful at this point to list a few mathematical facts 
concerning this $O(2N)$ spinor representation factor: In any representation
of (\ref{aacr}) we may define the antihermitean $O(2N)$ generators as 
\beq\label{O2Ngenerators}
M_{ij} = b_ib_j - b_jb_i\,. 
\eeq
In addition, let us define the hermitean operator 
$\Gam_5 = ib_1b_2\ldots b_{2N}$, which commutes with all group generators 
$M_{ij}$, but anticommutes with all generators of the Clifford
algebra $b_i$. Clearly, $\Gam_5^2 = 1$, and thus the possible eigenvalues of 
$\Gam_5$ are $\pm 1$. There is only a single irreducible representation of 
the {\em Clifford} algebra (\ref{aacr}), since $\Gam_5$ does not commute 
with the $b_i$, and any representation of the Clifford algebra (\ref{aacr}) 
must contain both ``left-handed'' isospinors with $\Gam_5=+1$, and 
``right-handed'' isospinors with $\Gam_5=-1$.

We would like to construct a unitary irreducible representation of this 
Clifford algebra on a positive norm Fock space. Thus, we have to group the 
Clifford generators $b_i$ into $N$ pairs of Grassmannian creation and 
annihilation operators which act on that space. This we achieve e.g., by 
pairing the $b_i$ according to (\ref{obviousnotations}), namely 
\beq\label{UNFockgenerators}
B_a = \frac{1}{2}(b_{2a-1} + ib_{2a})\,,\quad a=1,\ldots N\,,
\eeq
which satisfy the anticommutation relations 
\beq\label{UNcanonical}
\{B_a , B_b^\dgg\} =  \delta_{ab}\,,
\eeq
with all other anticommutators vanishing.  
An orthonormal basis for this space is then 
obtained by applying the creation operators $B_a^\dgg$ repeatedly on the 
Fock vacuum:
\beq\label{Fockbasis}
|0\rangle ,~ B_a^\dgg |0\rangle ,~ B_a^\dgg B_b^\dgg |0\rangle ,~ \ldots  
,~ B_1^\dgg B_2^\dgg \cdots B_N^\dgg |0\rangle\,.
\eeq
Thus, this irreducible 
representation of the Clifford algebra has dimension $2^N$.

However, since $\Gam_5$ commutes with all the $O(2N)$ generators $M_{ij}$,
these $2^N$ states transform reducibly under the $O(2N)$ group; the 
left-handed and right-handed states transform independently. For example, if 
we choose the Fock vacuum such that $\Gam_5 |0\rangle  = +|0\rangle $, then 
the right-handed states are those with an even number of $B_a^\dgg$'s acting 
on $|0\rangle $, and the left-handed states are those with an odd number of 
$B_a^\dgg$'s acting on $|0\rangle $. For later reference, let us demonstrate 
this reducibility in a concrete basis of the algebra, in which the $O(2N)$ 
Cartan subalgebra is generated by the $N$ generators 
\beq\label{Cartangenerators}
M_{12}, ~M_{34},~ \ldots ~M_{(2N-1)(2N)}\,.
\eeq
From (\ref{UNFockgenerators}) we see that $M_{(2a-1)(2a)} = 
2 b_{2a-1} b_{2a} = 
-2i [B_a^\dgg, B_a]$. Thus, the states (\ref{Fockbasis}) are simultaneous  
eigenstates of all the generators (\ref{Cartangenerators}), and the 
corresponding eigenvalues are essentially the components of the weight 
vectors of the representation. The step operators, on the other
hand, when acting on any of these states, change the number of Fock quanta 
of that state by $0$ or $\pm 2$ units (consider, e.g., 
$M_{13} = 2b_1b_3 = 2 (B_1 + B_1^\dgg) (B_2 + B_2^\dgg))$. In other words, 
the states with an even number of $B_a^\dgg$'s acting on $|0\rangle $ and the 
states with an odd number of $B_a^\dgg$'s acting on $|0\rangle $ form two 
disjoint (and irreducible) invariant subspaces under the action of the $O(2N)$ 
generators.

\subsubsection{The Antisymmetric Tensor Representations}
Suppose now that (\ref{diracmajoranaeq}), in a given solitonic background 
$\sigx$, has a pair of bound states at $\pm\om_n$. 
What are the irreducible $O(2N)$ factors in the multiplet of states
born by the soliton $\sigx$, which are acted upon by the operators $a_{n,i}$ 
and $a_{n,i}^\dgg$? To answer this question consider the following 
situation: Assume for simplicity that $\pm\om_n$ are the only bound states 
in (\ref{diracmajoranaeq}). Then choose $k$ {\em distinct} operators among the 
$2N$ $\xi_i$'s, and apply them to the vacuum state $|0\rangle $. The resulting 
state $\prod_{j=1}^{k}\xi_{i_j} |0\rangle $ will contain bound as well as 
scattering states.
Since we are interested only in static stable soliton states, we have to 
consider only those $a_{l,i}$'s which correspond to {\em bound} states
of (\ref{diracmajoranaeq}). Thus, projecting out all scattering states we 
are left with the state $(f_n^*)^k\,a_{n,i_1}^\dgg\ldots a_{n,i_k}^\dgg 
|0\rangle $ (where $(f_n^*)^k$ stands for the direct product of the k 
spinors). Clearly,
we can form $(2N)!/k!(2N-k)!$ bound states in this way, all of which 
have common energy $k\om_n$, and thus form an energy multiplet. We readily 
identify this multiplet as the antisymmetric tensor of $O(2N)$ of rank 
$k$, since all $ a_{n,i}^\dgg$ anticommute. 
In this construction, we can obviously take $1\leq k \leq 2N$, and thus 
obtain all $O(2N)$ antisymmetric tensors. However, in the context of 
studying extremal static $\sigx$ configurations in the GN model, the number
$\nu_n$ of quanta trapped in the bound states $\pm \om_n$, is part of the 
definition of $\sigx$, in which case one must take $k=\nu_n$.

For pedagogical clarity, let us derive this conclusion in a more pedestrian
way. To this end we have to invoke the notions of particles and antiparticles.
In order to differentiate ``particles'' from ``antiparticles'' in our model 
we have to pair the $2N$ Majorana fields in (\ref{expansion}) into $N$ Dirac 
fields $\psi_a$. (The fermion number operator is of course 
$\int\,dx\,:\sum_a\psi_a^\dgg\psi_a:\,.$) One possible pairing is according to 
(\ref{obviousnotations}), namely 
$\psi_a = \xi_{2a-1} + i\xi_{2a}, a=1,\ldots N$\footnote{This is by no means 
the only possible pairing. There is clearly a one to one correspondence 
between permutations of the $2N$ fields $\xi_i$ and their pairings into 
$N$ Dirac fermions, and thus $(2N)!$ ways to pair the $2N$ Majorana fields 
into $N$ Dirac fields, namely,  
$\psi_a = \xi_{P(2a-1)} + i\xi_{P(2a)} = f_0 (b_{P(2a-1)} + i b_{P(2a)}) +
\sum_{\om_n>0}\left\{f_n (a_{n,P(2a-1)} + ia_{n,P(2a)})
+ f_n^* (a_{n,P(2a-1)} - ia_{n,P(2a)})^\dgg\right\}\,,$
where $P$ is a permutation of $2N$ objects. Then clearly, the operator 
$a_{n,P(2a-1)} + ia_{n,P(2a)}$ may be thought of as annihilating a 
{\em particle} in the state $f_n(x,t)$, and the operator
$(a_{n,P(2a-1)} - ia_{n,P(2a)})^\dgg$ may be thought of as creating 
an {\em antiparticle} in the state $f_n^*(x,t)$. However, since the symmetry 
of the system in question is $O(2N)$ (rather than $U(N)$), all these 
pairings are physically equivalent. They simply correspond to different
orientations of the corresponding $U(N)$ subgroup in $O(2N)$. Thus, with no 
loss of generality, we choose to pair according to (\ref{diracops}), 
and define the notions of particles and antiparticles (or holes) 
accordingly.}, which gives rise to the Dirac field operators 
\beqra\label{diracops}
&&\psi_a = \xi_{2a-1} + i\xi_{2a} = f_0 (x) (b_{2a-1} + i b_{2a}) +
\nonumber\\
&& \sum_{\om_n>0}\left\{f_n(x,t) (a_{n,2a-1} + ia_{n,2a})
+ f_n^*(x,t) (a_{n,2a-1} - ia_{n,2a})^\dgg\right\}\,.
\eeqra

Suppose that the Dirac-Majorana equation (\ref{diracmajoranaeq})
has a pair of charge conjugate bound states at $\pm\om_n$. 
Consider the $(p_n,h_n)$ many-body configuration of $p_n$ particles and $h_n$ 
holes defined following (\ref{integral}). The energy of this state is simply 
$(p_n + h_n -N)\om_n$. It depends on the sum $\nu_n = p_n+h_n$ of numbers of 
particles and holes, and not on each one of them separately. Evidently, 
Fermi statistics implies that $0\leq \nu_n \leq 2N$. Thus, considering the 
set of all $(p_n,h_n)$ configurations with $\nu_n = p_n+h_n$ fixed, we 
obtain a multiplet  of $\sum_{k=0}^{\nu_n} C_N^k C_N^{\nu_n-k} = 
C^{\nu_n}_{2N} = (2N)!/\nu_n! (2N-\nu_n)!$ states with common energy 
$(\nu_n -N)\om_n$. This is, of course, precisely 
the result we obtained earlier directly from the $O(2N)$ symmetry,
without distinguishing between particles and antiparticles. The many-body 
wave functions of states in this multiplet are Slater determinants of 
$\nu_n$ single particle orbitals, with $2N$ possible orbitals (i.e., with a 
flavor index running through $1,\ldots 2N$). Thus, we readily identify this 
multiplet as the antisymmetric tensor representation of rank $\nu_n$ of 
$O(2N)$. Since $0\leq \nu_n \leq 2N$, all possible $O(2N)$ antisymmetric 
tensor multiplets may occur. Thus, in particular, both the antisymmetric 
tensors of rank $\nu_n$, and its dual tensor, of rank  $2N-\nu_n$, appear
in the spectrum. These two tensor multiplets have the same dimension and 
opposite energies $\pm(\nu_n -N)\om_n$. Note, however, that the extremal 
fermion bags, discussed in subsection 3.1, realize only half the 
antisymmetric tensors allowed by the $O(2N)$ symmetry, namely, only 
tensors of ranks in the range $0\leq \nu_n \leq N$ (recall (\ref{nurange})).

Referring, as usual, to the Dirac particles occupying the state at 
$+\om_n$ as fermions and to the Dirac holes in the state at $-\om_n$ as 
antifermions, then the $(p_n,h_n)$ configuration would 
carry fermion number $N_f^{(n)} =  p_n -h_n$. This is what we referred to 
as the {\em valence} fermion number $N_{f,val}^{(n)}$ in (\ref{valenceNfn}). 
Thus, states in our antisymmetric tensor multiplet carry different fermion 
numbers $N_f^{(n)} =  p_n -h_n = \nu_n - 2h_n$, which vary in the range 
$-\nu_n\leq N_f^{(n)} \leq \nu_n$ (if $\nu_n\leq N$), or  
$-(2N-\nu_n)\leq N_f^{(n)} \leq 2N- \nu_n$ (if $\nu_n > N$). Thus, the 
$N_f^{(n)}$ spectrum in the antisymmetric multiplet of rank $\nu_n$ coincides
with the $N_f^{(n)}$ spectrum in the dual antisymmetric multiplet of rank 
$2N - \nu_n$.

Similarly, according to our explicit construction of the 
spinorial representation in the previous subsection, we see that the states 
in that representation carry different fermion numbers in the range 
$0\leq N_f^{({\rm spinor})}\leq N$; just count the number of $B_a^\dgg$'s in a
given state. (In subsection 3.2 we referred to this quantity as the 
{\em valence} fermion number $n_0$.)
However, this is an intolerable assignment of fermion numbers in a 
multiplet appearing in the spectrum of a quantum field theory with charge 
conjugation invariance, such as the GN model. One may say that the theory 
preserves its invariance under charge conjugation through the phenomenon of 
fermion number fractionalization \cite{jackiwrebbi,niemi}, which shifts, as we
discussed in subsection 3.2 (recall (\ref{Nfzeromode})), the valence fermion 
numbers in the spinorial representation by $-N/2$, rendering the spectrum of 
fermion numbers in that representation symmetric: 
$-N/2 \leq N_f^{({\rm spinor})}\leq N/2$.
More formally, we can identify the fermion number operator 
$N_f^{({\rm spinor})}$ as the 
linear symmetric combination \cite{georgi} 
\beq\label{Nfoperator}
N_f^{({\rm spinor})} = {i\over 4} \sum_{a=1}^N M_{(2a-1)(2a)} = 
\sum_{a=1}^N B_a^\dgg B_a - {N\over 2}
\eeq
of the generators (\ref{Cartangenerators}) of the Cartan subalgebra of 
$O(2N)$, where we used (\ref{UNcanonical}). The splitting of 
$N_f^{({\rm spinor})}$ into valence and fractional parts is manifest.

\subsection{Multiplet Dimensions from the Partition Function}
The dimensions of degenerate energy multiplets occurring in the quantization 
of (\ref{auxiliaryD2N}) can be read off the partition function $Z(\beta) = 
{\rm Tr}\,\exp -\beta H $ in a straightforward manner. We consider here a 
topologically non-trivial $\sigx$, so that all  possible multiplets
contribute. The Hamiltonian 
(\ref{majoranahamiltonian}) is the sum of contributions of non-interacting
Grassmannian oscillators, and the partition function is a product over the 
contributions of individual modes. The modes at frequency $\om_n$  
contribute the factor $\prod_i{\rm Tr} \exp -\beta\om_n 
\,a_{n,i}^\dgg a_{n,i} = (1 + \exp -\beta\om_n)^{2N}$ to the partition 
function. The Fock space associated with the zero mode is $2^N$ dimensional, 
and thus contributes a factor $2^N$ to the partition function.

Gathering all these facts together, we thus obtain the formal expression
\beqra\label{partitionfunction}
Z(\beta) &=& 2^N\,\prod_{\om_n>0} \,(1 + e^{-\beta\om_n})^{2N}\nonumber\\
&=&  2^N\,\prod_{\om_n>0} \,\left( \sum_{\nu_n=0}^{2N} {(2N)!\over \nu_n! 
(2N-\nu_n)!}\, e^{-\nu_n\beta\om_n}\right)\,. 
\eeqra
Thus, a state in which there are $\nu_n$ quanta at frequency $\om_n$ 
contributes a factor $C^{\nu_n}_{2N} = (2N)!/\nu_n! (2N-\nu_n)!$ to the 
overall degeneracy. This is, of course, the dimension of the antisymmetric 
tensor multiplet of rank $\nu_n$. From the expansion of 
(\ref{partitionfunction}),
\beq\label{expandpartition}
Z(\beta) = \sum_{\{\nu_k\}} \left(2^N\,\prod_{\om_n>0} \,C^{\nu_n}_{2N}\right)
\,e^{-\beta\sum_{\om_l>0} \nu_l\om_l}\,,
\eeq
where the summation runs over sets of integers $0\leq \nu_k\leq 2N$, 
we see that the eigenstates of the Hamiltonian (\ref{majoranahamiltonian}) 
fall into all possible direct products of the $O(2N)$ antisymmetric tensors 
and the spinorial representation, 
which are built on the normal modes in the expansion (\ref{expansion}) of 
$\xi$.

It is somewhat amusing to obtain these results in the language of Dirac 
fields and the $U(N)$ group. In this language it is more convenient to 
consider the Hamiltonian without normal ordering, i.e., to include the zero
point energy (which we can subtract in the end). Thus, consider one of the 
$N$-fold degenerate normal modes at frequency $+\om_n$. If that state is 
empty the energy is $-\frac{\om_n}{2}$ (i.e., the zero point energy), 
and if it is occupied the energy is $+\frac{\om_n}{2}$. 
Thus, these modes contribute a factor $(\exp\frac{\beta\om_n}{2} + \exp 
-\frac{\beta\om_n}{2})^N = (2\cosh\frac{\beta\om_n}{2})^N $ to the partition 
function. The modes at $-\om_n$ 
obviously contribute the same factor. Thus, the states at frequency 
$\pm\om_n$ contribute altogether a factor $(2\cosh\frac{\beta\om_n}{2})^{2N} = 
(1 + \exp -\beta\om_n)^{2N}\,\exp N\beta\om_n$. The $N$-fold degenerate 
zero mode contributes a factor $(2\cosh\beta\cdot 0)^N = 2^N $,
which is, of course, the dimension of the spinor representation. Multiplying 
all these factors together, and subtracting the overall divergent contribution 
of the zero point energies, we obtain (\ref{partitionfunction}).

\pagebreak


\newpage
\setcounter{equation}{0}
\setcounter{section}{0}
\renewcommand{\theequation}{E.\arabic{equation}}
\renewcommand{\thesection}{Appendix E:}
\section{Derivation of the Mass Formula for the CCGZ Kink}
\vskip 5mm
\setcounter{section}{0}
\renewcommand{\thesection}{E}

In this Appendix we provide our own derivation of the mass formula 
$M_{kink} = {Nm\over\pi}$ (Eq. (\ref{massccgz})) of the CCGZ kink. The 
kink mass, as all other soliton masses in the GN model, ought to be 
proportional to $N$. In addition, due to dimensional considerations, it has 
to be proportional to the dynamical mass $m$. Thus, on very general grounds, 
\beq\label{mkink1}
M_{kink} = Nmc\,,
\eeq
where $c$ is a dimensionless pure number to be determined. In particular,
$c$ cannot depend on $m$. Therefore, 
\beq\label{cequation}
c = {1\over N} {\pa M_{kink}\over \pa m}\,.
\eeq
$M_{kink}$ is given by the renormalized energy functional (\ref{sigenergy}), 
evaluated at $\si_{kink} (x)  = m\,\tanh\,mx$.
Note that due to the subtracted vacuum term in (\ref{sigenergy}), 
that functional is also a function of $\si_{VAC} = m$, and of course, also 
of the bare coupling $g^2$. 
Thus, 
\beq\label{mkinkderivative}
{\pa M_{kink}
\over \pa m} = \int\limits_{-\infty}^\infty dx {\delta \ce 
\over \delta \sigx} {\pa \si_{kink} (x)\over \pa m} + 
{\pa \ce\over \pa \si_{VAC}} {\pa \si_{VAC} \over \pa m} + 
{\pa \ce\over \pa g^2} {\pa g^2\over \pa m}\,,
\eeq
where all derivatives of $\ce$ are evaluated at $\sigx = \si_{kink}(x)$. 
Now, $\si_{kink}(x)$ and $\si_{VAC}$ are extremal configurations. Thus, 
by definition\footnote{This simple argument can be verified by 
substituting the kink resolvents $B$ and $C$ (\ref{kinkresolvents})
into (\ref{varenergy1}), and by using the fact that 
$V_c = \si^2 + \si' = m^2 = \si_{VAC}^2$ for the kink. Then, the sum of the 
first two terms on the right hand side of (\ref{mkinkderivative}) produces a 
convergent integral in $x$, which is proportional to the bare gap equation 
(\ref{sigap}), and thus vanishes.}, $(\delta \ce /\delta \sigx)_{\si_{kink} 
(x)} = \pa \ce /\pa \si_{VAC} = 0$. It follows from this and from 
(\ref{sigenergy}) that 
\beq\label{mkinkderivative1}
{\pa M_{kink}\over \pa m} = {m\over g^4} {\pa g^2\over \pa m} = 
{1\over g^2} {\pa \log g^2\over \pa \log m}\,,
\eeq
where we used $ \int\limits_{-\infty}^\infty dx \,(\si_{kink}^2 - m^2) = -2m$.

What is the meaning of taking the derivative of the bare coupling constant 
$g^2$ with respect to the physically observable dynamical fermion mass $m$? 
The answer is that in order to produce the RG invariant mass $m$ at the 
large distance scale, we have to tune the bare coupling $g^2$ according to 
(\ref{mass}), namely, $\Lambda\,e^{-{\pi\over Ng^2\left(\Lambda\right)}} = m$.
The RG invariant mass $m$ thus parametrizes the RG trajectory of the bare 
coupling $g^2$ as a function of the cutoff $\Lambda$. Therefore, 
$(\pa g^2 / \pa m)\delta m$ measures the change in $g^2$ as we move to a
neighboring trajectory, keeping $\Lambda$ fixed. Thus, substituting 
$\pa \log g^2 / \pa \log m $ from (\ref{mass}) into (\ref{mkinkderivative1}),
followed by substituting the resulting $\pa M_{kink}/\pa m$ into 
(\ref{cequation}), we finally obtain 
\beq\label{cequationfinal}
c={1\over \pi},
\eeq
thereby proving (\ref{massccgz}).

\pagebreak


\newpage
\setcounter{equation}{0}
\setcounter{section}{0}
\renewcommand{\theequation}{F.\arabic{equation}}
\renewcommand{\thesection}{Appendix F:}
\section{An Alternative Proof of Eq. (\ref{potentiallystable})}
\vskip 5mm
\setcounter{section}{0}
\renewcommand{\thesection}{F}
In this Appendix we present an alternative proof of (\ref{potentiallystable}).
The idea is to consider the behavior of 
\beq\label{sinesum}
f(\Theta_1, \ldots, \Theta_K) = \sum_{n=1}^K \sin\Theta_n
\eeq
over the hyperplane 
\beq\label{constantNfmax}
\tilde\Sigma_{r\alpha}: \quad\quad\quad\quad \Th_1+\cdots + 
\Th_K = \alpha + \hp r\,,
\eeq
where $0\leq r\leq K$ is an integer, and $0\leq \alpha <\hp$. This hyperplane 
corresponds, of course, to solitons with a fixed value of the maximal 
fermion number $N_f^{max}(D_{\rm parent})$.

We will now prove that $f(\Theta_1, \ldots, \Theta_K)$ attains its absolute 
minimum on $\tilde\Sigma_{r\alpha}$ in the positive orthant, at the 
vertices of the intersection of $\tilde\Sigma_{r\alpha}$ and the hypercube 
$[0,\hp]^K$, namely, the points
\beq\label{Nfintersectionvertices}
\Theta_n^{(v)} = {\pi\over 2} \left( \delta_{nn_1} + \delta_{nn_2} + \ldots + 
\delta_{nn_r}\right) + \alpha\delta_{nn_{r+1}}\,,\quad (n=1, \ldots, K)\,,
\eeq
with all possible choices of $r+1$ coordinates $i_1, \ldots, i_{r+1}$ out of 
$K$. Once we have established that, the fact that the potentially stable 
solitons are given by (\ref{potentiallystable}) follows in a straightforward 
manner.

By standard constrained extremum analysis one can show that the local 
extremum of $f(\Theta_1, \ldots, \Theta_K)$ occurs at the symmetric point 
$\Th_n = {1\over K}(\alpha + \hp r), \forall n$, and that it is a maximum. 
Thus the minimum occurs at the boundary of the domain under
consideration. By the same argument, any local extremum on the boundary is 
again a maximum. Thus, repeating this analysis, we conclude that the 
absolute minimum occurs at the vertices 
(\ref{Nfintersectionvertices}) of the intersection of 
$\tilde\Sigma_{r\alpha}$ and the hypercube $[0,\hp]^K$, where\footnote{For 
$r=0$, (\ref{minvalue}) reduces to the well known inequality
$\sin (\sum_{n=1}^K \Theta_n) \leq \sum_{n=1}^K \sin\Theta_n$ where 
$\sum_{n=1}^K \Theta_n = \alpha <\hp$.} 
\beq\label{minvalue}
\min f_{|_{\tilde\Sigma_{r\alpha}}} = r  + \sin\alpha\,.
\eeq

An alternative elegant proof of this minimum behavior of 
(\ref{sinesum}), due to Raphael Yuster, goes as follows: Consider a sequence
$0\leq\Theta_1\leq \ldots \leq\Theta_K\leq\hp$, subjected to 
(\ref{constantNfmax}). Assume that for some $i$, 
$0 < \Th_i\leq \Th_{i+1} < \hp$. We will show that 
there exists another sequence of $\Th$'s, with the same sum, but with a 
lower sum of the sines. Thus, let  $\delta > 0$ be chosen such that 
$\Th_i - \delta > 0$ and $\Th_{i+1} + \delta < \hp$, i.e., $0\leq\delta\leq
\min\{\Th_i, \hp - \Th_{i+1}\}$. Modify the sequence under consideration 
by replacing $\Th_i$ by $\Th_i - \delta$ and $\Th_{i+1}$ by 
$\Th_{i+1} + \delta$, keeping the other $K-2$ terms unaltered. 
The new sequence thus obtained has the same sum as the original 
sequence, and thus defines another point on $\tilde\Sigma_{r\alpha}$.
We must show that $D(\delta) = f({\rm original~sequence})  - 
f({\rm new~sequence}) > 0$. Indeed, $D(\delta) =
\sin\Th_i - \sin (\Th_i - \delta) + \sin\Th_{i+1} - \sin (\Th_{i+1} + 
\delta)$. Clearly, $D(0) = 0$, and also $D\,'(\delta) > 0 $ in the relevant 
range of $\delta$. Thus, $D(\delta)$ increases monotonically with 
$\delta$, and reaches its maximum at $\delta_{\rm max} = 
\min\{\Th_i, \hp - \Th_{i+1}\}$, where, depending on the initial 
condition at $\delta = 0$, either $\Th_i - \delta_{\rm max} =0$ or 
$\Th_{i+1} + \delta_{\rm max} =\hp$. Thus, the sequence of $\Th$'s 
constrained to $\tilde\Sigma_{r\alpha}$, which minimizes (\ref{sinesum}), 
cannot have more than one element in the interior of $[0,\hp]$. Thus, due to 
(\ref{constantNfmax}), the absolute minimum is the sequence in which 
the $r$ largest $\Theta$'s are $\hp$, one $\Theta$ is $\alpha$ and the rest 
are zero, namely, the vertices (\ref{Nfintersectionvertices}).

Thus, a parent soliton corresponding to a point in the {\em interior} 
of the intersection of $\tilde\Sigma_{r\alpha}$ and the hypercube $[0,\hp]^K$,
can decay into a final state with quantum numbers corresponding to the points
(\ref{Nfintersectionvertices}), i.e., $L=K$ and $\th_n = \Th_n^{(v)}$ in 
(\ref{1KintoL}). In fact, such a parent soliton can also decay at least 
into the set of final states contained in small pockets above the vertices 
(\ref{Nfintersectionvertices}), which correspond to $L=K$ and 
$\th_n = \Th_n^{(v)} + \epsilon$ in (\ref{1KintoL}), with 
$$\epsilon << {\sum_{n=1}^K \sin\Theta_n - (r + \sin\alpha)\over 
K - r -1 + \cos\alpha}\,,$$ or into final states corresponding to $L>K$ 
in (\ref{1KintoL}), with $\th_i = \Th_i^{(v)}$ for $1\leq i\leq K$ and 
$\th_i = \epsilon$ for $K+1\leq i\leq L$, where 
$$\sin\epsilon < {\sum_{n=1}^K \sin\Theta_n - (r + \sin\alpha)\over 
L - K}\,.$$ 

On the other hand, the parent soliton which corresponds to the vertices 
(\ref{Nfintersectionvertices}) has no open channel to decay through. Hence it
is potentially stable. Indeed, if it could decay through a channel 
corresponding to $\th_1, \ldots, \th_L$, then, from the requirement that 
these parameters satisfy (\ref{1KintoL}), we would have 
\beqra\label{1KintoLappendix}
\sum_{i=1}^L \th_i &>& \alpha + \hp r \nonumber\\
\sum_{i=1}^L \sin\th_i &<& r + \sin\alpha\,.
\eeqra
Define the hyperplane 
\beq\label{hyperplaneappendix}
\Sigma_{r\alpha}: \quad\quad\quad\quad \th_1+\cdots + 
\th_L = \alpha + \hp r\,.
\eeq
From the analysis in this Appendix we know that the absolute minimum of 
$\sum_{i=1}^L \sin\th_i $ over the intersection of $\Sigma_{r\alpha}$ and the
hypercube $[0,\hp]^L$ is $r+\sin\alpha$. Thus, the points which satisfy the
second inequality in (\ref{1KintoLappendix}) are bounded by 
$\sum_{i=1}^L \th_i < \alpha + \hp r $, in contradiction with the first
inequality in (\ref{1KintoLappendix}). This completes our alternative 
proof of (\ref{potentiallystable}).

\pagebreak

{\bf Acknowledgments}~~~ 
I would like to thank Yitzhak Frishman, Marek Karliner, Yael Shadmi and 
Raphy Yuster for reading some parts of this Review. I am indebted to Yael 
Shadmi for making useful comments concerning Appendix D, and to 
Raphy Yuster for providing the alternative proof in Appendix F. This research 
has been supported in part by the Israeli Science Foundation.

\end{document}